\documentclass[11pt]{article}
\pdfoutput=1
\usepackage{putex}
\usepackage{comment}
\usepackage{graphicx}
\usepackage{caption}
\usepackage{amsmath}
\usepackage{array}
\usepackage{subcaption}
\usepackage{epstopdf}
\usepackage{enumerate}
\usepackage{cite}
\usepackage{youngtab}
\usepackage{tensor}
\usepackage{slashed}
\usepackage[export]{adjustbox}
\usepackage[aligntableaux=center]{ytableau}
\usepackage[utf8]{inputenc}
\usepackage{footmisc}
\usepackage[
      colorlinks=true,
      linkcolor=blue,
      urlcolor=blue,
      filecolor=black,
      citecolor=red,
      ]{hyperref}
\usepackage{cleveref}

\usepackage{tikz}

\usepackage{youngtab}

\newcommand{\RNum}[1]{\uppercase\expandafter{\romannumeral #1\relax}}

\newcommand {\be} {\begin {equation}}
\newcommand {\ee} {\end {equation}}

\newcommand {\bes} {\begin {equation*}}
\newcommand {\ees} {\end {equation*}}

\newcommand{\beq}{\begin{equation}}
\newcommand{\eeq}{\end{equation}}

\def\eqref#1{(\ref{#1})}

\def\ie{\begin{equation}\begin{aligned}}
\def\fe{\end{aligned}\end{equation}}

\newcommand{\ket}[1]{|#1\rangle}
\newcommand{\braket}[1]{\langle#1\rangle}

\newcommand{\half}{\frac{1}{2}}
 
\usepackage{esint}

\numberwithin{equation}{section}

\def\<{\langle}
\def\>{\rangle}

\def\comma{\,,}
\def\period{\,.}
\def\fn#1{\footnote{#1}}

\begin{document}
\preprint{PUPT-2646\\CERN-TH-2023-156}
\institution{PU}{Joseph Henry Laboratories, Princeton University, Princeton, NJ 08544, USA}
\institution{CERN}{Department of Theoretical Physics, CERN, 1211 Meyrin, Switzerland
}

\title{\huge Boundary reparametrizations and six-point functions \\ on the AdS$_2$ string}
\authors{Simone Giombi\worksat{\PU}, Shota Komatsu\worksat{\CERN}, Bendeguz Offertaler\worksat{\PU}, Jieru Shan\worksat{\PU}}
\abstract{We compute the tree-level connected six-point function of identical scalar fluctuations of the AdS$_2$ string worldsheet dual to the half-BPS Wilson line in planar ${\cal N}=4$ Super Yang-Mills. 
The calculation can be carried out analytically in the conformal gauge approach, where the boundary reparametrization mode of the string plays a crucial role. We also study the analytic continuation of the six-point function to an out-of-time-order configuration, which is related to a 3-to-3 scattering amplitude in flat space. As a check of our results, we also numerically compute the six-point function using the Nambu-Goto action in static gauge, finding agreement with the conformal gauge answer.}

\maketitle

\tableofcontents

\onehalfspacing
\section{Introduction}

In a conformal field theory (CFT), correlation functions of local operators are fully determined by the CFT data. Consequently, the most commonly studied correlators in CFTs are two- and three-point functions, which define the scaling dimensions and operator product expansion (OPE) coefficients that together constitute the CFT data, and four-point functions, which obey crossing relations that constrain the CFT data and form the basis of the conformal bootstrap. In comparison, higher-point functions have been explored relatively little. On the one hand, this is understandable since the higher-point functions involve extra complexity (e.g., they are functions of more independent cross-ratios of the positions of the external operators), while in principle not providing new information 
not already contained in the full set of lower-point functions. On the other hand, the {\it full set} of lower-point functions is usually out of reach, and hence one expects that higher-point functions are nonetheless useful observables that efficiently repackage certain information about the CFT. Recent explorations along these lines include studies of higher-point out-of-time-order correlators (OTOCs) as more refined diagnostics of quantum chaos \cite{Haehl:2017pak,Haehl:2017qfl,Haehl:2021tft}, studies of the decomposition of higher-point functions into conformal blocks of different topologies (see, e.g., \cite{Rosenhaus:2018zqn,Fortin:2019zkm,Fortin:2020bfq,Fortin:2020yjz,Buric:2020dyz,Poland:2021xjs}), and studies of structure constants of multiple spinning operators through the bootstrap analysis of higher point functions\cite{Bercini:2020msp,Antunes:2021kmm,Kaviraj:2022wbw,Poland:2023vpn}.

In this work, we study a certain six-point function in the Wilson line defect CFT in $\mathcal{N}=4$ super Yang-Mills (SYM). The half-BPS Wilson line in $\mathcal{N}=4$ SYM defines a rich example of a one-dimensional conformal defect that has been studied extensively in recent years using a variety of non-perturbative methods, including AdS/CFT \cite{Giombi:2017cqn}, integrability \cite{Drukker:2012de,Correa:2012hh,Kiryu:2018phb,Grabner:2020nis,Cavaglia:2021bnz,Cavaglia:2022qpg}, supersymmetric localization \cite{Correa:2012at,Giombi:2018qox,Giombi:2018hsx}, the conformal bootstrap \cite{Liendo:2018ukf,Ferrero:2021bsb}. Correlators in the one-dimensional CFT can be defined by inserting adjoint operators $O_i$ along the Wilson line. Concretely:
\begin{align}\label{eq:WL defect correlator}
    \braket{\braket{O_1(x_1)\ldots O_n(x_n)}}&=\frac{\braket{\text{P}O_1(x_1)\ldots O_n(x_n)e^{\int (iA_0+\Phi_6)dx}}}{\braket{\mathcal{W}}},
\end{align}
where $\mathcal{W}=\text{P }e^{\int (iA_0+\Phi^6)dx}$ is the half-BPS Wilson line without insertions. Here, $A_\mu$ is the gauge field, $\Phi_I$, $I=1,\ldots, 6$ are the six scalars of $\mathcal{N}=4$ SYM, the `P' indicates that the operators and intermediate Wilson line segments are path ordered, and we have taken the spacetime contour in $\mathbb{R}^4$ to lie along the $x^\mu=\delta^{0\mu}$ axis and the contour in $S^5$ embedded in $\mathbb{R}^6$ to sit at the point $\theta^I=\delta^{I6}$. 

Recently, \cite{Barrat:2021tpn,Barrat:2022eim} studied multi-point correlators on the Wilson line of arbitrary scalars (including the five protected scalars $\Phi_I$ with $I=1,\ldots,5$ and the unprotected scalar $\Phi_6$ that couples to the Wilson line) in the planar limit ($N\to \infty$ with $\lambda=g_{YM}^2N$ fixed) at weak coupling ($\lambda\ll 1$). In this paper, in a complementary direction to that work, we study the leading contribution at strong coupling ($\lambda \gg 1$) to the connected six point function of identical scalars,
\begin{align}\label{eq:six-point function conn}
    \braket{\braket{\Phi(x_1)\ldots \Phi(x_6)}}_{\rm conn.},
\end{align}
where $\Phi\equiv \Phi_1$ is one of the protected scalars.

At strong coupling in the planar limit, the Wilson line is dual to a classical fundamental string in AdS$_5\times S^5$ that is incident on the straight line on the boundary and has AdS$_2$ geometry \cite{Maldacena:1998im,Rey:1998ik,Drukker:1999zq}. Furthermore, insertions of local operators along the Wilson line are dual to small fluctuations of the string. One approach to computing the Wilson line correlators at strong coupling is to study the dual string in static gauge \cite{Giombi:2017cqn}, in which case the longitudinal modes of the string are non-dynamical and the transverse fluctuations of the string can be treated as scalar fields in AdS$_2$ governed by a tower of interactions. One can then compute the connected correlators perturbatively by evaluating Witten diagrams in AdS$_2$ \cite{Giombi:2017cqn,Bliard:2022xsm}, which at leading order are tree-level. For example, the leading connected six-point function includes contributions from both six-point contact diagrams and four-point exchange diagrams. This approach can be thought of as ``holography on the worldsheet'' and is similar to the standard supergravity computations of correlators of single trace operators in the four-dimensional CFT.

By contrast, in this work we build on the developments in \cite{Giombi:2022pas} and compute the six-point function in \eqref{eq:six-point function conn} by studying the dual string in conformal gauge. Working in conformal gauge gives rise to a dynamical reparametrization mode on the boundary of the string that mediates the interactions of the string transverse fluctuations. Perturbation theory in conformal gauge looks qualitatively rather different than in static gauge, and in fact more closely resembles perturbation theory in the Schwarzian theory \cite{Maldacena:2016hyu, Maldacena:2016upp,Haehl:2017pak,Qi:2019gny} that governs the boundary graviton in Jackiw-Teitelboim (JT) gravity, which is also represented by a reparametrization of the boundary of AdS$_2$. One advantage of the conformal gauge approach is that, at least in the simplest setting where only one transverse coordinate is turned on, one does not have to evaluate Witten diagrams with (possibly multiple) integrations of the bulk points in AdS$_2$. On the other hand, the effective action for the boundary reparametrization mode is not known in closed form and needs to be evaluated perturbatively in small fluctuations about its saddle point. The reparametrization action was derived to quadratic order in \cite{Polyakov:2000jg, Rychkov:2002ni,Makeenko:2010fj,Giombi:2022pas}, and we derive the cubic order correction in the present paper. From this perspective, the six-point function in \eqref{eq:six-point function conn} is interesting because it is the simplest correlator that is sensitive to the self-interaction of the reparametrization modes. As a check of the consistency of the conformal gauge and static gauge approaches, we will also set up the computation of the six-point function in \eqref{eq:six-point function conn} in static gauge and evaluate the relevant Witten diagrams numerically for a sample of external points, finding agreement with the analytic result from the conformal gauge approach.

When the defect correlators on the Wilson line at strong coupling are continued to Lorentzian configurations, they are related to scattering amplitudes on the string worldsheet. For example, the four-point function corresponds to the simplest $2$-to-$2$ scattering process. Thus, another motivation for studying the six-point function is that it should correspond to $3$-to-$3$ scattering, the simplest scattering process that can in principle probe the integrability of the worldsheet. As a simple instantiation of this idea, we will analytically continue the euclidean six-point function to an out-of-time-order configuration corresponding to high-energy $3$-to-$3$ scattering on the worldsheet, which is expected to be essentially determined by the scattering amplitude on the long string in flat space \cite{Dubovsky:2012wk}. This will provide another check of our result for the six point function via conformal gauge.

The rest of the paper is organized as follows. In Section \ref{sec:corr on AdS2 in conformal gauge}, we review the conformal gauge setup \cite{Giombi:2022pas} for the calculation of the boundary correlation functions. In Section \ref{sec:reparametrization action}, we discuss the reparametrization action arising from the longitudinal modes of the string, working up to cubic order in small fluctuations around the saddle point. We then present our derivation of the connected tree-level six-point function in Section \ref{sec:connected six-point function}. In Section \ref{sec:six-point-OTOC}, we study the analytic continuation of the six-point function to the out-of-time-order configuration, and discuss its relation to $3$-to-$3$ flat space scattering. In Section \ref{sec:static}, we compute the six-point function using the static gauge approach, and numerically verify agreement with the analytic conformal gauge result. We make some concluding remarks and discuss open problems and future directions in Section \ref{sec:conclusion}. Some additional details and discussions less essential to the main thread of the paper are relegated to appendices.

\section{Correlators on the AdS$_2$ string in conformal gauge}\label{sec:corr on AdS2 in conformal gauge}
We begin by reviewing the computation of boundary correlators on the AdS$_2$ string in the conformal gauge, which was worked out in \cite{Giombi:2022pas}.

\subsection{Classical string in conformal gauge}
Consider an open string $\Sigma$ in AdS$_2\times S^1$ (which can be a subspace of AdS$_5\times S^5$) that is incident on a curve $\gamma$ on the boundary. Let $x\in \mathbb{R},z\in[0,\infty)$ be Poincar\'e coordinates on AdS$_2$ and let $y\in [0,2\pi]$ be an angular coordinate on $S^1$ so that the target space metric is given by
\begin{align}\label{eq:ds^2 AdS_2 x S^1}
    ds^2=\frac{dx^2+dz^2}{z^2}+dy^2.
\end{align}
Let $\sigma^\alpha=(s,t)$ be the worldsheet coordinates, such that the boundary is at $s=0$. Then, the string can be represented by a map $\Sigma:(s,t)\mapsto(x(s,t),z(s,t),y(s,t))$ and the curve on the boundary by a map $\gamma:\alpha \mapsto (x=\alpha,y=\tilde{y}(\alpha))$. We take $\tilde{y}$ to be small, meaning that the boundary curve is approximately straight. 

The string partition function is given by a sigma model path integral whose boundary conditions are set by the function $\tilde{y}(\alpha)$ that specifies the shape of the boundary curve. At zero string coupling ($g_s=0$) the string worldsheet has the topology of a disk and at large string tension ($T_s\gg 1$) the path integral is dominated by its saddle point. We will focus on this regime, in which case the string partition function is determined by the action of the classical string:
\begin{align}\label{eq:classical partition function}
    Z[\tilde{y}]&\approx e^{-S_{\rm cl}[\tilde{y}]}.
\end{align}
In terms of the dual CFT, this is the planar strong coupling 
approximation for the expectation value of the Wilson operator $\mathcal{W}$ in $\mathcal{N}=4$ SYM whose contour is $\gamma$. As an explicit example, we may take the spacetime contour in $\mathbb{R}^4$ to be the straight line $x^\mu(x)=x\delta^{\mu0}$ and the contour in $S^5$ in embedding coordinates to be $\theta^I(x)=\cos{\tilde{y}(x)}\delta^{I6}+\sin{\tilde{y}(x)}\delta^{I1}$.

Transverse displacements of the boundary curve can be interpreted as insertions of local operators along the curve. Concretely, correlators of the boundary operators are defined by taking variational derivatives of the string partition function with respect to $\tilde{y}$:
\begin{align}\label{eq:string bndry correlators}
    \braket{y(x_1)\ldots y(x_n)}&=\frac{1}{Z}\frac{\delta^n Z}{\delta \tilde{y}(x_1)\ldots \delta\tilde{y}(x_n)}\bigg\rvert_{\tilde{y}=0}.
\end{align}
 Setting $\tilde{y}=0$ after taking the derivatives means that we take the curve without perturbations to be a straight line. The corresponding classical string without perturbations forms an AdS$_2$ subspace of AdS$_2\times S^1$. The string boundary correlator in \eqref{eq:string bndry correlators} is related to the Wilson line defect correlator defined in \eqref{eq:WL defect correlator} by the dictionary \cite{Giombi:2017cqn}
 \begin{align}
     \braket{y(x_1)\ldots y(x_n)}=\braket{\braket{\Phi(x_1)\ldots \Phi(x_n)}},
\end{align}
where $\Phi\equiv \Phi_1$. In the classical approximation, eq.~\eqref{eq:classical partition function}, the correlators are tree level. 

To compute the boundary correlators, we therefore need to determine the classical string action as a function of the boundary curve. The string action is:
\begin{align}\label{eq:polyakov action}
    S[x,z,y]&=\frac{T_s}{2}\int d^2\sigma \sqrt{h}h^{\alpha\beta}\left[\frac{\partial_\alpha x \partial_\beta x + \partial_\alpha z \partial_\beta z}{z^2}+\partial_\alpha y \partial_\beta y\right].
\end{align}
Here, $h_{\alpha\beta}$ is the auxiliary metric on the worldsheet. We need to pick a gauge to fix the diffeomorphism symmetry of the action. The classical action and the tree-level four-point functions were analyzed in static gauge in \cite{Giombi:2017cqn}. In this work, we continue the developments of \cite{Giombi:2022pas} and instead study the classical action and tree-level correlators in the conformal gauge, where we choose worldsheet coordinates such that the auxiliary metric is the AdS$_2$ metric: 
\begin{align}\label{eq:aux metric conformal gauge}
    h_{\alpha\beta}d\sigma^\alpha d\sigma^\beta&=\frac{ds^2+dt^2}{s^2}.
\end{align}
With this choice, the string action becomes
\begin{align}\label{eq:S=S_L+S_T+T_sA}
    S[x,z,y]&=S_L[x,z]+S_T[y]+T_sA,
\end{align}
where we have split the action into the contributions from the longitudinal modes $x,z$ and the transverse mode $y$, which are given explicitly by:
\begin{align}
    S_L[x,z]&=\frac{T_s}{2}\int d^2\sigma \left[\frac{\partial^\alpha x\partial_\alpha x + \partial^\alpha z \partial_\alpha z}{z^2}-\frac{2}{s^2}\right],\label{eq:longitudinal action}\\
    S_T[y]&=\frac{T_s}{2}\int d^2\sigma \partial^\alpha y \partial_\alpha y.\label{eq:transverse action}
\end{align}
We also separated out the area piece $T_sA = T_s\int \frac{d^2\sigma}{s^2}$ from the longitudinal action in \eqref{eq:S=S_L+S_T+T_sA} and \eqref{eq:longitudinal action} to make the latter finite. After suitable regularization, $A=0$.

The string being incident on the boundary curve $\gamma$ imposes the Dirichlet boundary conditions
\begin{align}\label{eq:longitudional bc}
    z(0,t)&=0, &x(0,t)&=\alpha(t), 
\end{align}
for the longitudinal modes and
\begin{align}\label{eq:transverse bc}
    y(0,t)&=\tilde{y}(\alpha(t)).
\end{align}
for the tranverse mode. Here, $\alpha(t)$ represents a general reparametrization of the boundary curve (i.e., a one-to-one map on $\mathbb{R}$, which we take to be positively oriented, $\dot{\alpha}>0$). Unlike in the static gauge, in the conformal gauge the boundary reparametrization is dynamical. This is necessary because the worldsheet coordinate transformation that puts the auxiliary metric into the conformal form will in general act non-trivially on the boundary (see for instance \cite{polyakov1987gauge, Cohen:1985sm, Rychkov:2002ni}).

A nice feature of the conformal gauge is that the longitudinal and transverse actions in \eqref{eq:S=S_L+S_T+T_sA} are decoupled. The longitudinal and transverse modes are only coupled through the reparametrization mode $\alpha(t)$ that appears in the boundary conditions and through the Virasoro constraint. This means that in the classical approximation we can proceed as follows: first, solve the equations of motion for $x$ and $z$ and determine the longitudinal action as a function of $\alpha(t)$; second, solve the equation of motion for $y$ and determine the transverse action as a function of $\tilde{y}(\alpha(t))$; and, third, fix $\alpha(t)$ by imposing the Virasoro constraint. This can be summarized by writing the classical action as:
\begin{align}\label{eq:S_cl = S_L + S_T + Virasoro}
    S_{\rm cl}[\tilde{y}]&=S_L[\alpha(t)]+S_T[\tilde{y}(\alpha(t))] \bigg\rvert_{\rm Virasoro},
\end{align}
where 
\begin{align}\label{eq:transverse action extremization}
    S_T[\tilde{y}(\alpha(t))]&=\frac{T_s}{2}\int d^2\sigma \partial^\alpha y \partial_\alpha y \bigg\rvert_{\substack{\text{extremize }y\\y(0,t)=\tilde{y}(\alpha(t))}},
\end{align}
is the ``on-shell'' transverse action and
\begin{align}\label{eq:longitudinal action extremization}
    S_L[\alpha]&= \frac{T_s}{2}\int d^2\sigma \left[\frac{\partial^\alpha x \partial_\alpha x + \partial^\alpha z\partial_\alpha z}{z^2}-\frac{2}{s^2}\right]\bigg\rvert_{\substack{\text{extremize }z,x\\ z(0,t)=0,\text{ }x(0,t)=\alpha(t)}}
\end{align}
is the ``on-shell'' longitudinal action, both of which we have expressed in terms of classical boundary value problems. Meanwhile, the Virasoro constraint is simply the equation of motion for the metric in \eqref{eq:polyakov action}, which imposes that the total stress tensor --- the sum of the longitudinal and the transverse stress tensors --- on the worldsheet is zero. As reviewed in \cite{Giombi:2022pas}, imposing the Virasoro constraint is also equivalent to extremizing over the reparametrization mode $\alpha$. Thus, another way to write \eqref{eq:S_cl = S_L + S_T + Virasoro} is 
\begin{align}\label{eq:S_cl = S_L + S_T + extremize}
    S_{\rm cl}[\tilde{y}]&=S_L[\alpha(t)]+S_T[\tilde{y}(\alpha(t))]\bigg\rvert_{\text{extremize }\alpha}.
\end{align}
As a final step, we promote the string partition function in \eqref{eq:classical partition function} to a path integral over boundary reparametrizations:
\begin{align}\label{eq:reparametrization path integral}
    Z[\tilde{y}]&=\int \mathcal{D}\alpha e^{-S_L[\alpha(t)]-S_T[\tilde{y}(\alpha(t))]}.
\end{align}
For the purposes of this paper, writing the partition function in this way is simply a convenient, aesthetic way to package the classical analysis in the conformal gauge. This is because we will work exclusively in the saddle point approximation to the path integral, in which case \eqref{eq:reparametrization path integral} reduces to \eqref{eq:classical partition function} together with \eqref{eq:S_cl = S_L + S_T + extremize}.\footnote{An important question is whether the reparametrization path integral can be defined precisely and computed beyond the saddle point approximation.}

In order to compute boundary correlators using \eqref{eq:reparametrization path integral}, we need to determine the on-shell transverse and longitudinal actions. The transverse action is simple because $y$ is a free massless scalar field on AdS$_2$. Classically, it satisfies the equation of motion $\partial^2 y =0$ with the boundary condition in eq.~(\ref{eq:transverse bc}), whose solution for $y(s,t)$ can be expressed using the boundary-to-bulk propagator, $K(s,t,t')=\frac{1}{\pi}\frac{s}{s^2+(t-t')^2}$. The action in \eqref{eq:transverse action extremization} takes the following closed form (see, e.g., \cite{Giombi:2022pas}):
\begin{align}\label{eq: transverse action explicit}
    S_T[\tilde{y}(\alpha(t))]&=-\frac{T_s}{2}\lim_{s\to 0}\int dt_1 dt_2 \partial_s K(s,t_1,t_2)\tilde{y}(\alpha(t_1))\tilde{y}(\alpha(t_2))\nonumber\\&=-\frac{T_s}{2\pi}\int dt dt' \frac{\tilde{y}(\alpha(t))\tilde{y}(\alpha(t'))}{|t-t'|^{2+\eta}}=\frac{T_s}{4\pi}\int dtdt'\frac{\left(\tilde{y}(\alpha(t))-\tilde{y}(\alpha(t'))\right)^2}{(t-t')^2}.
\end{align}
We have given two expressions for the transverse action in the second line. The first is defined via analytic continuation in the exponent $\eta\to 0$, and is more practical for many purposes. The second is manifestly finite. We can get from one to the other using integration by parts and the identity $\int dt |t|^{-2-\eta}=0$ in analytic regularization. This and other integrals we will use throughout the paper are collected in Appendix \ref{app:useful integrals}.

Meanwhile, the longitudinal modes $x,z$ satisfy non-linear equations of motion. It would be interesting to find a closed-form expression for the longitudinal action for arbitrary $\alpha$ (although it may be that \eqref{eq:longitudinal action extremization} is its most explicit representation). Instead, we will work perturbatively in section~\ref{sec:reparametrization action} and derive explicit expressions for the action at quadratic and cubic order in small fluctuations about the saddle point $\alpha(t)=t$. This will be sufficient to compute the tree-level six-point functions, as we will see in section~\ref{sec:connected six-point function}.

\subsection{Correlators in the reparametrization path integral}
We now recall how the boundary correlators defined in \eqref{eq:string bndry correlators} are expressed in terms of the reparametrization path integral in \eqref{eq:reparametrization path integral}. The function $\tilde{y}$ specifying the boundary curve only appears in the reparametrization path integral through the transverse action, $S_T$. To take variational derivatives, it is useful to change the integration variables in \eqref{eq: transverse action explicit} from $t$ and $t'$ to $x=\alpha(t)$ and $x'=\alpha(t')$, so that the transverse action becomes:
\begin{align}\label{eq:gfdert543e}
    S_T[\tilde{y}(\alpha(t))]&=-\frac{T_s}{2\pi}\int dx dx' \frac{\dot{\beta}(x)\dot{\beta}(x')}{|\beta(x)-\beta(x')|^{2+\eta}}\tilde{y}(x)\tilde{y}(x'),
\end{align}
where $\beta$ is the inverse of $\alpha$ (i.e., $\alpha(\beta(x))=x$ and $\beta(\alpha(t))=t$). Thus, for instance, taking two variational derivatives of the transverse action yields:
\begin{align}\label{eq:d^2S_T/dy^2}
    -\frac{\delta^2 S_T}{\delta \tilde{y}(x_1)\delta \tilde{y}(x_2)}&=\frac{T_s}{\pi} B(x_1,x_2),
\end{align}
where we have introduced the following bi-local object:
\begin{align}\label{eq:bi-local}
    B(x_1,x_2)&=\frac{\dot{\beta}(x_1)\dot{\beta}(x_2)}{(\beta(x_1)-\beta(x_2))^2}.
\end{align}
This object, which is familiar from Schwarzian quantum mechanics (see e.g., \cite{Maldacena:2016hyu,Kitaev:2017awl}), is a conformal two-point function of scaling dimension $\Delta=1$ ``dressed'' by the boundary reparametrization mode. From \eqref{eq:string bndry correlators}, \eqref{eq:reparametrization path integral} and \eqref{eq:d^2S_T/dy^2}, it follows that, for example, the two- and four-point 
boundary correlators are expressed in the language of the reparametrization integral as:
\begin{align}
    \braket{y(x_1)y(x_2)}&=\frac{T_s}{\pi}\frac{1}{Z}\int \mathcal{D}\alpha e^{-S_L[\alpha]} B(x_1,x_2),\\
    \braket{y(x_1)y(x_2)y(x_3)y(x_4)}&=\frac{T_s^2}{\pi^2}\frac{1}{Z}\int \mathcal{D}\alpha e^{-S_L[\alpha]} \bigg[B(x_1,x_2)B(x_3,x_4)+B(x_1,x_3)B(x_2,x_4)\nonumber\\&\hspace{4cm}+B(x_1,x_4)B(x_2,x_3)\bigg]\,,
\end{align}
and similarly for higher-point functions. 
We see that the longitudinal action plays the role of an effective action for the boundary reparametrization mode, and will therefore use the terms ``longitudinal action'' and ``reparametrization action'' interchangeably. We should also note that we can equivalently identify either $\alpha$ or $\beta\equiv \alpha^{-1}$ as the boundary reparametrization mode that is integrated over in the path integral. If we work with $\alpha$, then its inverse $\alpha^{-1}$ appears in the dressed two-point function $B(x_1,x_2)$ in \eqref{eq:bi-local}. If we work with $\beta$, then its inverse $\beta^{-1}$ appears in the longitudinal action in \eqref{eq:longitudinal action extremization}. In section 6 of \cite{Giombi:2022pas}, we found it convenient to switch from $\alpha$ to $\beta$ in the reparametrization path integral in order to make the comparison with the Schwarzian theory more direct, but in the present paper we will continue working with $\alpha$. 

Instead of studying the correlators of the $y$ operators directly, it is convenient as a book-keeping trick to introduce the ``fictitious'' operators $U$, $V$, $W$ with conformal dimensions $\Delta_U=\Delta_V=\Delta_W=1$, whose two-, four-, and six-point functions we define to be:\footnote{The correlators in \eqref{eq:<UU>}-\eqref{eq:<UUVVWW>} can arise in the path integral in \eqref{eq:reparametrization path integral} if we let the transverse action be equal to the sum of three effective actions for three independent modes $U$, $V$ and $W$: $S_T[\tilde{y}(\alpha(t)]\to \frac{\pi}{T_s}[S_T[\tilde{u}(\alpha(t))]+S_T[\tilde{v}(\alpha(t))]+S_T[\tilde{w}(\alpha(t))]]$. Then $U$, $V$, and $W$ are inserted in the correlators by taking variational derivatives with respect to $\tilde{u}$, $\tilde{v}$ and $\tilde{w}$, respectively. These effective actions for the three modes would arise, for instance, if we studied the open string in AdS$_2\times S^1\times S^1\times S^1$.}
\begin{align}
    \braket{U(x_1)U(x_2)}&=\frac{1}{Z}\int \mathcal{D}\alpha e^{-S_L[\alpha]}B(x_1,x_2)\label{eq:<UU>}\\
    \braket{U(x_1)U(x_2)V(x_3)V(x_4)}&=\frac{1}{Z}\int \mathcal{D} \alpha e^{-S_L[\alpha]} B(x_1,x_2)B(x_3,x_4)\label{eq:<UUVV>}\\
    \braket{U(x_1)U(x_2)V(x_3)V(x_4)W(x_5)W(x_6)}&=\frac{1}{Z}\int \mathcal{D} \alpha e^{-S_L[\alpha]} B(x_1,x_2)B(x_3,x_4)B(x_5,x_6)\label{eq:<UUVVWW>}
\end{align}
The advantage of working with $U$, $V$ and $W$ is that, because we can consider distinct pairs of operators in the correlators, it reduces the number of terms in expressions for the correlators. (We have also absorbed the factor $\frac{T_s}{\pi}$ appearing in the transverse action in \eqref{eq:gfdert543e} into the operators.) We can always translate the $U$, $V$, and $W$ correlators back to correlators for the transverse direction $y$ by permuting over the external points. For example, for the six-point function,
\begin{align}
    \braket{y(x_1)y(x_2)y(x_3)y(x_4)y(x_5)y(x_6)}&=\frac{T_s^3}{\pi^3}\bigg[\braket{U(x_1)U(x_2)V(x_3)V(x_4)W(x_5)W(x_6)}\nonumber\\&\hspace{3cm}+14\text{ more permutations}\bigg].
\end{align}
In addition, we are going to focus on the leading (i.e., tree-level) contribution to the connected part of the six-point function, $\braket{UUVVWW}_{\rm c}$, which is related to the full six-point function by:
\begin{align}\label{eq:full, connected, disconnected six-point function}
\braket{U_1U_2V_3V_4W_5W_6}&=\braket{U_1U_2}\braket{V_3V_4}\braket{W_5W_6}+\braket{U_1U_2W_5W_6}_{\rm c}\braket{V_3V_4}+\braket{V_3V_4W_5W_6}_{\rm c}\braket{U_1U_2}\nonumber\\&+\braket{U_1U_2V_3V_4}_{\rm c}\braket{W_5W_6}+\braket{U_1U_2V_3V_4W_5W_6}_{\rm c},
\end{align}
where $U_n\equiv U(x_n)$, etc.

Finally, one lesson of \cite{Giombi:2022pas} worth reviewing is that the AdS$_2$ string in conformal gauge has two $SL(2,\mathbb{R})$ symmetries, which have different implications for computing correlators via the reparametrization path integral. Let $f(t)=\frac{at+b}{ct+d}$ with $a,b,c,d\in \mathbb{R}$ and $ad-bc=1$ denote a general $SL(2,\mathbb{R})$ transformation. The first $SL(2,\mathbb{R})$ symmetry arises due to the isometries of AdS$_2$, acts on the target space coordinates $x$ and $z$ as $x+iz\to f(x+iz)$, and is physical--- meaning that it gives rise to Ward identities for the correlators. The second $SL(2,\mathbb{R})$ symmetry arises due to the residual transformations of the worldsheet coordinates that leave the auxiliary metric invariant up to a Weyl rescaling, acts on the worldsheet coordinates $t$ and $s$ as $t+is\to f(t+is)$ and is gauged--- meaning that it needs to be gauge fixed when we integrate over reparametrizations. 

More concretely, the action of the physical $SL(2,\mathbb{R})$ symmetry on the reparametrization mode and its inverse is $\alpha(t)\to f(\alpha(t))$, $\beta(x)\to \beta(f^{-1}(x))$, and the action of the gauged $SL(2,\mathbb{R})$ symmetry is $\alpha(t)\to \alpha(f(t))$, $\beta(x)\to f^{-1}(\beta(x))$. The reparametrization action in \eqref{eq:longitudinal action extremization} is invariant under both transformations. The dressed two-point function in \eqref{eq:bi-local} (and the transverse action in \eqref{eq:gfdert543e}) is also invariant under the gauge $SL(2,\mathbb{R})$ transformation, while under the physical $SL(2,\mathbb{R})$ it transforms covariantly as $B(x_1,x_2)\to \dot{f}^{-1}(x_1)\dot{f}^{-1}(x_2)B(f^{-1}(x_1),f^{-1}(x_2))$. This property implies the Ward identity
\begin{align}
    \braket{y(x_1)y(x_2)\ldots y(x_{n})}&=\dot{f}(x_1)\ldots \dot{f}(x_n)\braket{y(f(x_1))y(f(x_2))\ldots y(f(x_n))},
\end{align}
which is consistent with the interpretation of these correlators as defining a 1d defect CFT. The Ward identity fixes the form of two- and three-point functions, the four-point functions up to a function of a single cross ratio, and the six-point function up to a function of three cross ratios.

\subsection{Mapping between the line and circle}

In our discussion of the AdS$_2$ string thus far, we have used Poincar\'e half-plane coordinates $x$ and $z$ on the AdS$_2$ target space and $t$ and $s$ on the string worldsheet (see \eqref{eq:ds^2 AdS_2 x S^1} and \eqref{eq:aux metric conformal gauge}), and have viewed the boundary of the string as the line, $\mathbb{R}$. We can equivalently use Poincar\'e disk coordinates $r\in [0,\infty)$ and $\theta\in [0,2\pi]$ on the AdS$_2$ target space so that the target space metric is
\begin{align}\label{eq:ds^2 target space}
    ds^2&=\frac{dr^2+d\theta^2}{\sinh^2{r}}+dy^2,
\end{align}
and disk coordinates $\sigma\in [0,\infty)$ and $\tau\in[0,2\pi]$ on the worldsheet so that the auxiliary metric is:
\begin{align}\label{eq:hyperbolic disk metric}
    h_{\alpha\beta}d\sigma^\alpha d\sigma^\beta=\frac{d\sigma^2+d\tau^2}{\sinh^2{\sigma}}.
\end{align}
In these coordinates, we view the boundary of the string as the circle, $S^1$.

Everything that we do on the half-plane can be repeated on the disk, but the actions and correlators on the disk can also be deduced from those on the half-plane by using a coordinate transformation to map the half-plane to the disk on both the target space and the worldsheet. This is achieved by the pair of transformations:
\begin{align}
    e^{-\sigma+i\tau}&=-\frac{t+i(s-1)}{t+i(s+1)},&e^{-r+i\theta}&=-\frac{x+i(z-1)}{x+i(z+1)}.   
\end{align}
In particular, when restricted to the boundary this identifies 
\begin{align}\label{eq:t, tau, x, theta on line and circle}
    t&=\tan{\frac{\tau}{2}}, &x&=\tan{\frac{\theta}{2}}.
\end{align}
If $\alpha(t)$ and $\beta(x)$ are the reparametrization mode and its inverse on the line, and $\tilde{\alpha}(\tau)$ and $\tilde{\beta}(\theta)$ are the reparametrization mode and its inverse on the circle, then they are related by
\begin{align}\label{eq:alpha beta on line and circle}
    \alpha(t)&=\tan\left(\frac{\tilde{\alpha}(\tau)}{2}\right), &\beta(x)&=\tan\left(\frac{\tilde{\beta}(\theta)}{2}\right).
\end{align}
It follows then that the reparametrization actions on the line and circle satisfy
\begin{align}\label{eq: S_L on line and circle}
    S_L^{\rm circle}[\tilde{\alpha}(\tau)]&=S_L^{\rm line}[\alpha(t)], 
\end{align}
and the dressed two-point functions satisfy
\begin{align}\label{eq: B on line and circle}
    B^{\rm circle}(\theta_1,\theta_2)&=\frac{dx_1}{d\theta_1}\frac{dx_2}{d\theta_2}B^{\rm line}(x_1,x_2)=\frac{\dot{\tilde{\beta}}(\theta_1)\dot{\tilde{\beta}}(\theta_2)}{\big[2\sin\big(\frac{\tilde{\beta}(\theta_1)-\tilde{\beta}(\theta_2)}{2}\big)\big]^2}.
\end{align}
(The $\frac{dx}{d\theta}$ prefactors are the usual conformal factors for operators with dimension $\Delta=1$; alternatively, they come from changing integration variables from $x$ to $\theta$ in \eqref{eq:gfdert543e}). Finally, if the correlators on the circle are defined in analogy with \eqref{eq:string bndry correlators} by taking variational derivatives of the string partition function with respect to $\tilde{y}(\tau)$ instead of $\tilde{y}(t)$, which inserts copies of $B^{\rm circle}$ instead of $B^{\rm line}$, then the correlators on the circle and line satisfy
\begin{align}
    \braket{y(\theta_1)\ldots y(\theta_n)}_{\rm circle}=\frac{dx_1}{d\theta_1}\ldots \frac{dx_n}{d\theta_n}\braket{y(x_1),\ldots y(x_n)}_{\rm line}.
\end{align}
Translating between correlators on the line and circle then reduces to interchanging euclidean and chordal distances: $x_{12}\leftrightarrow \sin{\frac{\theta_{12}}{2}}$. 

\section{Reparametrization action}\label{sec:reparametrization action}
In this section, we study the on-shell longitudinal action in detail. We begin by noting the classical equations of motion for the longitudinal modes that follow from \eqref{eq:longitudinal action extremization}:
\begin{align}
    0&=z(\ddot{x}+x'')-2(\dot{z}\dot{x}+z'x'), &0&=z(\ddot{z}+z'')+(\dot{x}^2+{x'}^2)-(\dot{z}^2+{z'}^2).\label{eq:x,z eom}
\end{align}
Here, $\dot{f}\equiv \partial_t f$ and $f'\equiv \partial_sf$. The equations of motion are supplemented by the boundary conditions in \eqref{eq:longitudional bc}. 

We do not know the general solution to \eqref{eq:x,z eom}. However, a family of solutions, which are the saddle points of the reparametrization path integral about which we will consider small fluctuations in our perturbative analysis, are given by the $SL(2,\mathbb{R})$ transformations on the upper half plane:
\begin{align}\label{eq:SL(2,R) x+iz}
    x(s,t)+iz(s,t)&=\frac{a(t+is)+b}{c(t+is)+d},
\end{align}
where $a,b,c,d\in \mathbb{R}$ and $ad-bc=1$. These satisfy the boundary conditions with the reparametrization mode given by: 
\begin{align}\label{eq:SL(2,R) alpha(t)}
    \alpha(t)=\frac{at+b}{ct+d},
\end{align}
which is an $SL(2,\mathbb{R})$ transformation on the line. These solutions are the saddle points of \eqref{eq:reparametrization path integral} because extremizing over $\alpha$ in \eqref{eq:S_cl = S_L + S_T + extremize} without transverse modes (i.e., with $\tilde{y}=0$) imposes the Virasoro constraint on only the longitudinal modes. As explained in \cite{Giombi:2022pas}, the longitudinal stress tensor being zero forces $x+iz$ to be a holomorphic function of $t+is$, which, combined with the requirement that the function be invertible, implies that it is an $SL(2,\mathbb{R})$ transformation.

It is also useful to note the behavior of $x$ and $z$ near the boundary for a general $\alpha(t)$:
\begin{align}\label{eq:x and z near the boundary}
    x(s,t)&=\alpha(t)-\frac{\ddot{\alpha}(t)}{2}s^2+O(s^3),&z(s,t)&=\dot{\alpha}(t)+O(s^3)
\end{align}
As explained in \cite{Polyakov:2000jg,Polyakov:2000ti}, this follows from \eqref{eq:x,z eom} after expanding $z(s,t)$ and $x(s,t)$ as series in $s$. As shown in \cite{Giombi:2022pas}, one consequence of \eqref{eq:x and z near the boundary} is that the on-shell longitudinal action in \eqref{eq:longitudinal action extremization} is finite.

Now we proceed to the perturbative analysis about the saddle point. We will expand around the solution $\alpha(t)=t$, $x(s,t)=t$, $z(s,t)=s$, which corresponds to the gauge choice $a=d=1$ and $b=c=0$ in \eqref{eq:SL(2,R) alpha(t)}. (This choice gives rise to three $SL(2,\mathbb{R})$ zero modes that we will have to handle appropriately in the path integral later.) Thus, we write the boundary reparametrization as
\begin{align}
    \alpha(t)=t+\lambda \epsilon(t),
\end{align}
where $\epsilon(t)$ is an arbitrary function on the real line that vanishes sufficiently quickly as $t\to \pm \infty$, and $\lambda$ is a small book-keeping parameter. We then expand the longitudinal modes in powers of $\lambda$,
\begin{align}
    x(s,t)&=t+\lambda x_1(s,t)+\lambda^2 x_2(s,t)+\ldots,\label{eq: x series}\\
    z(s,t)&=s+\lambda z_1(s,t)+\lambda^2 z_2(s,t)+\ldots,\label{eq: z series}
\end{align}
and also expand the Lagrangian of the reparametrization action in \eqref{eq:longitudinal action} in powers of $\lambda$:
\begin{align}\label{eq:action expansion}
    \mathcal{L}&=\frac{\partial_\beta x \partial^\beta x + \partial_\beta z \partial^\beta z}{2z^2}-\frac{1}{s^2}=\underbrace{\lambda \mathcal{L}_1+\lambda^2 \mathcal{L}_2 +\ldots}_{\approx 0\text{ by e.o.m.}}+ \lambda^2 \mathcal{L}_{1^2}+\lambda^3\left(\mathcal{L}_{1^3}+\mathcal{L}_{1,2}\right)+\ldots
\end{align}
Here, the notation $\mathcal{L}_{1^a,2^b,\ldots}$ denotes the part of the Lagrangian that is of $a$th order in $x_1$ and $z_1$, $b$th order in $x_2$ and $z_2$, etc. As noted in \eqref{eq:action expansion}, the contributions of $\mathcal{L}_1$, $\mathcal{L}_2$, etc. to the action are zero by the equations of motion because they represent first order variations about the saddle point. Thus, the expansion of the longitudinal action is given by:
\begin{align}
    S_L[t+\lambda \epsilon(t)]&=\lambda^2 S_{L,2}[\epsilon]+\lambda^3 S_{L,3}[\epsilon]+\ldots,
\end{align}
where
\begin{align}
    S_{L,2}[\epsilon]&=T_s\int ds dt \mathcal{L}_{1^2},\label{eq:S_{L,2}}\\
    S_{L,3}[\epsilon]&=T_s\int ds dt (\mathcal{L}_{1^3}+\mathcal{L}_{1,2}),\label{eq:S_{L,3}}
\end{align}
and so on. For our analysis of the tree-level six-point functions, it is sufficient to determine the action to cubic order in $\lambda$. The explicit expressions for contributions to the Lagrangian that we will need are:
\begin{align}
    \mathcal{L}_{1^2}&=-\frac{2z_1z_1'}{s^3}+\frac{3z_1^2}{s^4}-\frac{2z_1\dot{x}_1}{s^3}+\frac{\dot{x}_1^2+{x_1'}^2+\dot{z}_1^2+{z_1'}^2}{2s^2},\label{eq:L_{1^2}}\\
    \mathcal{L}_{1^3}&= -\frac{4z_1^3}{s^5}+\frac{3z_1^2\dot{x}_1}{s^4}-\frac{z_1\dot{x}_1^2}{s^3}-\frac{z_1\dot{z}_1^2}{s^3}-\frac{z_1{x_1'}^2}{s^3}+\frac{3z_1^2z_1'}{s^4}-\frac{z_1{z_1'}^2}{s^3},\label{eq:L_{1^3}}\\
    \mathcal{L}_{1,2}&=\frac{6z_1z_2}{s^4}-\frac{2z_2z_1'}{s^3}-\frac{2z_1z_2'}{s^3}-\frac{2z_2\dot{x}_1}{s^3}-\frac{2z_1\dot{x}_2}{s^3}+\frac{1}{s^2}\left(\dot{x}_1\dot{x}_2+\dot{z}_1\dot{z}_2+x_1'x_2'+z_1'z_2'\right).\label{eq:L_{1,2}}
\end{align}

We also note that the boundary condition in \eqref{eq:longitudional bc} perturbatively becomes $x_1(0,t)=\epsilon(t)$, $z_1(0,t)=0$ at first order and $x_n(0,t)=z_n(0,t)=0$ for $n\geq 2$, and the general behavior near the boundary in \eqref{eq:x and z near the boundary} becomes
\begin{align}\label{eq:x_1 z_1 boundary behavior}
    x_1(s,t)&=\epsilon(t)-\frac{\ddot{\epsilon}(t)}{2}s^2+O(s^3),&z_1(s,t)&=\dot{\epsilon}(t)s+O(s^3),
\end{align}
at first order and
\begin{align}\label{eq:x_n z_n boundary behavior}
    x_n(s,t)&=O(s^3),&z_n(s,t)&=O(s^3),
\end{align}
for $n\geq 2$.

This completes the set-up of the perturbative analysis. Now we carry out the details, starting by finding the longitudinal modes to linear order and the longitudinal action to quadratic order, which was worked out in \cite{Giombi:2022pas}. The equations of motion in \eqref{eq:x,z eom} at first order are
\begin{align}
    0&=s(\ddot{x}_1+x_1'')-2(x_1'+\dot{z}_1), &0&=s(\ddot{z}_1+z_1'')+2(\dot{x}_1-z_1'),\label{eq:eom linear}
\end{align}
with  the boundary conditions $x_1(0,t)=\epsilon(t)$, $z_1(0,t)=0$. These can be solved using ``boundary-to-bulk'' propagators for $x$ and $z$:
\begin{align}\label{eq:x1 and z1}
    x_1(s,t)&=\int dt' K_x(s,t,t')\epsilon(t'),&z_1(s,t)&=\int dt' K_z(s,t,t')\epsilon(t').
\end{align}
where
\begin{align}\label{eq:Kx Kz}
    K_x(s,t,t')&=\frac{4}{\pi}\frac{s^3(s^2-(t-t')^2)}{(s^2+(t-t')^2)^3},&K_z(s,t,t')&=-\frac{8}{\pi}\frac{s^4(t-t')}{(s^2+(t-t')^2)^3}.
\end{align}
These propagators solve \eqref{eq:eom linear}, become sharply peaked at $t=t'$ as $s\to 0$, and satisfy $\int dt' K_x(s,t,t')=1$ and $\int dt' K_z(s,t,t')=0$ for any $s$. Finally, the quadratic Lagrangian in \eqref{eq:L_{1^2}} can be written as a sum of total derivatives which means the quadratic action reduces to an integral over the boundary at $s=0$, whose finiteness is guaranteed by the behavior in \eqref{eq:x_1 z_1 boundary behavior}. The explicit expression for the action takes the following bi-local form \cite{Giombi:2022pas} (see also \cite{Polyakov:2000jg, Rychkov:2002ni,Makeenko:2010fj}):
\begin{align}
    S_{L,2}[\epsilon]&=-\frac{T_s}{4}\lim_{s\to 0}\int dt_1 dt_2 \partial_s^3 K_x(s,t_1,t_2)\epsilon(t_1)\epsilon(t_2)\nonumber\\
    &=\frac{6T_s}{\pi}\int dt_1 dt_2 \frac{\epsilon(t_1)\epsilon(t_2)}{|t_{12}|^{4+\eta}}=\frac{T_s}{2\pi}\int dt_1 dt_2 \frac{(\dot{\epsilon}(t_1)-\dot{\epsilon}(t_2))^2}{t_{12}^2}.\label{eq:quadratic rep action}
\end{align}
To get from the first to second line we used $\partial_s^3 K_x(0,t_1,t_2)=-\frac{24}{\pi}\frac{1}{t_{12}^4}$ and exchanged the $s\to 0$ regulator for the more convenient analytic regulator $\eta\to 0$. In the second line, we give both the analytically regularized expression and a manifestly finite expression for the action. The quadratic action also takes a simple form in Fourier space. Writing
\begin{align}\label{eq:eps(t) = Fourier eps(w)}
    \epsilon(t)=\int \frac{d\omega}{2\pi}e^{-i\omega t}\epsilon(\omega),
\end{align}
and applying the identity in \eqref{eq:analytic reg. integral line 2} in Appendix~\ref{app:useful integrals} yields:
\begin{align}\label{eq:quadratic rep action fourier}
    S_{L,2}[\epsilon]&=\frac{T_s}{2\pi}\int d\omega \epsilon(\omega)\epsilon(-\omega)|\omega|^3.
\end{align}

Next, we turn to the cubic order action in \eqref{eq:S_{L,3}}. A nice simplification of the analysis is that we do not need to solve for $x_2$ and $z_2$ because the integral of $\mathcal{L}_{1,2}$-- the only term in the cubic order Lagrangian that contains $x_2$ and $z_2$--- is zero:
\begin{align}\label{eq:int L_{1,2}=0}
    \int ds dt \mathcal{L}_{1,2}=0.
\end{align}
To see this, we note that eq.~\eqref{eq:L_{1,2}} can be rewritten as:
\begin{align}
    \mathcal{L}_{1,2}&=-2\partial_s\left(\frac{z_1z_2}{s^3}\right)-2\partial_t\left(\frac{z_1x_2}{s^3}\right)+\partial_t\left(\frac{\dot{x}_1x_2+\dot{z}_1z_2}{s^2}\right)+\partial_s\left(\frac{x_1'x_2+z_1'z_2}{s^2}\right)\nonumber\\&-\frac{z_2}{s^3}\left(s(\ddot{z}_1+z_1'')+2\dot{x}_1-2z_1'\right)+\frac{x_2}{s^3}\left(s(\ddot{x}_1+x_1'')-2\dot{z}_1-2x_1'\right).\label{eq:L_{1,2} rewritten}
\end{align}
The total derivative terms in the first line in \eqref{eq:L_{1,2} rewritten} all individually integrate to zero. In particular, there are no contributions from the boundary at $s=0$ because $x_2$ and $z_2$ vanish sufficiently quickly near the boundary as guaranteed by \eqref{eq:x_1 z_1 boundary behavior}-\eqref{eq:x_n z_n boundary behavior}. Meanwhile, the second line in \eqref{eq:L_{1,2} rewritten} vanishes due to the first order equations of motion in \eqref{eq:eom linear}. Eq.~\eqref{eq:int L_{1,2}=0} therefore follows.\footnote{At higher orders, the same reasoning implies that $\int ds dt \mathcal{L}_{1,n}=0$ for $n\geq 3$ and therefore we only need to solve for $x_1$, $x_2$, $\ldots x_{n-1}$ and $z_1$, $z_2$, $\ldots z_{n-1}$ to determine the reparametrization action to order $\lambda^{n+1}$.}

Thus, the cubic action only receives contributions from the first term in \eqref{eq:S_{L,3}}, which is a function of only $x_1$ and $z_1$, whose explicit expressions are given in \eqref{eq:x1 and z1}. One way to try to derive an explicit expression for the cubic action in terms of $\epsilon(t)$ would be to substitute the integral expressions in \eqref{eq:x1 and z1} into \eqref{eq:L_{1^3}} and then integrate over the worldsheet coordinates $s$ and $t$, which would ostensibly give rise to a tri-local form for the cubic action that involves three integrations over the boundary. However, this is not straightforward to implement because of subtleties with interchanging the worldsheet and boundary integrals. Instead, we will work in Fourier space, and will end up finding a bi-local expression for the cubic action in position space. 

We begin by noting the Fourier transforms of the $x$ and $z$ boundary-to-bulk propagators:
\begin{align}
    \int dt'K_x(s,t,t')e^{-i\omega t'}&=e^{-i\omega t -|\omega|s}(1+s|\omega|+s^2\omega^2),\label{eq:Kx fourier}
    \\
    \int dt'K_z(s,t,t')e^{-i\omega t'}&=-ie^{-i\omega t-|\omega|s}s\omega(1+s|\omega|).\label{eq:Kz fourier}
\end{align}
Then, for example, the contribution of the first term in \eqref{eq:L_{1^3}} to the cubic action is:
\begin{align}
    S_{L,3}[\epsilon]&\ni -4\int ds dt \frac{z_1^3}{s^5}=-4\int \frac{ds dt}{s^5}\int \prod_{i=1}^3 \bigg(dt_i K_z(s,t,t_i)\epsilon(t_i)\bigg)\nonumber\\&=-4\int \prod_{i=1}^3\bigg( \frac{d\omega_i}{2\pi} \epsilon(\omega_i)\bigg) \int \frac{ds dt}{s^5} \int \prod_{j=1}^3 \bigg(dt_j e^{-i\omega_j t_j} K_z(s,t,t_j)\bigg)\nonumber\\&=
    -4i\int \prod_{i=1}^3 \left(\frac{d\omega_i}{2\pi}\epsilon(\omega_i)\right) 2\pi \delta(\omega_1+\omega_2+\omega_3) \int \frac{ds}{s^2} \prod_{j=1}^3 \bigg(e^{-|\omega_i|s}\omega_i(1+s|\omega_i|)\bigg).\label{eq:example Fourier transforms}
\end{align}
To get to the second line, we used \eqref{eq:eps(t) = Fourier eps(w)} and interchanged the order of integration. To get to the third line, we first evaluated the $t_i$ integrals using \eqref{eq:Kz fourier} and then the $t$ integral to get the energy-conserving delta function. Although the remaining integral over $s$ is divergent near $s=0$, we know that the reparametrization action is finite (see the comment below \eqref{eq:x and z near the boundary}). Indeed, once we apply the same steps as in \eqref{eq:example Fourier transforms} to all of the terms in \eqref{eq:L_{1^3}} and sum up their contributions, the integral over $s$ is finite. The final result for the cubic order action ultimately simplifies to:
\begin{align}\label{eq:action cubic in x1, z1}
    S_{L,3}[\epsilon]&=T_s\int ds dt \mathcal{L}_{1^3}=T_s\int d \omega_1 d\omega_2 d\omega_3 \epsilon(\omega_1)\epsilon(\omega_2)\epsilon(\omega_3) f(\omega_1,\omega_2,\omega_3)2\pi \delta(\omega_1+\omega_2+\omega_3),
\end{align}
where
\begin{align}
    f(\omega_1,\omega_2,\omega_3)&=-\frac{i}{48\pi^3}\left(\text{sgn}(\omega_1)\omega_1^4+\text{sgn}(\omega_2)\omega_2^4+\text{sgn}(\omega_3)\omega_3^4\right).
\end{align}
Here, $\text{sgn}(x)=1$ if $x>0$ and $-1$ if $x<0$. Using the delta function to evaluate one of the $\omega$ integrals, this can also be written as
\begin{align}\label{eq:reparametrization action to cubic order Fourier}
    S_{L,3}[\epsilon]&=\frac{iT_s}{8\pi^2}\int d\omega_1 d\omega_2\epsilon(\omega_1)\epsilon(\omega_2)\epsilon(-\omega_1-\omega_2)|\omega_1+\omega_2|^4\text{sgn}(\omega_1+\omega_2).
\end{align}

Given the cubic action in Fourier space, it is also possible to deduce its form in position space:
\begin{align}\label{eq:reparametrization action to cubic order}
    S_{L,3}[\epsilon]=-\frac{12 T_s}{\pi}\int dt_1 dt_2 \frac{\epsilon(t_1)^2\epsilon(t_2)}{|t_{12}|^{4+\eta}t_{12}}.
\end{align}
To check this, we write $\epsilon(t)$ in Fourier space as in \eqref{eq:eps(t) = Fourier eps(w)} and evaluate the integral over $t_{12}$ using \eqref{eq:analytic reg. integral line 3}, in which case \eqref{eq:reparametrization action to cubic order} reproduces \eqref{eq:reparametrization action to cubic order Fourier}. Finally, integrating \eqref{eq:reparametrization action to cubic order} by parts three times, we can also write the cubic action in position space in the following manifestly finite form:
\begin{align}\label{eq:SL_3 finite}
    S_{L,3}[\epsilon]&=-\frac{T_s}{4\pi}\int dt_1 dt_2\frac{(\epsilon(t_1)^2-\epsilon(t_2)^2)(\dddot{\epsilon}(t_1)-\dddot{\epsilon}(t_2))}{t_{12}^2}. 
\end{align}

From \eqref{eq:quadratic rep action} and \eqref{eq:reparametrization action to cubic order}, we see that the longitudinal action is bilocal both at quadratic and cubic order. In appendix \ref{app:conf bilocal rep action} we show that one can write an (essentially unique) action for $\alpha$ that is non-perturbatively bilocal and has the $SL(2,\mathbb{R})\times SL(2,\mathbb{R})$ symmetry required for the longitudinal action. It is interesting to note that that action, given in eq.~\eqref{eq:bilocal action}, reproduces \eqref{eq:quadratic rep action} and \eqref{eq:reparametrization action to cubic order} when expanded in $\epsilon$ to cubic order. However, we believe that the bilocal action in \eqref{eq:bilocal action} is not the same as the longitudinal action of the AdS$_2$ string and that the two differ in the expansion in $\epsilon$ starting at fourth order.\footnote{As evidence, it is relatively straightforward to find explicit solutions to the longitudinal equations of motion in \eqref{eq:x,z eom} for certain specific choices of $\epsilon(t)$ (e.g., $\epsilon(t)=\lambda/(1+t^2)$, $\epsilon(t)=\lambda t/(1+t^2)$) up to at least third order in $\lambda$. Thus, for these values of $\epsilon(t)$, one can evaluate the on-shell longitudinal action up to fourth order in $\lambda$ by substituting the perturbative solutions into \eqref{eq:longitudinal action}. The result can be compared with the result for the bilocal action in \eqref{eq:bilocal action}, and they are found to differ at fourth order. This seems to imply that the longitudinal action is not bilocal at higher orders. \label{fn:bilocal action}}

\paragraph{Reparametrization action on the circle}

We could (and did) repeat the above analysis to determine the reparametrization action on the circle, but it is simpler to deduce it from the reparametrization action on the line using \eqref{eq:alpha beta on line and circle}-\eqref{eq: B on line and circle}. First, from \eqref{eq:t, tau, x, theta on line and circle}, we see that expanding around $\alpha(t)=t$ on the line corresponds to expanding around $\alpha(\tau)=\tau$ on the circle. If $\alpha(t)=t+\epsilon(t)$ and $\alpha(\tau)=\tau+\tilde{\epsilon}(\tau)$, then $\epsilon(t)$ and $\tilde{\epsilon}(\tau)$ are related by:
\begin{align}
    \epsilon(t)&=\frac{\tilde{\epsilon}(\tau)}{2\cos\left(\frac{\tau}{2}\right)^2}+\frac{\sin\left(\frac{\tau}{2}\right)\tilde{\epsilon}(\tau)^2}{4\cos\left(\frac{\tau}{2}\right)^3}+O(\tilde{\epsilon}^3),&\tilde{\epsilon}(\tau)&=\frac{2\epsilon(t)}{1+t^2}-\frac{2t\epsilon(t)^2}{(1+t^2)^2}+O(\epsilon^3).
\end{align}
Substituting this into \eqref{eq:quadratic rep action} and \eqref{eq:reparametrization action to cubic order}, it follows from \eqref{eq: S_L on line and circle} that the reparametrization action on the circle to cubic order is given by:
\begin{align}\label{eq:SL circle cubic order}
    S_L&=S_{L,2}+S_{L,3}+O(\tilde{\epsilon}^4),
\end{align}
where
\begin{align}
    S_{L,2}&=\frac{6T_s}{\pi}\int_{-\pi}^\pi d\tau_1 d\tau_2 \frac{\tilde{\epsilon}(\tau_1)\tilde{\epsilon}(\tau_2)}{|2\sin\left(\frac{\tau_1-\tau_2}{2}\right)|^{4+\eta}},\label{eq:S_{L,2} circle}\\
    S_{L,3}&=-\frac{12T_s}{\pi}\int_{-\pi}^\pi d\tau_1 d\tau_2 \frac{\sin(\tau_1-\tau_2)\tilde{\epsilon}(\tau_1)^2\tilde{\epsilon}(\tau_2)}{|2\sin\left(\frac{\tau_1-\tau_2}{2}\right)|^{6+\eta}}.\label{eq:S_{L,3} circle}
\end{align}
Using integration  by parts and the identity $\int d\tau |\sin{\frac{\tau}{2}}|^{-2+\eta}\to 0$ as $\eta\to 0$, we can also write the quadratic and cubic terms in manifestly finite form as follows:
\begin{align}
    S_{L,2}&=\frac{T_s}{2\pi}\int_{-\pi}^\pi d\tau_1 d\tau_2 \frac{(\dot{\epsilon}(\tau_1)-\dot{\epsilon}(\tau_2))^2-(\tilde{\epsilon}(\tau_1)-\tilde{\epsilon}(\tau_2))^2}{[2\sin\left(\frac{\tau_1-\tau_2}{2}\right)]^2},\\S_{L,3}&=-\frac{T_s}{4\pi}\int_{-\pi}^\pi d\tau_1 d\tau_2 \frac{(\dddot{\epsilon}(\tau_1)-\dddot{\epsilon}(\tau_2))(\tilde{\epsilon}(\tau_1)^2-\tilde{\epsilon}(\tau_2)^2)+(\dot{\epsilon}(\tau_1)-\dot{\epsilon}(\tau_2))(\tilde{\epsilon}(\tau_1)^2-\tilde{\epsilon}(\tau_2)^2)}{[2\sin\left(\frac{\tau_1-\tau_2}{2}\right)]^2}.
\end{align}

Finally, we can also write the reparametrization action in terms of its Fourier modes by substituting $\tilde{\epsilon}(\tau)=\sum_n \tilde{\epsilon}_n e^{-in\tau}$ into \eqref{eq:S_{L,2} circle} and \eqref{eq:S_{L,3} circle} and evaluating the $\tau_1$ and $\tau_2$ integrals using \eqref{eq:analytic reg. integral circle 2}. The result is:
\begin{align}\label{eq:S_L + O(eps^4) fourier circle}
    S_{L,2}&=2\pi T_s\sum_{n\in \mathbb{Z}}|n|(n^2-1)\tilde{\epsilon}_n \tilde{\epsilon}_{-n},&S_{L,3}&=-i\pi T_s\sum_{n,m \in \mathbb{Z}}\tilde{\epsilon}_m \tilde{\epsilon}_n \tilde{\epsilon}_{p}\text{sgn}(p)p^2(p^2-1).
\end{align}
where $p=-m-n$ in the cubic term.

\section{Connected six-point function}\label{sec:connected six-point function}
In this section, we will compute the connected contribution to the tree-level six point function by expanding about the saddle point of the reparametrization path integral. To proceed, we let $\alpha(t)=t+\epsilon(t)$ in \eqref{eq:<UUVVWW>}, expand in powers of $\epsilon(t)$, and keep the leading diagrams in which the three pairs $U_1U_2$, $V_3V_4$ and $W_5W_6$ are fully connected. 

We need the expansion in powers of $\epsilon$ of both the action and the dressed two-point function. The expansion of the action at the quadratic and cubic order was worked out in the previous section. The quadratic action determines the propagator for $\epsilon$ and the cubic action defines an interaction vertex. We use the $\epsilon$ propagator to perform Wick contractions between the $\epsilon$'s appearing in the dressed two-point function and in the cubic interaction vertex. We will denote correlators that are weighted only by the quadratic action in the reparametrization path integral as follows:
\begin{align}
    \braket{\ldots}_0\equiv \frac{1}{Z}\int \mathcal{D}\epsilon e^{-S_{L,2}[\epsilon]}\left(\ldots\right).
\end{align}

From \eqref{eq:quadratic rep action fourier}, it follows that the $\epsilon$ propagator in Fourier space is:
\begin{align}\label{eq:<eps(w_1)eps(w_2)>}
    \braket{\epsilon(\omega_1)\epsilon(\omega_2)}_0&=\frac{\pi}{T_s}\frac{1}{|\omega_1|^3}\delta(\omega_1+\omega_2),
\end{align}
which in position space becomes:
\begin{align}\label{eq:<eps(t_1)eps(t_2)> 1}
    \braket{\epsilon(t_1)\epsilon(t_2)}_0&=\frac{1}{4\pi T_s}\int d\omega \frac{e^{-i\omega t_{12}}}{|\omega|^3}.
\end{align}
This integral diverges at $\omega=0$, which reflects fact that the reparametrization action has three gauge $SL(2,\mathbb{R})$ zero modes, $\epsilon(t)=1,t,t^2$, which we should not be integrating over in the reparametrization path integral. The zero modes can be more carefully gauge fixed on the circle, where the perturbation about the saddle point, $\epsilon(t)$, can be written as a discrete sum of Fourier modes and the gauge zero modes can be cleanly isolated from the physical modes. On the line, we handle the zero modes more pragmatically. We analytically regulate \eqref{eq:<eps(t_1)eps(t_2)> 1} by replacing $3\to 3+\eta$, in which case we can evaluate the integral using \eqref{eq:analytic reg. integral line 2} and then expand in $\eta$. Keeping only the divergent and finite terms as $\eta\to 0$, we find \cite{Giombi:2022pas}
\begin{align}\label{eq:<eps(t_1)eps(t_2)> 2}
    \braket{\epsilon(t_1)\epsilon(t_2)}_0&=\frac{1}{T_s}\left[a+bt_{12}^2+\frac{1}{8\pi }t_{12}^2\log(t_{12}^2)\right].
\end{align}
Here, $a=0$ and $b=\frac{1}{4\pi}(\frac{1}{\eta}-\frac{3}{2}+\gamma_E)$. These coefficients are gauge/regularization dependent, but they drop out in the computation of any gauge $SL(2,\mathbb{R})$ invariant observable.

Next, we need the expansion of the dressed two-point function in \eqref{eq:bi-local}. Recall that $\beta(x)$ is the inverse of $\alpha(t)=t+\epsilon(t)$. Thus, when $\epsilon$ is small, the inverse has the expansion $\beta(x)=x-\epsilon(x)+\epsilon(x)\dot{\epsilon}(x)+O(\epsilon^3)$. It follows that the dressed two-point function (which we normalize for convenience by the two-point function without dressing, $1/x_{12}^2$),
\begin{align}
    \mathcal{B}(x_1,x_2)\equiv x_{12}^2 B(x_1,x_2),
\end{align}
has the following expansion:
\begin{align}
    \mathcal{B}(x_1,x_2)&=1+\mathcal{B}_1(x_1,x_1)+\mathcal{B}_2(x_1,x_2)+O(\epsilon^3),
\end{align}
where 
\begin{align}
    \mathcal{B}_1(x_1,x_2)&=-\dot{\epsilon}_1-\dot{\epsilon}_2+\frac{2\epsilon_{12}}{x_{12}},\label{eq:B1 line}\\
    \mathcal{B}_2(x_1,x_2)&=\frac{3}{x_{12}^2}\epsilon_{12}^2+\frac{2}{x_{12}}(-2\epsilon_1\dot{\epsilon}_1+2\epsilon_2\dot{\epsilon}_2+\dot{\epsilon}_1\epsilon_2-\dot{\epsilon}_2\epsilon_1)+\dot{\epsilon}_1^2+\dot{\epsilon}_2^2+\dot{\epsilon}_1\dot{\epsilon}_2+\ddot{\epsilon}_1\epsilon_1+\ddot{\epsilon}_2\epsilon_2.\label{eq:B2 line}
\end{align}
Here, we use the shorthand $\epsilon_i\equiv \epsilon(x_i)$ and $\epsilon_{ij}\equiv \epsilon(x_i)-\epsilon(x_j)$. 

\begin{figure}[t]
\centering
\begin{minipage}{0.49\hsize}
\centering
\includegraphics[clip, height=3.5cm]{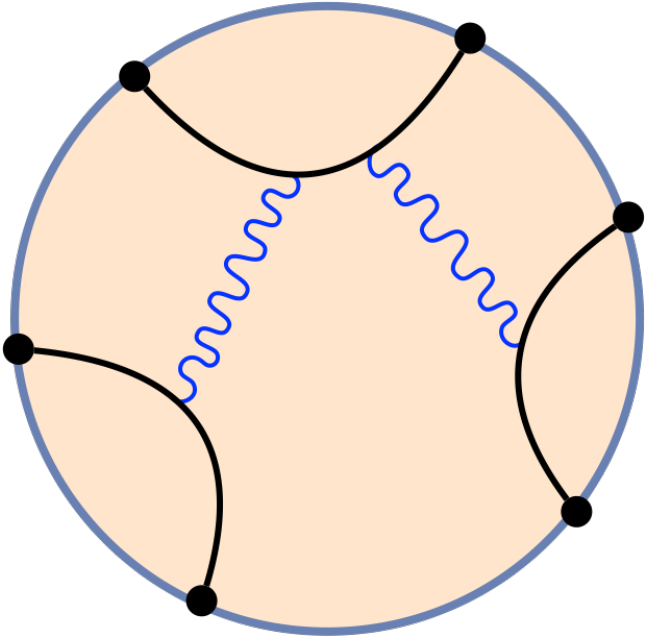}\\
{\bf a.}
\end{minipage}
\begin{minipage}{0.49\hsize}
\centering
\includegraphics[clip, height=3.5cm]{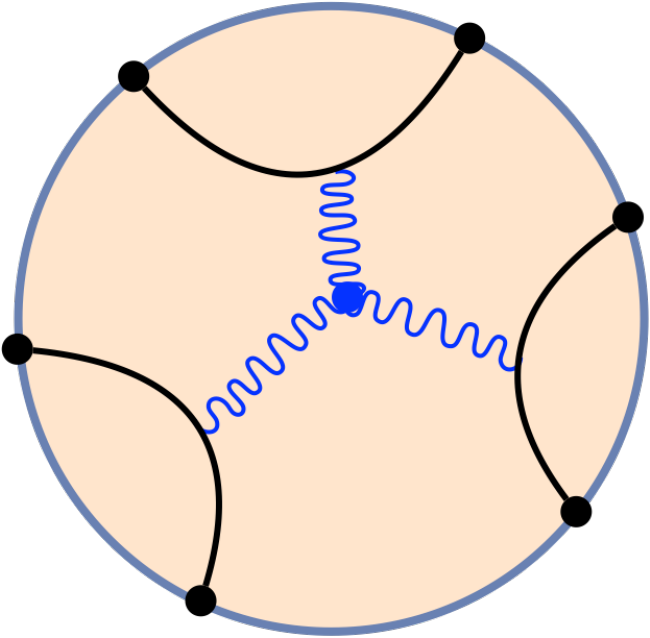}\\
{\bf b.} 
\end{minipage}
\caption{There are two general classes of reparametrization diagrams contributing to the connected six point function. \textbf{a.} In the first class, the reparametrization mode propagators connect the three dressed two-point functions without interacting. \textbf{b.} In the second class, the reparametrization propagators connect the three dressed two-point functions to a three-point interaction vertex. Each propagator brings a factor of $1/T_s$ and each (blue) interaction vertex brings a factor of $T_s$, so both classes of diagrams are of order $1/T_s^2$.}
\label{fig:six point reparametrization diagrams}
\end{figure}

The expansion of the dressed two-point functions and the interaction terms in the reparametrization action in \eqref{eq:<UUVVWW>} gives the following contributions to the connected tree-level six-point function:
\begin{align}
    &\frac{\langle U_1U_2V_3V_4 W_5W_6\rangle_{\rm c}}{\braket{U_1U_2}\braket{V_3V_4}\braket{W_5W_6}}=\braket{\mathcal{B}(x_1,x_2)\mathcal{B}(x_3,x_4)\mathcal{B}(x_5,x_6)e^{-S_{L,3}+\ldots}}_{0,\text{ c}}\nonumber\\&=
    \bigg\langle(1+\mathcal{B}_1(x_1,x_2)+\mathcal{B}_2(x_1,x_2)+\ldots)(1+\mathcal{B}_1(x_3,x_4)+\mathcal{B}_2(x_3,x_4)+\ldots)\nonumber\\&\quad\times(1+\mathcal{B}_1(x_5,x_6)+\mathcal{B}_2(x_5,x_6)+\ldots)(1-S_{L,3}+\ldots)\bigg\rangle_{0,\text{c}}\label{eq:xTHnnKJsO1}\\&=\langle \mathcal{B}_2(x_1,x_2) \mathcal{B}_1(x_3,x_4) \mathcal{B}_1(x_5,x_6)\rangle_{0,\text{c}}+(U\leftrightarrow V)+(U\leftrightarrow W)\nonumber\\&\quad-\langle \mathcal{B}_1(x_1,x_2)\mathcal{B}_1(x_3,x_4)\mathcal{B}_1(x_5,x_6)S_{L,3}\rangle_{0,\text{c}}.\label{eq:dfy52134rft5}
\end{align}

In getting from \eqref{eq:xTHnnKJsO1} to \eqref{eq:dfy52134rft5}, we kept only the leading contributions in $1/T_s$ in which the three pairs of points, $x_1,x_2$, $x_3,x_4$ and $x_5,x_6$, are fully connected. In particular, in the first line of \eqref{eq:dfy52134rft5} the subscript ``c'' indicates that we sum over all Wick contractions in which one of the two $\epsilon$'s in each term in $\mathcal{B}_2(x_1,x_2)$ is contracted with the $\epsilon$'s in each term in $\mathcal{B}_1(x_3,x_4)$, and the other $\epsilon$ in each term in $\mathcal{B}_2(x_1,x_2)$ is contracted with the $\epsilon$'s in each term in $\mathcal{B}_1(x_5,x_6)$. In other words, we do not include self-contractions between the $\epsilon$'s in $\mathcal{B}_2(x_1,x_2)$, since these would contribute to the disconnected part of the six-point function. Similarly, in the fourth line of \eqref{eq:dfy52134rft5}, the subscript ``c'' indicates that we sum over all Wick contractions in which the three $\epsilon$'s in $S_{L,3}$ are each contracted with the $\epsilon$'s in $\mathcal{B}_1(x_1,x_2)$, $\mathcal{B}_1(x_3,x_4)$, and $\mathcal{B}(x_5,x_6)$. Namely, we do not include self-contractions between the $\epsilon$'s in $S_{L,3}$. Thus, there are two types of contributions to the connected six-point function, the first of which involves only contractions using the propagator in \eqref{eq:<eps(t_1)eps(t_2)> 2} while the second of which also involves one three-point interaction of the reparametrization mode. The two classes of diagrams contributing to the leading connected six-point function are summarized in Figure~\ref{fig:six point reparametrization diagrams}.

In the second class of diagrams, the basic object we need to compute is the three-point contact diagram with three external $\epsilon$'s. It is convenient to evaluate it in Fourier space:
\begin{align}\label{eq:E(th1,th2,th3)}
    \braket{\epsilon(x_1)\epsilon(x_2)\epsilon(x_3) S_{L,3}}_{0,\text{c}}=\int \frac{d\omega_a}{2\pi}\frac{d\omega_b}{2\pi}\frac{d\omega_c}{2\pi}e^{-i\omega_ax_1-i\omega_bx_2-i\omega_cx_3}\braket{\epsilon(\omega_a)\epsilon(\omega_b)\epsilon(\omega_c)S_{L,3}}_{0,\text{c}}.
\end{align}
Given the cubic action in \eqref{eq:reparametrization action to cubic order Fourier} and given the $\epsilon$ propagator in \eqref{eq:<eps(w_1)eps(w_2)>}, it follows that 
\begin{align}
    \braket{\epsilon(\omega_a)\epsilon(\omega_b)\epsilon(\omega_c)S_{L,3}}_{0,\text{c}}&=\frac{iT_s}{4\pi^2}\int d\omega_1 d\omega_2 |\omega_1+\omega_2|^4\text{sgn}(\omega_1+\omega_2)\\&\quad\times\left[\braket{\epsilon(\omega_a)\epsilon(\omega_1)}_0\braket{\epsilon(\omega_b)\epsilon(\omega_2)}_0\braket{\epsilon(\omega_c)\epsilon(-\omega_1-\omega_2)}_0+(a\leftrightarrow c)+(b\leftrightarrow c)\right]\nonumber\nonumber\\&=\frac{i\pi}{4T_s^2}\int d\omega_1 d\omega_2 \frac{\omega_1+\omega_2}{|\omega_1|^3|\omega_2|^3}\big[\delta(\omega_a+\omega_1)\delta(\omega_b+\omega_2)\delta(\omega_c-\omega_1-\omega_2)\nonumber\\&\hspace{7cm}+(a\leftrightarrow c)+(b\leftrightarrow c)\big]
\end{align}
Using the $\delta$-functions to evaluate the integrals over $\omega_a$, $\omega_b$ and $\omega_c$, \eqref{eq:E(th1,th2,th3)} becomes:
\begin{align}\label{eq:<eps(x_1)eps(x_2)eps(x_3)S_{L,3}> 2}
    \braket{\epsilon(x_1)\epsilon(x_2)\epsilon(x_3) S_{L,3}}_{0,\text{c}}&=\frac{i}{32\pi^2 T_s^2}\int d\omega_1 d\omega_2 \frac{\omega_1+\omega_2}{|\omega_1|^3|\omega_2|^3}\left[e^{-i\omega_1 x_{31}-i\omega_2x_{32}}+(1\leftrightarrow 3)+(2\leftrightarrow 3)\right]\nonumber\\&=-\frac{1}{2}\partial_3 \left[\braket{\epsilon(x_1)\epsilon(x_3)}_0\braket{\epsilon(x_2)\epsilon(x_3)}_0\right]+(1\leftrightarrow 3)+(2\leftrightarrow 3).
\end{align}
The three-point contact diagram thus ``factorizes'' in terms of the propagator in \eqref{eq:<eps(t_1)eps(t_2)> 1}.\footnote{The clean factorization of the double integral over $\omega_1$ and $\omega_2$ in \eqref{eq:<eps(x_1)eps(x_2)eps(x_3)S_{L,3}> 2} relies on the fact that $|\omega_1+\omega_2|\text{sgn}(\omega_1+\omega_2)=\omega_1+\omega_2$. If we were more explicit about using analytic regularization to handle the $SL(2,\mathbb{R})$ zero modes--- by replacing $3\to 3+\eta$ in \eqref{eq:<eps(w_1)eps(w_2)>}--- then we would instead find $|\omega_1+\omega_2|^{1-\eta}\text{sgn}(\omega_1+\omega_2)$ in the integrand of the double integral, which does not simplify for nonzero $\eta$. One might worry then that \eqref{eq:<eps(x_1)eps(x_2)eps(x_3)S_{L,3}> 2} might receive corrections when the limit $\eta\to0$ is handled more carefully. However, in Appendix~\ref{app:rep mode 3-pt function circle}, we show that we arrive at the same result for the reparametrization mode three-point interaction working with hyperbolic disk coordinates on the worldsheet, where we have more explicit control over the zero modes.}

In order to combine the diagrams that have a cubic interaction with the diagrams without the cubic interaction, it is useful to re-package \eqref{eq:<eps(x_1)eps(x_2)eps(x_3)S_{L,3}> 2} as 
\begin{align}\label{eq:<eps(x_1)eps(x_2)eps(x_3)S_{L,3}> 3}
    \braket{\epsilon(x_1)\epsilon(x_2)\epsilon(x_3)S_{L,3}}_{0,\text{c}}=&-\frac{1}{2}\braket{\epsilon(x_1)\dot{\epsilon}(x_1)\epsilon(x_2)\epsilon(x_3)}_{0,\text{c}}-\frac{1}{2}\braket{\epsilon(x_1)\epsilon(x_2)\dot{\epsilon}(x_2)\epsilon(x_3)}_{0,\text{c}}\nonumber\\&-\frac{1}{2}\braket{\epsilon(x_1)\epsilon(x_2)\epsilon(x_3)\dot{\epsilon}(x_3)}_{0,\text{c}}
\end{align}
Here, the subscript ``c'' indicates that we include only Wick contractions in which the three points $x_1$, $x_2$ and $x_3$ are fully connected; self-contractions at the same point are excluded. The three-point interaction vertex thus effectively ``doubles'' each of the external $\epsilon$'s. We can apply this to the three-point contact interaction of the linear dressed two-point functions in the fourth line of \eqref{eq:dfy52134rft5}. Given the explicit expression \eqref{eq:B1 line}, it follows that 
\begin{align}
    \braket{\mathcal{B}_1(x_1,x_2)\mathcal{B}_1(x_3,x_4)\mathcal{B}_1(x_5,x_6)S_{L,3}}_{0,\text{c}}&=\braket{\tilde{\mathcal{B}}_2(x_1,x_2)\mathcal{B}_1(x_3,x_4)\mathcal{B}_1(x_5,x_6)}_{0,\text{c}}\nonumber\\&+\braket{\mathcal{B}_1(x_1,x_2)\tilde{\mathcal{B}}_2(x_3,x_4)\mathcal{B}_1(x_5,x_6)}_{0,\text{c}}\nonumber\\&+\braket{\mathcal{B}_1(x_1,x_2)\mathcal{B}_1(x_3,x_4)\tilde{\mathcal{B}}_2(x_5,x_6)}_{0,\text{c}}
\end{align}
where we define $\tilde{\mathcal{B}}_2$, the ``doubled'' version of $\mathcal{B}_1$ in accordance with \eqref{eq:<eps(x_1)eps(x_2)eps(x_3)S_{L,3}> 3} to be:
\begin{align}
    \tilde{\mathcal{B}}_2(x_1,x_2)&=\frac{1}{2}\left[\ddot{\epsilon}_1\epsilon_1+\epsilon_1^2+\ddot{\epsilon}_2\epsilon_2+\epsilon_2^2\right]-\frac{1}{x_{12}}\left[\epsilon_1\dot{\epsilon}_1-\epsilon_2\dot{\epsilon}_2\right].
\end{align}
Because $\tilde{B}_2$ is also quadratic in $\epsilon$, it is natural to combine it with $\mathcal{B}_2$ into an ``effective'' second-order correction to the dressed two-point function:
\begin{align}
    \mathcal{B}_2^{\rm eff}(x_1,x_2)=\mathcal{B}_2(x_1,x_2)-\tilde{\mathcal{B}}_2(x_1,x_2).
\end{align}
The minus sign in front of $\tilde{B}_2$ comes from the fourth line in \eqref{eq:dfy52134rft5}.

Given the steps taken thus far, we can write the connected six-point function in \eqref{eq:dfy52134rft5} as:
\begin{align}\label{eq:fr24iOTFOw}
    \frac{\langle U_1U_2V_3V_4 W_5W_6\rangle_{\rm c}}{\braket{U_1U_2}\braket{V_3V_4}\braket{W_5W_6}}&=\langle \mathcal{B}^{\rm eff}_2(x_1,x_2) \mathcal{B}_1(x_3,x_4) \mathcal{B}_1(x_5,x_6)\rangle_{0,\text{c}}+(U\leftrightarrow V)+(U\leftrightarrow W)
\end{align}
This is progress compared to \eqref{eq:dfy52134rft5} because all we need to do now is contract the $\epsilon$'s appearing in $\mathcal{B}_2^{\rm eff}$ and $\mathcal{B}_1$ using the propagator in \eqref{eq:<eps(t_1)eps(t_2)> 2}. 

In order to organize the contractions, it is helpful to write $\mathcal{B}_2^{\rm eff}$ as the square of the linear dressed two-point function, plus some remainder:
\begin{align}
    \mathcal{B}_2^{\rm eff}(x_1,x_2)&=\frac{1}{2}\mathcal{B}_1(x_1,x_2)^2+\mathcal{B}_2^{\rm rem}(x_1,x_2),
\end{align}
where the remainder is given explicitly by:
\begin{align}
    \mathcal{B}_2^{\rm rem}(x_1,x_2)&=\frac{1}{2}\ddot{\epsilon}_1\epsilon_1+\frac{1}{2}\ddot{\epsilon}_2\epsilon_2-\frac{\dot{\epsilon}_1\epsilon_1+\dot{\epsilon}_2\epsilon_2}{x_{12}}+\frac{(\epsilon_1-\epsilon_2)^2}{x_{12}^2}\nonumber\\&=-\frac{1}{2}\left(\epsilon_1\partial_{x_1}+\epsilon_2\partial_{x_2}\right)\mathcal{B}_1(x_1,x_2).
\end{align}
In this way, \eqref{eq:fr24iOTFOw} becomes:
\begin{align}\label{eq:mbDxGk7bQG}
    \frac{\langle U_1U_2V_3V_4 W_5W_6\rangle_{\rm c}}{\braket{U_1U_2}\braket{V_3V_4}\braket{W_5W_6}}&=\langle \mathcal{B}^{\rm rem}_2(x_1,x_2) \mathcal{B}_1(x_3,x_4) \mathcal{B}_1(x_5,x_6)\rangle_{0,\text{c}}\nonumber\\&\quad+\frac{1}{2}\langle \mathcal{B}_1(x_1,x_2)^2 \mathcal{B}_1(x_3,x_4) \mathcal{B}_1(x_5,x_6)\rangle_{0,\text{c}}\nonumber\\&\quad+(U\leftrightarrow V)+(U\leftrightarrow W)
\end{align}

Now we note that the second line in \eqref{eq:mbDxGk7bQG} factorizes:
\begin{align}\label{eq:<B_1B_1>_0}
    \langle \mathcal{B}_1(x_1,x_2)^2 \mathcal{B}_1(x_3,x_4) \mathcal{B}_1(x_5,x_6)\rangle_{0,\text{c}}&=2\braket{\mathcal{B}_1(x_1,x_2)\mathcal{B}_1(x_3,x_4)}_0\braket{\mathcal{B}_1(x_1,x_2)\mathcal{B}_1(x_5,x_6)}_0,
\end{align}
Each of the two factors on the r.h.s. is just the tree-level contribution to the connected four-point function--- namely, 
\begin{align}\label{eq:connected four-point function}
    \frac{\braket{U(x_i)U(x_j)V(x_k)V(x_l)}_c}{\braket{U(x_i)U(x_j)}\braket{V(x_k)V(x_l)}}=\braket{\mathcal{B}_1(x_i,x_j)\mathcal{B}_1(x_k,x_l)}_{0}+O(1/T_s^2)\equiv G(\xi_{ijkl})+O(1/T_s^2),
\end{align}
and similarly for the $U,W$ and $V,W$ four-point functions. Here, 
\begin{align}
    \xi_{ijkl}\equiv \frac{x_{ik}x_{jl}}{x_{il}x_{jk}}.
\end{align}
is a conformally invariant cross ratio and, given \eqref{eq:B1 line} and the propagator in \eqref{eq:<eps(t_1)eps(t_2)> 1}, one finds \cite{Giombi:2022pas}
\begin{align}\label{eq:G(xi)}
    G(\xi)&=-\frac{1}{4\pi T_s}\left(4+\frac{1+\xi}{1-\xi}\log(\xi^2)\right).
\end{align}
This result respects both the gauge and physical $SL(2,\mathbb{R})$ symmetries: i.e., it depends on the positions only through the cross-ratio and is independent of the gauge-dependent coefficients $a$ and $b$ in \eqref{eq:<eps(t_1)eps(t_2)> 2}.  

Meanwhile, the remainder term in the first line in \eqref{eq:mbDxGk7bQG} becomes:
\begin{align}
    \braket{\mathcal{B}_2^{\rm rem}(x_1,x_2)\mathcal{B}_1(x_3,x_4)\mathcal{B}_1(x_5,x_6)}_{0,\text{c}}=&-\frac{1}{2}\braket{\epsilon_1\mathcal{B}_1(x_3,x_4)}\partial_{x_1}\braket{\mathcal{B}_1(x_1,x_2)\mathcal{B}_1(x_5,x_6)}\\&-\frac{1}{2}\braket{\epsilon_2\mathcal{B}_1(x_3,x_4)}\partial_{x_2}\braket{\mathcal{B}_1(x_1,x_2)\mathcal{B}_1(x_5,x_6)}+(3,4\leftrightarrow 5,6).\nonumber
\end{align}
The (derivative of the) four-point function in \eqref{eq:<B_1B_1>_0} appears again as one component of the result. Meanwhile, using the propagator in \eqref{eq:<eps(t_1)eps(t_2)> 2}, we find that the remaining terms are given by:
\begin{align}\label{eq:tgft543edf54}
    \braket{\epsilon(x_i)\mathcal{B}_1(x_k,x_l)}&=\frac{1}{4\pi T_s}\left(2x_i-x_k-x_l+\frac{x_{ik}x_{il}}{x_{kl}}\log\left(\frac{x_{ik}^2}{x_{il}^2}\right)\right).
\end{align}
(This result is also independent of the gauge dependent coefficients $a$ and $b$.) Then we can use the chain rule and the fact that 
\begin{align}
    \partial_{x_1}\xi_{1256}&=\frac{x_{56}}{x_{15}x_{16}}\xi_{1256},&\partial_{x_2}\xi_{1256}&=-\frac{x_{56}}{x_{25}x_{26}}\xi_{1256},
\end{align}
to write
\begin{align}\label{eq:t5434543}
    &\braket{\mathcal{B}_2^{\rm rem}(x_1,x_2)\mathcal{B}_1(x_3,x_4)\mathcal{B}_1(x_5,x_6)}_{\rm 0, c}\nonumber\\&=-\frac{1}{8\pi T_s}\biggr[\left(2x_1-x_3-x_4+\frac{x_{13}x_{14}}{x_{34}}\log\left(\frac{x_{13}^2}{x_{14}^2}\right)\right)\frac{x_{56}}{x_{15}x_{16}}\nonumber\\&\hspace{2cm}-\left(2x_2-x_3-x_4+\frac{x_{23}x_{24}}{x_{34}}\log\left(\frac{x_{23}^2}{x_{24}^2}\right)\right)\frac{x_{56}}{x_{25}x_{26}}\biggr]\xi_{1256}G'(\xi_{1256})\nonumber\\&\quad+(3,4\leftrightarrow 5,6).
\end{align}
Although this expression is not particularly nice, there are simplifications after we sum over the permutations of $x_1,x_2$, $x_3,x_4$ and $x_5,x_6$, in accordance with \eqref{eq:mbDxGk7bQG}. For example, consider all the terms in \eqref{eq:mbDxGk7bQG} coming from the $\mathcal{B}_2^{\rm rem}$ terms that multiply $\xi_{1256}G'(\xi_{1256})$. These are given by the square brackets in \eqref{eq:t5434543} plus the same thing with $1,2\leftrightarrow 5,6$. First, we note:
\begin{align}
    &(2x_1-x_3-x_4)\frac{x_{56}}{x_{15}x_{16}}-(2x_2-x_3-x_4)\frac{x_{56}}{x_{25}x_{26}}\nonumber\\&+(2x_5-x_3-x_4)\frac{x_{12}}{x_{15}x_{25}}-(2x_6-x_3-x_4)\frac{x_{12}}{x_{16}x_{26}}=0
\end{align}
Thus, the ``$2x_i-x_j-x_k$'' terms in \eqref{eq:t5434543} cancel after summing over permutations.

Second, we denote the remaining terms multiplying $\xi_{1256}G'(\xi_{1256})$ by $F_{34;1256}$, where
\begin{align}\label{eq:curly F}
    F_{ij;klmn}\equiv &-\frac{1}{8\pi T_s}\bigg[\frac{x_{ki}x_{kj}x_{mn}}{x_{ij}x_{km}x_{kn}}\log\left(\frac{x_{ki}^2}{x_{kj}^2}\right)-\frac{x_{li}x_{lj}x_{mn}}{x_{ij}x_{lm}x_{ln}}\log\left(\frac{x_{li}^2}{x_{lj}^2}\right)\nonumber\\&+\frac{x_{im}x_{jm}x_{kl}}{x_{ij}x_{km}x_{lm}}\log\left(\frac{x_{im}^2}{x_{jm}^2}\right)-\frac{x_{in}x_{jn}x_{kl}}{x_{ij}x_{kn}x_{ln}}\log\left(\frac{x_{in}^2}{x_{jn}^2}\right)\bigg].
\end{align}
It is easy to check that this is conformally invariant. Note that it has the permutation symmetries $F_{ij;klmn}=F_{ij;mnkl}=F_{ji;klmn}=-F_{ij;lkmn}$. 

When we sum all the contributions to \eqref{eq:mbDxGk7bQG}, we arrive at our final result for the tree-level connected six-point function:
\begin{align}\label{eq:6-pt function result}
    &\frac{\langle U_1U_2V_3V_4 W_5W_6\rangle_{\rm c}}{\braket{U_1U_2}\braket{V_3V_4}\braket{W_5W_6}}=G(\xi_{1234})G(\xi_{1256})+G(\xi_{1234})G(\xi_{3456})+G(\xi_{1256})G(\xi_{3456})\nonumber\nonumber\\&\hspace{3cm}+F_{12;3456}\xi_{3456}G'(\xi_{3456})+F_{34;1256}\xi_{1256}G'(\xi_{1256})+F_{56;1234}\xi_{1234}G'(\xi_{1234}).
\end{align}
If desired, the normalized six-point function can be expressed in terms of three independent cross-ratios $\chi_i=\frac{x_{1i}x_{56}}{x_{15}x_{i6}}$ for $i=2,3,4$ by setting $x_1=0$, $x_5=1$, and $x_6=\infty$, in which case $\chi_i=x_i$.

\section{Six-point OTOC and scattering on the worldsheet}
\label{sec:six-point-OTOC}

In this section, we will analytically continue the euclidean six-point function that we computed in the previous section to a Lorentzian configuration in which the operators are out of time order. Like four-point out-of-time-order correlators (OTOCs) \cite{Larkin1969QuasiclassicalMI,KitaevTalks,Shenker:2013pqa,Shenker:2014cwa,Maldacena:2015waa}, higher-point OTOCs can serve as diagnostics of quantum chaos \cite{Haehl:2017pak,Haehl:2017qfl,Haehl:2021tft}. In holographic theories, they have an interpretation in the bulk in terms of high energy multi-particle scattering or, relatedly, in terms of particles propagating on backgrounds with multiple shockwaves \cite{Shenker:2013pqa,Shenker:2013yza,Shenker:2014cwa}. Studying the six-point OTOC on the AdS$_2$ string and comparing it with the behavior expected from a scattering analysis provides one check of our result for the connected six-point function and of the conformal gauge method.\footnote{Our set-up in this section is somewhat similar to the one in \cite{Haehl:2021tft}, which studied various six-point OTOCs in JT gravity and in Einstein gravity in AdS$_3$.}

\begin{figure}
    \centering
    \begin{minipage}{0.49\textwidth}
        \centering
        \vspace{0.2cm}
        \includegraphics[width=0.6\textwidth]{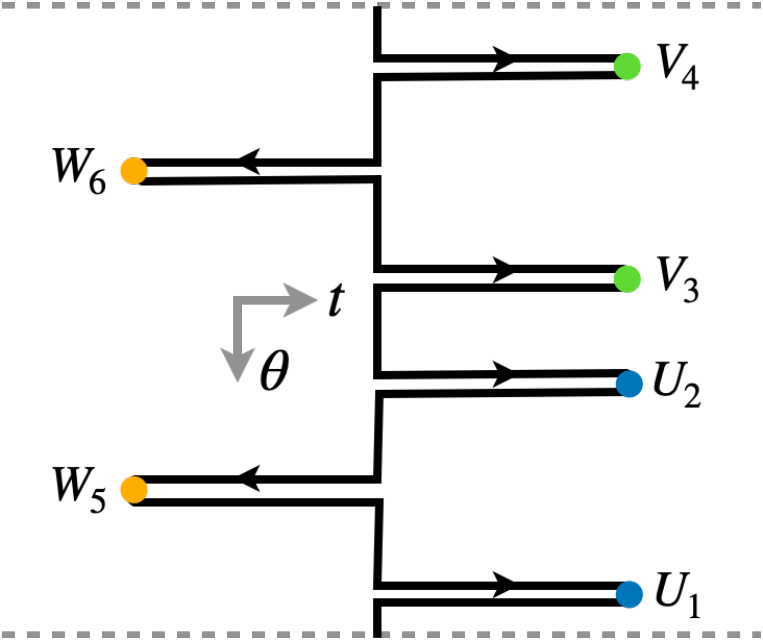}\\
        \vspace{0.2cm}
        
        {\bf a. }
    \end{minipage}
    \begin{minipage}{0.49\textwidth}
        \centering
        \includegraphics[width=0.6\textwidth]{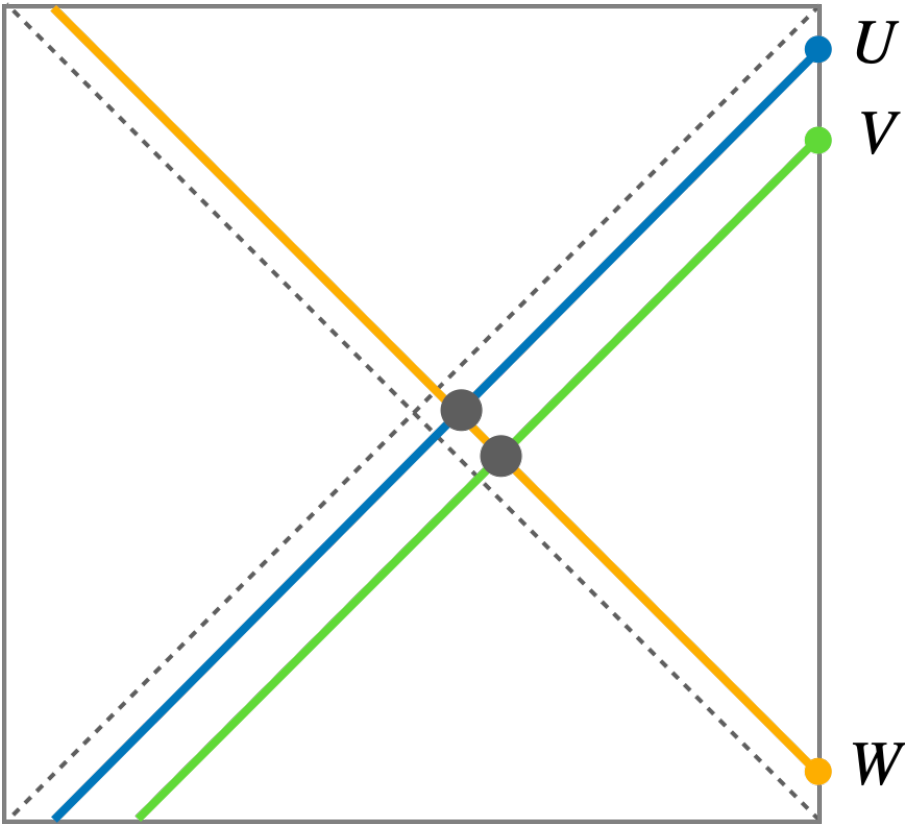}\\
        {\bf b. }
    \end{minipage}
    \caption{\textbf{a.} The contour specifying the ordering and positions on the thermal cylinder of the operators in the six-point OTOC. \textbf{b.} The OTOC has the physical interpretation of a $3$-to-$3$ scattering process between a left moving particle produced by $W$ and two right-moving particles produced by $U$ and $V$.}
    \label{fig:OTOC contour and scattering process}
\end{figure}

We start with the six-point function on the euclidean circle with the following ordering of operators:
\begin{align}\label{eq:6 pt OTOC}
    \braket{U_1W_5U_2V_3W_6V_4},
\end{align}
where $U_i\equiv U(\theta_i)$, etc. We will work for simplicity with the unit circle so that the euclidean time is an angle $\theta\sim \theta+2\pi$. This corresponds to an inverse temperature $\beta=2\pi$. We then analytically continue the positions of the operators, $\theta_i$, to the following values
\begin{equation}\label{eq:euclid times analytic cont}
    \begin{aligned}
        \theta_1&=\delta_1+iT_1,&\theta_2&=\delta_2+iT_1,&\theta_3&=\delta_3+ i T_2,\\\theta_4&=\delta_4+ i T_2,&\theta_5&=\delta_5+iT_3,&\theta_6&=\delta_6+iT_3,
    \end{aligned}
\end{equation}
where the $\delta_i$ are fixed and chosen such that $\delta_4<\delta_6<\delta_3<\delta_2<\delta_5<\delta_1$ up to cyclic permutations, and we tune $T_1,T_2,T_3$ from zero to non-zero real values. For concreteness we will choose the $\delta_i$ such that the six operators are spaced equally around the circle--- e.g.,\footnote{Different choices of $\delta_i$ with the same cyclic ordering are related by trivial analytic continuations, so one can switch freely between them. For example, while \eqref{eq:symmetric config} is particularly simple, other choices can make the physical interpretation of the OTOC clearer--- e.g., $|\delta_i|\ll 1$, which  corresponds to an almost purely Lorentzian configuration in which the small non-zero $\delta_i$ serve mainly to determine the order of operators and to regulate divergences due to operators being coincident.}
\begin{align}\label{eq:symmetric config}
    \delta_1&=\frac{4\pi}{3}, &\delta_2&=\frac{2\pi}{3}, &\delta_3&=\frac{\pi}{3}, &\delta_4&=-\frac{\pi}{3},& \delta_5&=\pi, &\delta_6&=0.
\end{align}
This configuration is depicted in Figure~\ref{fig:OTOC contour and scattering process}a. We will focus on the simplifying regime when the Lorentzian time differences between the operators are large. In particular, we will take the double-scaling limit
\begin{align}\label{eq:config 1 ds limit}
    T_{13},T_{23}&\to \infty, &T_s&\to \infty, &\frac{e^{T_{13}}}{T_s},\frac{e^{T_{23}}}{T_s}:\text{ fixed}.
\end{align}
This can be achieved by, for instance, sending $U$ and $V$ to late times in the future ($T_1,T_2\to \infty$) and $W$ to early times in the past ($T_3\to -\infty$). The OTOC in this configuration has the interpretation of high energy $3$-to-$3$ scattering involving a left-moving particle created by $W$ interacting with two right-moving particles created by $U$ and $V$ on an AdS$_2$ background. See Figure~\ref{fig:OTOC contour and scattering process}b. There are other interesting configurations that one can consider, and we will comment on one in particular in the conclusion.

\subsection{OTOC as a scattering amplitude on the AdS$_2$ string}
Because scattering is a Lorentzian phenomena, a natural first step of our discussion of the OTOC as a scattering amplitude is to analytically continue the euclidean AdS$_2$ string to Lorentzian signature --- rather than continuing just the final correlator as in \eqref{eq:6 pt OTOC}-\eqref{eq:euclid times analytic cont}. We start with the euclidean AdS$_2$ string embedded in euclidean AdS$_3$. Let the metric in AdS$_3$ be $ds^2=\frac{dx_0^2+dx_1^2+dz^2}{z^2}$, and let the AdS$_2$ string worldsheet be the hemisphere $x_0^2+x_1^2+z^2=1$. It is incident on the circle on the boundary with unit radius given by $x_0(\theta)=\sin{\theta}$, $x_1(\theta)=\cos{\theta}$.

\begin{figure}[t!]
    \centering
    \begin{minipage}{0.49\textwidth}
        \centering
        \includegraphics[width=0.6\textwidth]{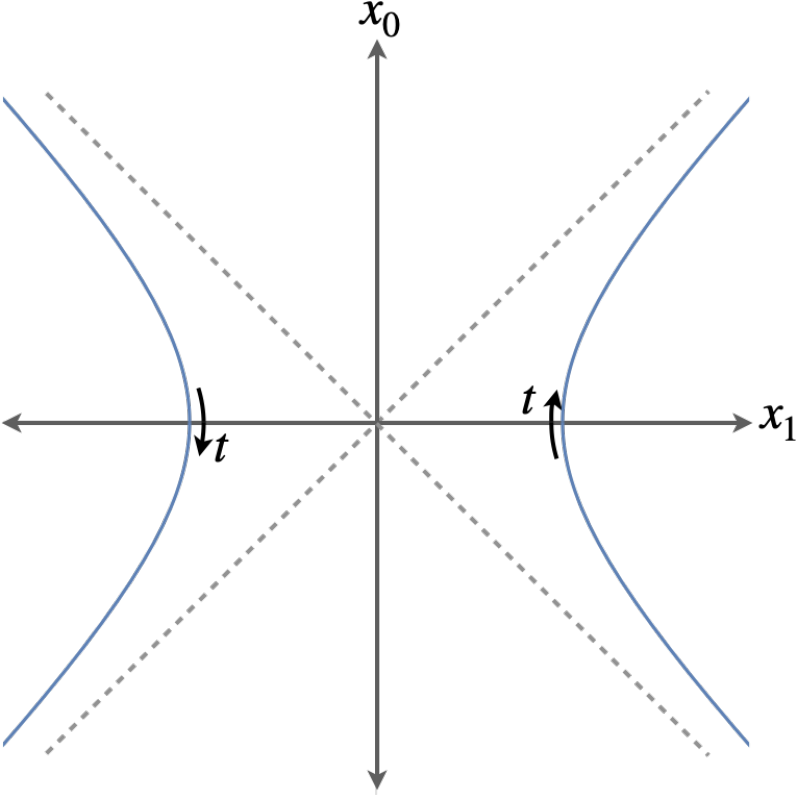}\\
        {\bf a. }
    \end{minipage}
    \begin{minipage}{0.49\textwidth}
        \centering
        \vspace{0.7cm}
        
        \includegraphics[width=0.8\textwidth]{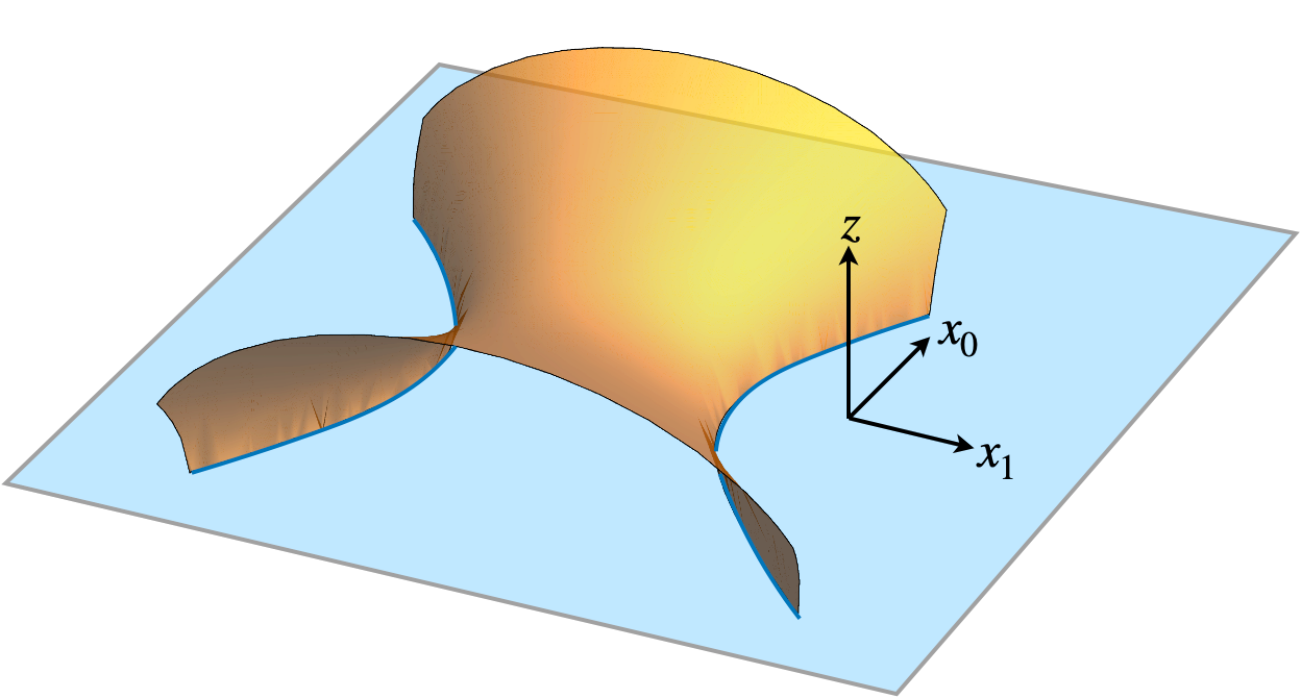} \\
        \vspace{0.7cm}
        
        {\bf b. }
    \end{minipage}
    \caption{{\bf a. } The worldlines of two uniformly accelerating quarks in Minkowski space. {\bf b. } The classical surface in the bulk incident on the two quark trajectories on the boundary is a hyperboloid.}
    \label{fig:two quarks accelerating}
\end{figure}

Now we send $x_0\to ix_0$, $\theta \to i t$. The metric in the Poincar\'e patch of Lorentzian AdS$_3$ becomes $ds^2=\frac{-dx_0^2+dx_1^2+dz^2}{z^2}$, the Lorentzian AdS$_2$ string worldsheet becomes the hyperboloid $-x_0^2+x_1^2+z^2=1$ and it is incident on the boundary curve $x_0(t)=\sinh{t}$, $x_1(t)=\cosh{t}$. This boundary curve defines the worldline of a quark undergoing constant proper acceleration $a=1$ with proper time $t$. See Figure~\ref{fig:two quarks accelerating}. The second branch of the hyperbola on the boundary--- $x_0(t)=-\sinh{t}$, $x_1(t)=-\cosh{t}$--- defines the worldline of an antiquark undergoing constant proper acceleration $a=-1$, and it can also be accessed by continuing $\tau\to it+\pi$. Thus, one can also analytically continue the euclidean correlators on the circle to ``two sided'' configurations on the hyperbola, but we will focus on one-sided configurations for concreteness.

\begin{figure}[t!]
\centering
\begin{minipage}{0.49\hsize}
\centering
\includegraphics[clip, height=4cm]{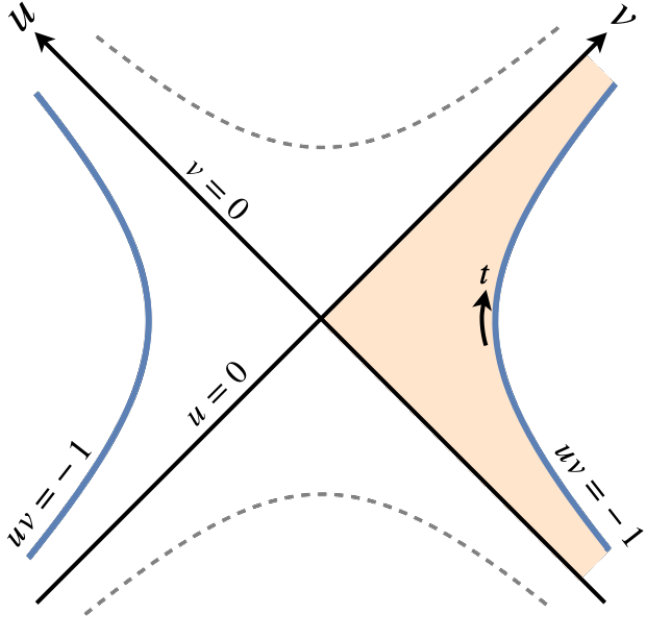}\\
{\bf a. }
\end{minipage}
\begin{minipage}{0.49\hsize}
\centering
\includegraphics[clip, height=4cm]{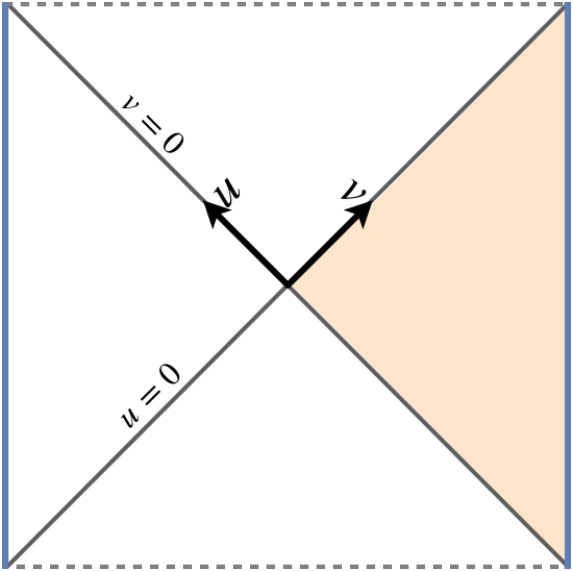}\\
{\bf b.} 
\end{minipage}
\caption{The \textbf{a.} Kruskal and \textbf{b.} Penrose diagrams for the AdS$_2$ ``black hole.'' The shaded wedge denotes the region causally accessible to the right accelerating quark.}
\label{fig:AdS2 BH Kruskal and Penrose}
\end{figure}

It will be useful to introduce Kruskal light-cone coordinates $u$ and $v$ that cover the entire worldsheet:
\begin{align}\label{eq:huytrfde345r}
    x_0&=\frac{u+v}{1-uv},&x_1&=-\frac{u-v}{1-uv},&z&=\frac{1+uv}{1-uv}.
\end{align}
In these coordinates, the induced metric is $ds^2=-\frac{4dudv}{(1+uv)^2}$ and the AdS boundary is at $uv=-1$. One the right boundary, $u$ and $v$ are related to the proper time $t$ by $u=-e^{-t}$ and $v=e^{t}$. Finally, we note that the proper time is periodic with $t\sim t+2\pi i$, so that the Unruh temperature felt by the accelerating quark is $\beta^{-1}=\frac{1}{2\pi}$. This set-up is sometimes called the AdS$_2$ wormhole or the holographic EPR pair \cite{Xiao:2008nr,Jensen:2013ora,Sonner:2013mba}. The Kruskal and Penrose diagrams for the worldsheet are given in Figure~\ref{fig:AdS2 BH Kruskal and Penrose}.

Next, we will interpret the six-point OTOC as a scattering process on the AdS$_2$ background. The basic idea is that acting with $U$, $V$ and $W$ along the contour of the Wilson operator creates particles that propagate on the string worldsheet, and the six-point OTOC can be written as an overlap of an in state and an out state--- i.e., as an S-matrix. Additional ingredients in the following analysis include the fact that, because $U$ and $V$ are separated from $W$ by large times, the particles they create are highly boosted and the scattering process on the string worldsheet is high energy and localized to an essentially flat region of AdS$_2$. Moreover, we will assume that the $3$-to-$3$ S-matrix factorizes into a product of two $2$-to-$2$ $S$-matrices, one for the interaction between $W$ and $U$ and one for the interaction between $W$ and $V$, as depicted in Figure~\ref{fig:OTOC contour and scattering process}b.\footnote{
Unlike a $3$-to-$3$ scattering matrix of massive particles in an integrable theory, which factorizes into three $2$-to-$2$ scattering matrices, the $3$-to-$3$ scattering matrix corresponding to the string six-point function will factorize into two $2$-to-$2$ scattering matrices because the scattering particles are massless. The masslessness can be seen either as an exact property because the scalars on the Wilson line have dimension $\Delta=1$ corresponding to $m^2=0$ in AdS$_2$, or as an approximation that should also be valid for other operators (like the displacement operators with $\Delta=2$ corresponding to $m^2=2$) because the scattering particles are highly boosted.} The OTOC in this limit is thus fully determined by the $2$-to-$2$ S-matrix for scattering on the string worldsheet in flat space that was studied in \cite{Dubovsky:2012wk}, and by the wavefunctions for the in and out states.

\begin{figure}
    \centering
    \begin{minipage}{0.49\textwidth}
        \centering
        \includegraphics[width=0.6\textwidth]{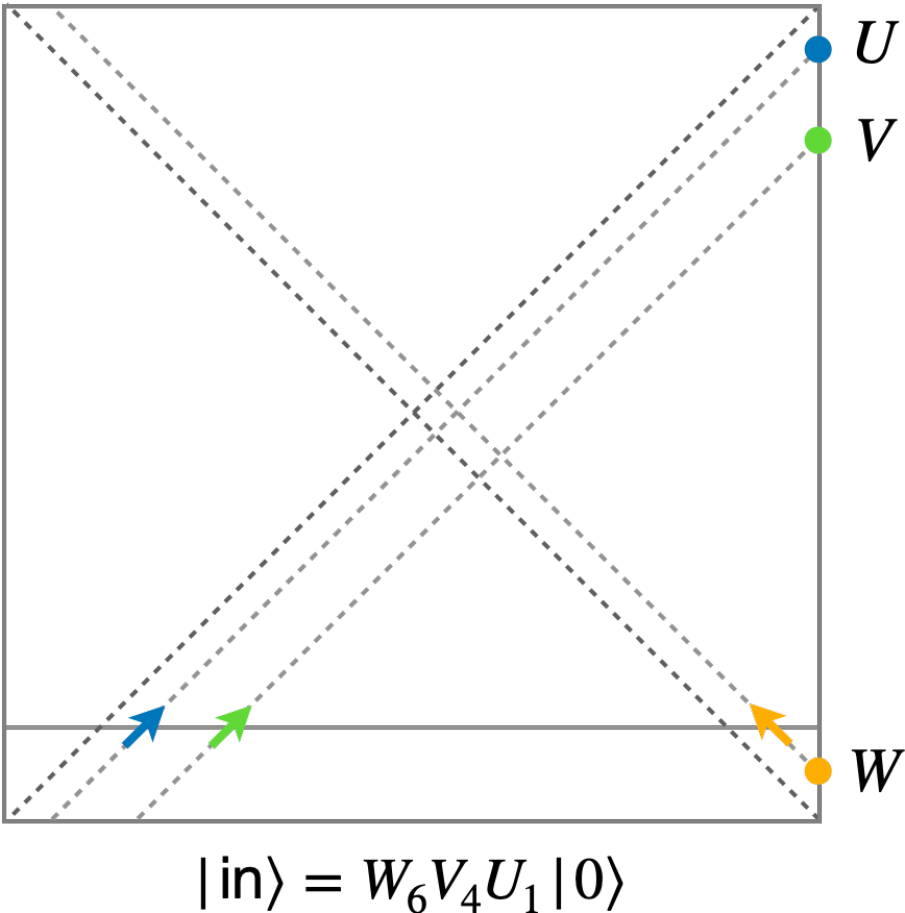}
    \end{minipage}
    \begin{minipage}{0.49\textwidth}
        \centering
        \includegraphics[width=0.6\textwidth]{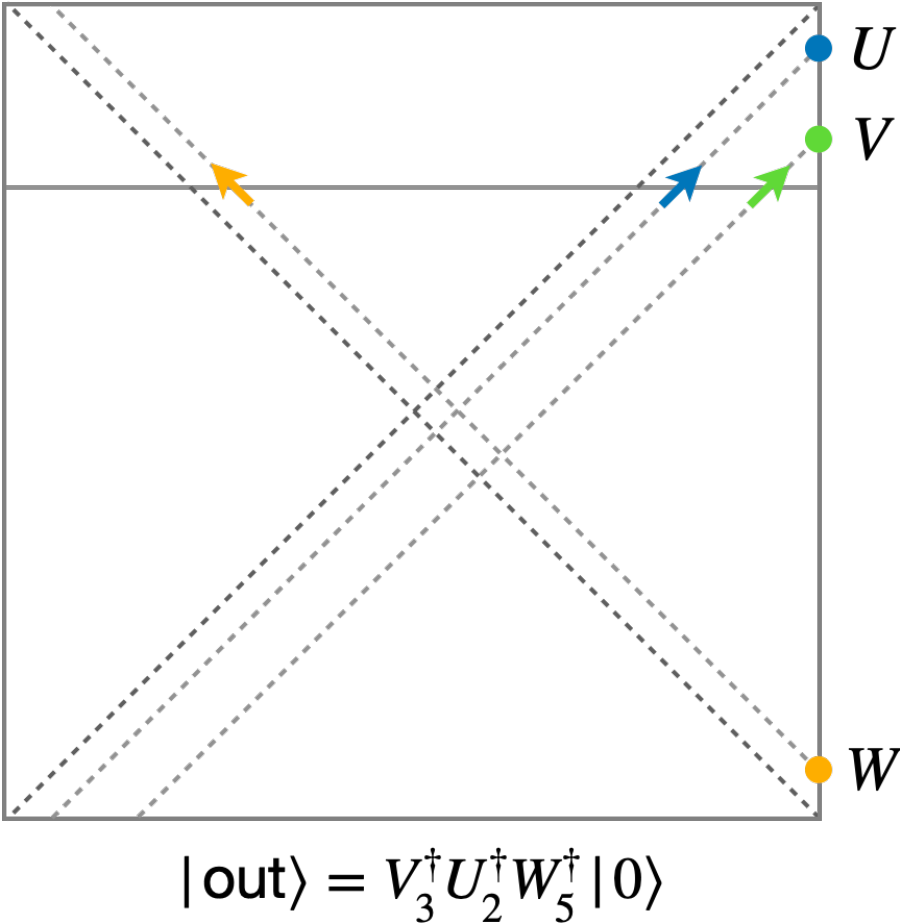}  
    \end{minipage}
    \caption{In and out scattering states for the six-point OTOC.}
    \label{fig:OTOC 6pt In/Out}
\end{figure}

To implement the above procedure in detail,\footnote{See \cite{Shenker:2014cwa} for the four-point scattering analysis in Einstein gravity and \cite{deBoer:2017xdk,Giombi:2022pas} for its adaptation to the worldsheet.} we first write the correlator as an overlap between two states:
\begin{align}\label{eq:rtgvdert54es}
    \braket{W_5U_2V_3W_6V_4U_1}&=\braket{\text{out}|\text{in}}.
\end{align}
Here, we have used a cyclic permutation to move $U(\theta_1)$ to the rightmost position in the correlator in \eqref{eq:6 pt OTOC} and we have defined the states
\begin{align}\label{eq:in/out states}
    \ket{\rm in}&=W_6V_4U_1\ket{0},&\ket{\rm out}&=V_3^\dagger U_2^\dagger W_5^\dagger\ket{0}.
\end{align}
These are naturally interpreted as ``in'' and ``out'' states for three particles propagating in AdS$_2$, which we can understand heuristically as follows (see \cite{Shenker:2014cwa}). In setting up the ``in'' state, we start with the vacuum $\ket{0}$ at $t=0$, evolve forward to time $T_1>0$ and act with the operator $U$ to create a particle on the boundary, evolve to time $T_2>0$ and create a second particle on the boundary, and, finally, evolve all the way back to time $T_3<0$ and act with $W$ to create a third particle on the boundary. We assume that the particles move along lightlike trajectories as we evolve forward and backward in time and that they propagate freely unless their trajectories cross. Given the order in which we act with $U$, $V$ and $W$ on the vacuum state, the trajectories of the three particles do not cross while setting up the state, so the end result of this procedure is to produce three separated particles in the far past, two of which are right-moving particles (carrying positive lightcone momentum $p^u$) in the bottom left of the Penrose diagram and one of which is a left-moving particle (carrying positive lightcone momentum $p^v$) in the bottom right of the Penrose diagram.\footnote{
One might ask whether it is sensible to define asymptotic in and out states on the two dimensional worldsheet that involve two or more massless particles moving in the same direction, given that the particles do not become well-separated in either the infinite past or the infinite future and therefore ostensibly interact ``forever.'' This is a reasonable objection, but for the purpose of performing a check of the analytically continued six-point function by comparing it with a scattering amplitude on the worldsheet, it appears to be valid to ignore the interactions between the parallel-moving $U$ and $V$ particles. Intuitively, this seems to be related to the fact that $U$ and $V$ are in relative time order in \eqref{eq:rtgvdert54es}. Nonetheless, it would be good to have a better understanding of the subtleties associated with massless scattering in two dimensions. A related discussion can be found in section 2 of \cite{Dubovsky:2012wk}.} An analogous interpretation can be given to the out state. The in and out states are depicted in Figure~\ref{fig:OTOC 6pt In/Out}.

Next, we write the in and out states in terms of in and out  Kruskal momentum eigenstates:
\begin{align}
    \ket{\rm in}&=\int dp_1^v dp_4^v dp_6^u \Phi_{\Delta_U}(p_1^v,t_1)\Phi_{\Delta_V}(p_4^v,t_4)\Psi_{\Delta_W}(p_6^u,t_6)\ket{p_1^v,p_4^v,p_6^u}_{\rm in},\label{eq:in state in Kruskal momentum basis}\\\
    \ket{\rm out}&=\int dp_2^v dp_3^v dp_5^u \Phi_{\Delta_U}(p_2^v,t_2^*)\Phi_{\Delta_V}(p_3^v,t_3^*)\Psi_{\Delta_W}(p_5^u,t_5^*)\ket{p_2^v,p_3^v,p_5^u}_{\rm out}.\label{eq:out state in Kruskal momentum basis}
\end{align}
Here, $\Phi_{\Delta}(p^v,t)$ is the momentum-space wavefunction for a right-moving particle with lightcone momentum $p^v$ that is created by a local operator of conformal dimension $\Delta$ acting at lorentzian time $t$ on the right AdS$_2$ boundary; likewise $\Psi_{\Delta}(p^u,t)$ is the wavefunction for a left-moving particle with lightcone momentum $p^u$. We use $t=-i\theta$ to denote the Lorentzian time, and in \eqref{eq:out state in Kruskal momentum basis} we used $O(t)^\dagger=O(t^*)$ if $O$ is Hermitian. Finally, we take the momentum eigenstates to be normalized as
\begin{align}
    {_{\rm in}\braket{p|q}_{\rm in}} = {_{\rm out}\braket{p|q}_{\rm out}} = p\delta(p-q).
\end{align}

Taking the inner product of \eqref{eq:in state in Kruskal momentum basis} and \eqref{eq:out state in Kruskal momentum basis}, the OTOC in \eqref{eq:euclid times analytic cont} can be expressed as the following amplitude:
\begin{align}\label{eq:out|in momentum integral rep}
    \braket{\rm out|in}&=\int \prod_i dp_i\Phi_{\Delta_U}(p_2^v,t_2^*)^*\Phi_{\Delta_V}(p_3^v,t_3^*)^*\Psi_{\Delta_W}(p_5^u,t_5^*)^* \text{ }  _{\rm out}\braket{p_2^v,p_3^v,p_5^u|p_1^v,p_4^v,p_6^u}_{\rm in}\nonumber\\&\hspace{2cm}\times\Phi_{\Delta_U}(p_1^v,t_1)\Phi_{\Delta_V}(p_4^v,t_4)\Psi_{\Delta_W}(p_6^u,t_6)
\end{align}
To evaluate this expression, we need to know the momentum-space S-matrix and the wavefunctions. The wavefunctions are simply the Fourier transforms of the boundary-to-bulk propagator with respect to $u$ on the $v=0$ horizon and with respect to $v$ on the $u=0$ horizon, respectively. Explicitly, they are \cite{deBoer:2017xdk,Lam:2018pvp,Giombi:2022pas}
\begin{align}\label{eq:Phi Psi wavefunctions}
    \Psi_\Delta(p^u,t_i)&=\theta(p^u)\frac{(2ip^uv_i)^\Delta}{\sqrt{\Gamma(2\Delta)}p^u}e^{2ip^u v_i},&\Phi_\Delta(p^v,t_i)&=\theta(p^v)\frac{(2ip^vu_i)^\Delta}{\sqrt{\Gamma(2\Delta)}p^v}e^{2ip^v u_i}.
\end{align}
where $(u_i,v_i)=(-e^{-t_i},e^{t_i})$ for the points on the boundary.

Furthermore, the momentum space S-matrix takes a very simple form, assuming that the two high energy scattering events are governed by the flat-space S-matrix derived in \cite{Dubovsky:2012wk}. In that work, the $2$-to-$2$ S-matrix for two particles with momenta $p^u$ and $p^v$ was found to be a phase with exponent proportional to the center-of-mass energy $s=4p^up^v$:
\begin{align}
    \ket{p^u,p^v}_{\rm out}&=e^{-i\ell_s^2 p^u p^v}\ket{p^u,p^v}_{\rm in}.
\end{align}
In our case, we have a $3\to 3$ scattering event, which we assume factorizes into two $2$-to-$2$ scattering events:
\begin{align}
    \ket{p_2^v,p_3^v,p_5^u}_{\rm out}=e^{-i\ell_s^2p_2^v p_5^u}e^{-i\ell_s^2p_3^v p_5^u}\ket{p_2^v,p_3^v,p_5^u}_{\rm in}.
\end{align}
It follows that
\begin{align}\label{eq:out|in momentumspace}
    _{\rm out}\braket{p_2^v,p_3^v,p_5^u|p_1^v,p_4^v,p_6^u}_{\rm in}&=e^{i\ell_s^2p_1^v p_5^u}e^{i\ell_s^2p_3^v p_5^u}p_1^v p_3^v p_5^u \delta(p_1^v-p_2^v)\delta(p_3^v-p_4^v)\delta(p_5^u-p_6^u).
\end{align}
Substituting \eqref{eq:Phi Psi wavefunctions} and \eqref{eq:out|in momentumspace} into \eqref{eq:out|in momentum integral rep}, we find the scattering representation of the OTOC to be
\begin{align}\label{eq:outinrep6pt}
    \braket{\rm out|in}&=(4u_1u_2)^{\Delta_U}(4u_3u_4)^{\Delta_V}(4v_5v_6)^{\Delta_W}\int_0^\infty dp_1^v dp_3^v dp_5^u \biggr[\frac{(p_1^v)^{2\Delta_U-1}}{\Gamma(2\Delta_U)}\frac{(p_3^v)^{2\Delta_V-1}}{\Gamma(2\Delta_V)}\frac{(p_5^u)^{2\Delta_W-1}}{\Gamma(2\Delta_W)}\nonumber\\&\hspace{3cm}\times e^{2ip_1^v(u_1-u_2)}e^{2ip_3^v(u_4-u_3)}e^{2ip_5^u(v_6-v_5)}e^{i\ell_s^2p_1^v p_5^u}e^{i\ell_s^2p_3^v p_5^u}\biggr].
\end{align}
Changing the momentum integration variables to $p=-2ip_1^v(u_1-u_2)$, $q=-2ip_3^v(u_4-u_3)$, $\phi=-2ip_5^u(v_6-v_5)$, and also normalizing the six-point function by the two-point functions,
\begin{align}
    \braket{U_1U_2}&=\frac{1}{[2\sin\left(\frac{it_{12}}{2}\right)]^{2\Delta_U}}=\left[-\frac{u_1u_2}{(u_1-u_2)^2}\right]^{\Delta_U},
\end{align}
with analogous expressions for $\braket{V_3V_4}$ and $\braket{W_5W_6}$, we arrive at the following expression for the normalized OTOC:
\begin{align}\label{eq:thgfrty65redcvbt}
     \frac{\braket{U_1W_5U_2V_3W_6V_4}}{\braket{U_1U_2}\braket{V_3V_4}\braket{W_5W_6}}&=\int_0^\infty dp dq d\phi \frac{p^{2\Delta_U-1}q^{2\Delta_V-1}\phi^{2\Delta_W-1}}{\Gamma(2\Delta_U)\Gamma(2\Delta_V)\Gamma(2\Delta_W)}e^{-p-q-\phi-\kappa_1p\phi-\kappa_2q\phi}.
\end{align}
Here, we have introduced
\begin{align}
    \kappa_1&=\frac{-i\ell_s^2}{4(u_1-u_2)(v_5-v_6)}=-\frac{i\ell_s^2e^{(t_1+t_2-t_5-t_6)/2}}{16\sinh\left(\frac{t_{12}}{2}\right)\sinh\left(\frac{t_{56}}{2}\right)},\label{eq:kappa1}\\\kappa_2&=\frac{i\ell_s^2}{4(u_3-u_4)(v_5-v_6)}=\frac{i\ell_s^2e^{(t_3+t_4-t_5-t_6)/2}}{16\sinh\left(\frac{t_{34}}{2}\right)\sinh\left(\frac{t_{56}}{2}\right)}.\label{eq:kappa2}
\end{align}
For the configuration in \eqref{eq:euclid times analytic cont} with the $\delta_i$ given in \eqref{eq:symmetric config} and with $t_i=-i\theta_i$, $\kappa_1$ and $\kappa_2$ become:
\begin{align}\label{eq:kappa1 kapp2}
    \kappa_1&=\frac{e^{T_{13}}}{8\sqrt{3}T_s},&\kappa_2&=\frac{e^{ T_{23}}}{8\sqrt{3}T_s}. 
\end{align}

We can evaluate the integrals over $p$ and $q$ in \eqref{eq:thgfrty65redcvbt} to get:
\begin{align}\label{eq:six-point function scattering result}
    \frac{\braket{U_1W_5U_2V_3W_6V_4}}{\braket{U_1U_2}\braket{V_3V_4}\braket{W_5W_6}}&=\frac{1}{\Gamma(2\Delta_W)}\int_0^\infty d\phi \phi^{2\Delta_W-1}(1+\kappa_1 \phi)^{-2\Delta_U}(1+\kappa_2\phi)^{-2\Delta_V}e^{-\phi}.
\end{align}
To get the behavior of the OTOC at strong-coupling, we can expand in $\kappa_1$ and $\kappa_2$ and evaluate the integrals order by order.\footnote{Another regime in which we can evaluate \eqref{eq:Om4GiwLXdm} analytically is when $\Delta_U,\Delta_V\gg 1$. Then we can use a saddle point approximation to derive:
\begin{align*}
    \frac{\braket{U_1W_5U_2V_3W_6V_4}}{\braket{U_1U_2}\braket{V_3V_4}\braket{W_5W_6}}&\approx \frac{1}{(1+2\kappa_1\Delta_U+2\kappa_2\Delta_V)^{2\Delta_W}}. 
\end{align*}
This expression has a nice geometric interpretation in terms of the length of the geodesic connecting the $W$ operators on the two boundaries of an AdS$_2$ background with two parallel shockwaves sourced by the $U$ and $V$ operators. The two parallel shockwaves behave like a single shockwave with a total shift proportional to $4\kappa_1\Delta_U+4\kappa_2\Delta_V$ along the $v$ direction. For example, compare the above expression with eq.~(28) in \cite{Shenker:2013pqa} or eq.~(4.3) of \cite{Murata:2017rbp}.\label{fn:shockwave interp}} When $\Delta_U=\Delta_V=\Delta_W=1$, the result is:
\begin{equation}\label{eq:Om4GiwLXdm}
    \begin{aligned}
     \frac{\braket{U_1W_5U_2V_3W_6V_4}}{\braket{U_1U_2}\braket{V_3V_4}\braket{W_5W_6}}&=1-4\kappa_1-4\kappa_2+24\kappa_1\kappa_2+\ldots\\&=1-\frac{e^{T_{13}}}{2\sqrt{3}T_s}-\frac{e^{T_{23}}}{2\sqrt{3}T_s}+\frac{e^{T_{13}+T_{23}}}{8T_s^2}+\ldots
     \end{aligned}
\end{equation}
This can be compared with the perturbative expansion of the six-point function in $1/T_s$. In particular, we will see below that the analytic continuation of the connected six-point function in \eqref{eq:6-pt function result} will reproduce the $\kappa_1\kappa_2$ term. But first we will take a small detour and discuss how the scattering result in \eqref{eq:thgfrty65redcvbt} (to all orders in $\kappa_1$ and $\kappa_2$) can also be derived from the reparametrization path integral.

\subsection{Double-scaled OTOC from an eikonal resummation} Section~\ref{sec:connected six-point function} showed us that knowing the reparametrization action to cubic order is sufficient to compute the leading (tree-level) contribution to the boundary six-point function. We will now show that, if we focus on the six-point OTOC in the double scaling limit in \eqref{eq:config 1 ds limit}, then we can also use the reparametrization path integral in \eqref{eq:reparametrization path integral} to resum all the non-interacting diagrams involving exchanges of the reparametrization mode between the dressed two-point functions of the transverse modes and, moreover, this ``eikonal resummation'' reproduces the full scattering result in \eqref{eq:thgfrty65redcvbt}. The following argument in the case of the six-point OTOC is essentially the same as the analogous argument in the case of the four-point OTOC, which is presented in section 6.3 of \cite{Giombi:2022pas}. 

The eikonal resummation is achieved by working with the exact dressed two-point function in \eqref{eq:bi-local} while truncating the reparametrization action to the quadratic order. The normalized six-point function is then:
\begin{align}
    \frac{\braket{U_1W_5U_2V_3W_6V_4}_{\rm eik}}{\braket{U_1U_2}\braket{V_3V_4}\braket{W_5W_6}}&=\int \mathcal{D}\epsilon e^{-S_2[\epsilon]}\mathcal{B}(x_1,x_2)^{\Delta_U}\mathcal{B}(x_3,x_4)^{\Delta_V}\mathcal{B}(x_5,x_6)^{\Delta_W}.
\end{align}
(Here we use the approximation $\braket{U_1U_2}=x_{12}^{-2\Delta_U}+O(T_s^{-1})$, and likewise for $\braket{V_3V_4}$ and $\braket{W_5W_6}$, with the corrections becoming negligible once we take the double scaling limit in \eqref{eq:config 1 ds limit}). We can put the components of the dressed two-point function into an exponential as follows:
\begin{align}
    \mathcal{B}(x_i,x_j)^{\Delta}&=\left[\frac{(1+\dot{\epsilon}_i)(1+\dot{\epsilon}_j)}{(1+\epsilon_{ij}/x_{ij})^2}\right]^{\Delta}=\left(\frac{\partial}{\partial \alpha_i}\frac{\partial}{\partial \alpha_j}\right)^{\Delta}e^{\alpha_i(1+\dot{\epsilon}_i)+\alpha_j(1+\dot{\epsilon}_j)}\int_0^\infty dp \frac{p^{2\Delta-1}}{\Gamma(2\Delta)} e^{-p(1+\epsilon_{ij}/x_{ij})}.
\end{align}
This representation--- combined with the identity $\braket{e^{f(\epsilon)}}=e^{\frac{1}{2}\braket{f(\epsilon)^2}}$ if the action is quadratic and $f$ is linear in $\epsilon$ --- implies that the six-point function in the eikonal approximation can be written as
\begin{align}
    \frac{\braket{U_1W_5U_2V_3W_6V_4}_{\rm eik}}{\braket{U_1U_2}\braket{V_3V_4}\braket{W_5W_6}}&=\left(\frac{\partial}{\partial \alpha_1}\frac{\partial}{\partial \alpha_2}\right)^{\Delta_U}\left(\frac{\partial}{\partial \alpha_3}\frac{\partial}{\partial \alpha_4}\right)^{\Delta_V}\left(\frac{\partial}{\partial \alpha_5}\frac{\partial}{\partial \alpha_6}\right)^{\Delta_W}\biggr[e^{\prod_{i=1}^6 \alpha_i}\times \nonumber\\&\int dpdqd\phi\frac{p^{2\Delta_U-1}q^{2\Delta_V-1}\phi^{2\Delta_W-1}}{\Gamma(2\Delta_U)\Gamma(2\Delta_V)\Gamma(2\Delta_W)}e^{-p-q-\phi+X}\biggr]\label{eq:gytrew456543}
\end{align}
where 
\begin{align}\label{eq:X}
    X&=\frac{1}{2}\big\langle\big(-p\frac{\epsilon_{12}}{x_{12}}-q\frac{\epsilon_{34}}{x_{34}}-\phi\frac{\epsilon_{56}}{x_{56}}+\sum_{i=1}^6 \alpha_i \dot{\epsilon}_i\big)^2\big\rangle_0.
\end{align}
The explicit expression for $X$ can be found by performing the Wick contractions with the propagator in \eqref{eq:<eps(t_1)eps(t_2)> 2}. This gives rise to many different terms, some involving self-contractions (which should be regulated with a short distance cut-off $\delta$ by, e.g., setting $\langle\dot{\epsilon}(0)^2\rangle_0=\frac{1}{8\pi T_s}\log{\delta^2}$), and the result is invariant under neither the physical nor the gauge $SL(2,\mathbb{R})$ symmetries (which is a consequence of us keeping only the quadratic part of the action). However, all these complications wash out once we analytically continue to the OTOC configuration at late times. More precisely, we map the points on the thermal cylinder in \eqref{eq:euclid times analytic cont} to the complex plane using $x_i=\tan{\frac{\theta_i}{2}}$ and take $T_1\sim T_2\to \infty$, $T_3\to -\infty$ and $T_s\to \infty$ with $e^{T_{12}}/T_s$ and $e^{T_{13}}/T_s$ held fixed.\footnote{It is important in getting from \eqref{eq:X} to \eqref{eq:X after ds} that we send $T_3\to -\infty$ at the same time that we send $T_1\sim T_2\to \infty$, but the relative rate at which the times are sent to $\pm \infty$ does not matter.} In this limit, $x_1,x_2,x_3,x_4$ become coincident and $x_5,x_6$ become coincident with $|x_{12}|,|x_{13}|,|x_{14}|,|x_{23}|,|x_{24}|,|x_{34}|\sim e^{-T_1}$ and $|x_{56}|\sim e^{T_3}$ and the distances between the other pairs of points approaching non-zero values. Consequently, the only contributions to \eqref{eq:X} that survive in the double scaling limit are:
\begin{align}\label{eq:X after ds}
    X&=\frac{p\phi}{x_{12}x_{56}}\braket{\epsilon_{12}\epsilon_{56}}_0+\frac{q\phi}{x_{34}x_{56}}\braket{\epsilon_{34}\epsilon_{56}}_0.
\end{align}
Performing the contractions using the propagator in \eqref{eq:<eps(t_1)eps(t_2)> 2}, we have for the first term:
\begin{align}
    \frac{1}{x_{12}x_{56}}\braket{\epsilon_{12}\epsilon_{56}}_0&=\frac{1}{T_s x_{12}x_{56}}\bigg[\frac{1}{8\pi}(x_{15}^2\log{x_{15}^2}+x_{26}^2\log{x_{26}^2}-x_{16}^2\log{x_{16}^2}-x_{25}^2\log{x_{25}^2})\nonumber\\&\hspace{3cm}+b(x_{15}^2+x_{26}^2-x_{16}^2-x_{25}^2)\bigg]\nonumber\\&=\frac{1}{8\pi T_s}\frac{x_{15}x_{26}}{x_{12}x_{56}}\log\left(\frac{x_{15}^2x_{26}^2}{x_{16}^2x_{25}^2}\right)+\ldots,
\end{align}
where we have put the leading term in a conformally invariant form and $\ldots$ denotes terms that vanish in the double scaling limit. Likewise, the second term in \eqref{eq:X after ds} becomes
\begin{align}
    \frac{1}{x_{34}x_{56}}\braket{\epsilon_{34}\epsilon_{56}}_0&=\frac{1}{8\pi T_s}\frac{x_{35}x_{46}}{x_{34}x_{56}}\log\left(\frac{x_{35}^2x_{46}^2}{x_{36}^2x_{45}^2}\right)+\ldots.
\end{align}
Letting $x_i=\tan{\frac{\theta_i}{2}}$ and performing the analytic continuation specified by \eqref{eq:euclid times analytic cont}-\eqref{eq:config 1 ds limit}, we find:
\begin{align}
    X&\to -\kappa_1 p\phi-\kappa_2 q\phi,
\end{align}
where $\kappa_1$ and $\kappa_2$ are given in \eqref{eq:kappa1 kapp2}. Noting that the $\alpha_i$ derivatives become trivial and the Schwinger parameters play the role of the light cone momenta, we see that \eqref{eq:gytrew456543} precisely reproduces \eqref{eq:thgfrty65redcvbt}.

\subsection{Explicit check of the six-point OTOC}

To compare the strong coupling expansion of the scattering result for the six-point OTOC in \eqref{eq:Om4GiwLXdm} with the analytic continuation of the six-point function, we need the full six-point function in \eqref{eq:full, connected, disconnected six-point function}. This includes a fully disconnected piece, three partially connected pieces, and the fully connected piece. After normalizing by the product of the three two-point functions, the fully disconnected piece gives $1$, the leading contribution to the partially connected pieces are given by \eqref{eq:connected four-point function}, and the leading contribution to the fully connected piece is given in \eqref{eq:6-pt function result}. Then, to perform the analytic continuation, we first map from the line to the circle by replacing $x_{ij}\to \sin{\frac{1}{2}\theta_{ij}}$ and letting the cross ratios be $\xi_{ijkl}=\frac{\sin{\frac{1}{2}\theta_{ik}}\sin{\frac{1}{2}\theta_{jl}}}{\sin{\frac{1}{2}\theta_{il}}\sin{\frac{1}{2}\theta_{jk}}}$, then fix the $\theta_i$ to be as in \eqref{eq:euclid times analytic cont}-\eqref{eq:symmetric config}, and finally take the limit in \eqref{eq:config 1 ds limit}. In particular, we need to determine the behavior of  $G(\xi_{ijkl})$, $G'(\xi_{ijkl})$ and $F_{ij;klmn}$ in this limit.

Given the configuration of points specified by \eqref{eq:euclid times analytic cont} and \eqref{eq:symmetric config}, the three conformally invariant cross ratios are:
\begin{align}
    \xi_{1234}&=1+\frac{3}{-1+2\cosh(T_{12})},&\xi_{1256}&=1+\frac{6}{-3+2i\sqrt{3}\sinh{T_{13}}},&\xi_{3456}&=1+\frac{6}{-3-2i\sqrt{3}\sinh{T_{23}}}
\end{align}
As $T_{13}$ and $T_{23}$ increase from $0$, $\xi_{1256}$ moves from $-1$ to $1$ clockwise and $\xi_{3456}$ moves from $-1$ to $1$ counterclockwise around the origin, and at late times they are given by
\begin{align}
    \xi_{1256}&\sim 1-2\sqrt{3}ie^{-T_{13}},&\xi_{3456}&\sim 1+2\sqrt{3}ie^{-T_{23}}.
\end{align}
The leading behavior of the connected four-point functions given by \eqref{eq:G(xi)} in the double scaling limit in \eqref{eq:config 1 ds limit} is:
\begin{align}
    G(\xi_{1234})&\sim O(1/T_s), &G(\xi_{1256})&\sim -\frac{e^{T_{13}}}{2\sqrt{3}T_s},&G(\xi_{3456})&\sim -\frac{e^{T_{23}}}{2\sqrt{3}T_s}
\end{align}
Similarly, we find that the leading behavior of the derivatives is:
\begin{align}
    G'(\xi_{1234})&\sim O(1/T_s), &G'(\xi_{1256})&\sim \frac{ie^{2T_{13}}}{12T_s},&G'(\xi_{3456})&\sim -\frac{ie^{2T_{23}}}{12T_s},
\end{align}
And, being careful about smoothly continuing to the correct branches of the various logarithms in \eqref{eq:curly F}, we find that the leading behaviors of the $F$ functions are given by:
\begin{align}
    F_{56;1234}&\sim O(T_s^0),&F_{34;1256}&\sim -\frac{i}{4T_s}e^{-T_{12}},&F_{12;3456}&\sim \frac{i}{4T_s}e^{T_{12}},
\end{align}
Finally, taking $T_{13},T_{23},T_s\to \infty$ with $\frac{e^{T_{13}}}{T_s}$ and $\frac{e^{T_{23}}}{T_s}$ fixed yields 
\begin{align}\label{eq:41B26qrb4X}
    \frac{\braket{U_1W_5U_2V_3W_6V_4}}{\braket{U_1U_2}\braket{V_3V_4}\braket{W_5W_6}}=1&+\left(-\frac{1}{2\sqrt{3}}\frac{e^{T_{13}}}{T_s}+\ldots\right)+\left(-\frac{1}{2\sqrt{3}}\frac{e^{T_{23}}}{T_s}+\ldots\right)\nonumber\\&+\frac{1}{8}\frac{e^{T_{13}+T_{23}}}{T_s^2}+\ldots
\end{align}
The first line gives the leading contributions to the OTOC coming the disconnected diagrams (i.e., the terms in the first line in \eqref{eq:full, connected, disconnected six-point function}), and the second line gives the contribution from the tree-level connected piece, given in \eqref{eq:6-pt function result}. We see that \eqref{eq:41B26qrb4X} precisely matches \eqref{eq:Om4GiwLXdm} in the overlapping terms.

\subsection{Relationship with the flat-space limit of AdS/CFT}
To conclude this section, let us make brief comments on the relation with the flat-space limit of AdS/CFT discussed in the literature, which also expresses the CFT correlation functions in terms of the scattering amplitudes in flat space. The flat space limit of the correlation functions in AdS has been analyzed using various approaches (see e.g.\cite{Okuda:2010ym,Penedones:2010ue,Paulos:2016fap,Dubovsky:2017cnj,Hijano:2019qmi,Komatsu:2020sag,Li:2021snj,Cordova:2022pbl,vanRees:2022zmr}). The details of the limit depend on whether the particles are massless or massive, and for massless scatterings, the derivation was explained in detail in section 2.1 of \cite{Okuda:2010ym} for AdS$_{D\geq 3}$, under the assumption that the scattering event happens locally at a single point in the AdS spacetime. In what follows, we explain how their derivation can be generalized to AdS$_{D=2}$ and how it is related to the formula used in this paper and \cite{Shenker:2014cwa,Giombi:2022pas,deBoer:2017xdk}. We focus on the $2$-to-$2$ scattering for simplicity but the argument can be readily generalized to higher-point scatterings.

The starting point of their analysis the following rewriting of the bulk-to-boundary propagator $G_{B\partial}$\footnote{Here we set $R$ and $\ell_s$ in \cite{Okuda:2010ym} to unity and took $d=1$.}
\beq\label{eq:mellinbulkboundary}
G_{B\partial}(X,P)\propto\int_{0}^{\infty}\frac{d\beta}{\beta}\beta^{\Delta}e^{-2i \beta P\cdot X}\comma
\eeq
where $X$ and $P$ are the embedding coordinate representations for the bulk and the boundary points respectively (see \cite{Okuda:2010ym} for details). We next assume that the dominant contribution to the correlation function comes from a small region near the center of AdS (denoted by $X_{\ast}$), and approximate the correlation function as\fn{The formula \eqref{eq:AdSLSZ} can be viewed as the AdS version of the LSZ formula. In flat space, correlation functions are given by the convolution of amputated Green's functions with the leg factors, which are the Fourier transforms of the propagators $\sqrt{Z}/(p^2+m^2)$. Here the leg factors are replaced with their AdS counterparts $G_{B\partial}$.}
\beq\label{eq:AdSLSZ}
\langle \mathcal{O}_1\mathcal{O}_2\mathcal{O}_3\mathcal{O}_4\rangle \sim \int_{R^{1,1}} \prod_{i=1}^{4}dY_i G_{B\partial}(X_{\ast}+Y_i,P_i)G_{\rm flat}(Y_1,Y_2,Y_3,Y_4)\period
\eeq
Here $Y_i$'s parametrize a small region near $X_{\ast}$ which can be approximated by $R^{1,1}$, and $G_{\rm flat}$ is the {\it amputated} Green's function in flat space.
We then use the formula \eqref{eq:mellinbulkboundary} to get
\beq\label{eq:flatlimit1}
\langle \mathcal{O}_1\mathcal{O}_2\mathcal{O}_3\mathcal{O}_4\rangle\sim \int_0^{\infty}\left(\prod_{j=1}^{4}d\beta_j\beta_j^{\Delta_j-1}e^{-2i \beta_j P_j \cdot X_{\ast}} \right)\mathcal{S}\comma
\eeq
with
\beq
\mathcal{S}\equiv \int_{R^{1,1}}\prod_{j=1}^{4}dY_je^{-2i\beta_jP_j\cdot Y_j}G_{\rm flat}(Y_1,Y_2,Y_3,Y_4)\period
\eeq

To evaluate $\mathcal{S}$, we now express the embedding coordinates $X$ and $P$ in terms of the Kruskal coordinates of AdS$_2$ (see \cite{Giombi:2022pas} for a review),
\beq
ds^2 = -\frac{4 du dv}{(1+uv)^2}\comma
\eeq
as follows:
\beq
X=\left(\frac{1-uv}{1+uv}\comma\frac{u+v}{1+uv}\comma\frac{u-v}{1+uv}\right)\comma\qquad P_j =\left(1,\frac{u_j+v_j}{2},\frac{u_j-v_j}{2}\right)\period
\eeq
Here $u_j$ and $v_j$ parametrizes the boundary of AdS$_2$ and satisfy $u_jv_j=-1$, and the center of AdS$_2$ corresponds to $u=v=0$, namely
\beq
X_{\ast}=\left(1,0,0\right)\period
\eeq
Since $Y_j$'s describe a small neighborhood of $X_\ast$, they only have two non-vanishing components;
\beq
Y_j \sim \left(0, \delta u+\delta v,\delta u-\delta v\right)\period
\eeq
In addition, boundary points at early or late times correspond to either $u_j \to \pm \infty$ or $v_j\to \pm \infty$, and the embedding coordinates can be approximated as
\beq
P_j \sim \frac{1}{2}(0,\bar{p}_j^{0},\bar{p}_j^{1})\qquad \text{with}\qquad (\bar{p}_j^{0},\bar{p}_j^{1})=u_j\left(1,1\right)\quad \text{or}\quad v_j\left(1,-1\right)\period
\eeq
Thus, dot products $P_j\cdot Y_j$'s, which were originally defined in the embedding space $R^{2,1}$, can be viewed as inner products in $R^{1,1}$.  Then, $\mathcal{S}$ can be identified with the S-matrix $S_{\rm flat}$ in $R^{1,1}$ and the equation \eqref{eq:flatlimit1} can be rewritten as
\beq
\langle \mathcal{O}_1\mathcal{O}_2\mathcal{O}_3\mathcal{O}_4\rangle\sim \int_0^{\infty}\left(\prod_{j=1}^{4}d\beta_j\beta_j^{\Delta_j-1}e^{2i \beta_j } \right)S_{\rm flat}(p_j=-2\beta_j \bar{p}_j)\period
\eeq
Here we are using a convention in which the momenta $p_j$ are all incoming. To make contact with the analysis in \cite{Shenker:2014cwa,Giombi:2022pas,deBoer:2017xdk}, we take $p_{2,4}$ ($p_{1,3}$) to be incoming (outgoing) and $p_{1,2}$ ($p_{3,4}$) to be left-moving (right-moving). This corresponds to choosing
\beq
\begin{aligned}
p_1&=2\beta_1 u_1 (1,1)=-\frac{2\beta_1}{v_1} (1,1) \qquad &&(u_1>0,v_1<0) \comma\\
p_2&=-2\beta_2 u_2 (1,1)=\frac{2\beta_2}{v_2}(1,1) \qquad &&(u_2<0,v_2>0)\comma\\
p_3&=2\beta_3 v_3 (1,-1)=-\frac{2\beta_3}{u_3} (1,-1)\qquad &&(u_3<0,v_3>0) \comma\\
p_4&=-2\beta_4 v_4 (1,-1)=\frac{2\beta_4}{u_4} (1,-1)\qquad &&(u_4>0,v_4<0)\period
\end{aligned}
\eeq
We then change the variable of integration to $p_j$ to get\fn{Here we omitted the prefactor independent of the integration variables, such as $u_j^{\Delta_j}$ and $v_j^{\Delta_j}$.}
\beq
\begin{aligned}
&\langle \mathcal{O}_1\mathcal{O}_2\mathcal{O}_3\mathcal{O}_4\rangle\sim \\
&v_1^{\Delta_1}v_2^{\Delta_2}u_3^{\Delta_3}u_4^{\Delta_4}\int_0^{\infty}\left(\prod_{j=1}^{4}dp_j^{0}\left(p_j^{0}\right)^{\Delta_j-1}\right)e^{-i (p_1^{0}v_1-p_2^{0}v_2+p_3^{0}u_3-p_4^{0}u_4) } S_{\rm flat}(p_2,p_4; p_1,p_3)\comma
\end{aligned}
\eeq
This reproduces the formula for the four-point OTOC used in \cite{Shenker:2014cwa,Giombi:2022pas,deBoer:2017xdk} (up to numerical prefactors which we did not keep track of in the discussion above). Essentially the same analysis can be applied to the six-point OTOC and will reproduce the formulas in eq. \eqref{eq:out|in momentum integral rep}-\eqref{eq:Phi Psi wavefunctions} or \eqref{eq:outinrep6pt}.

Let us also point out that the argument here and the argument in \cite{Okuda:2010ym} both rely on a rather strong assumption that the dominant contribution comes from a small region near $X_{\ast}$. Although this may seem like a physically reasonable assumption, it would be desirable to prove it or clarify when it can be justified. This is particularly important in view of recent works \cite{Komatsu:2020sag,Cordova:2022pbl,vanRees:2022zmr}, which showed that similar assumptions do not hold for the scattering of massive particles in certain kinematics. We leave this for future investigation.
 
\section{Connected six-point function in static gauge} \label{sec:static}
In this section, we compute the connected six-point function using the Nambu-Goto action in static gauge. This serves as an additional check of the conformal gauge result. The four-point function was computed in static gauge in \cite{Giombi:2017cqn}, and we will follow a similar procedure here.

\subsection{Action and propagators in static gauge}
To study the correlation function of identical scalars, we restrict the fluctuations of the string in AdS$_5 \times S^5$ to an AdS$_2 \times S^1$ subsector. Then the Nambu-Goto action in static gauge reduces to \cite{Giombi:2022pas}
\begin{align}
	S[y] = T_s \int d^2 \sigma \sqrt{\det [g_{\alpha\beta} + \partial_\alpha y \partial_\beta y]}.
\end{align}
To find the six-point functions, we can expand the action in powers of $y$ up to sixth order, treating $y$ as a small fluctuation:
\begin{align}
	S[y] = T_s \int d^2 \sigma \sqrt{g} L_{2n}.
\end{align}
We find the vertices
\begin{align}
	L_0 = 1, \quad L_2 = \half g^{\mu\nu} \partial_\mu y \partial_\nu y, \quad 
	L_4 = -\frac{1}{8} (g^{\mu\nu} \partial_\mu y \partial_\nu y)^2, \quad 
	L_6 = \frac{1}{16} (g^{\mu\nu} \partial_\mu y \partial_\nu y)^3. \label{Eq:vertex}
\end{align}

In AdS$_2$, the bulk-boundary propagator has the form \cite{Witten:1998ads}
\begin{align}
    K_\Delta (x';x,z) =C_{\Delta} \Tilde{K}_\Delta (x';x,z) =C_{\Delta} \left (\frac{z}{z^2 + (x-x')^2}\right) ^{\Delta},
\end{align}
where $C_\Delta = \frac{\Gamma(\Delta)}{2\sqrt{\pi} \Gamma(\Delta + \half)}$. To deal with the derivatives in vertices, one can use the identity \cite{Freedman:1999dfmmr}
\begin{multline*}
    g^{\mu \nu} \partial_\mu \Tilde{K}_{\Delta_1} (x_1; x, z) \partial_\nu \Tilde{K}_{\Delta_2} (x_2; x, z) \\
    = \Delta_1 \Delta_2 [\Tilde{K}_{\Delta_1} (x_1; x, z) \Tilde{K}_{\Delta_2} (x_2; x, z) - 2 x_{12}^2 \Tilde{K}_{\Delta_1 +1} (x_1; x, z) \Tilde{K}_{\Delta_2 + 1}(x_2; x, z)]. \label{Eq:DerivIdentity}
\end{multline*}

For the exchange diagrams, we also need the bulk-bulk propagator,
\begin{align}
    G^\Delta (a,b) = C_\Delta \frac{1}{2u}~{_2 F_1} (\Delta, \Delta, 2\Delta,-\frac{2}{u}), \quad\quad 
    u = \frac{(z_a - z_b)^2 + (x_a - x_b)^2}{2 z_a z_b}.
\end{align}
For $\Delta = 1$, the hypergeometric function reduces to an elementary function, 
\begin{align}
    G^{\Delta = 1}(a, b) = C_{\Delta = 1} \frac{1}{4} \log{\left(1+\frac{2}{u} \right)}.
\end{align}

\subsection{Computation of correlators}
At tree level, the connected six-point function is given by the sum of contact diagrams and exchange diagrams,
\begin{align}
    \langle y(x_1) ... y(x_6) \rangle _{\rm conn.}
    = \langle y(x_1) ... y(x_6) \rangle_{\rm contact} + \langle y(x_1) ... y(x_6) \rangle_{\rm exchange}.
\end{align}

\begin{figure}[t]
\centering
\begin{minipage}{0.49\hsize}
\centering
\includegraphics[clip, height=3.5cm]{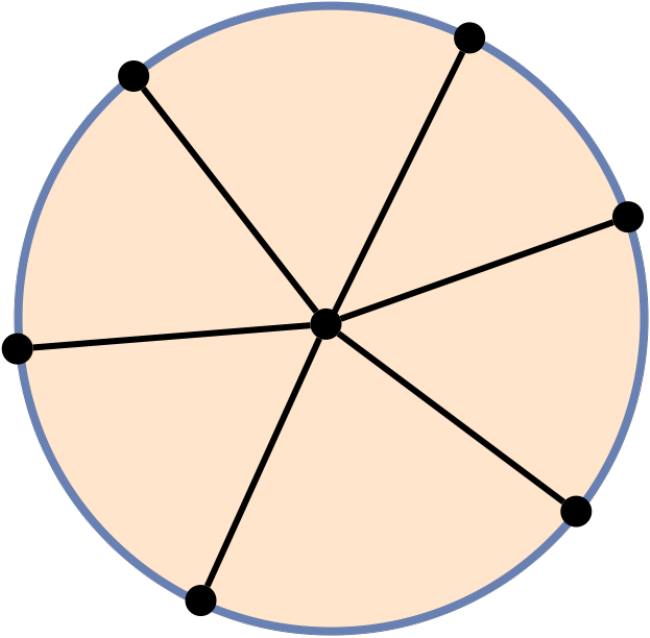}\\
{\bf a.}
\end{minipage}
\begin{minipage}{0.49\hsize}
\centering
\includegraphics[clip, height=3.5cm]{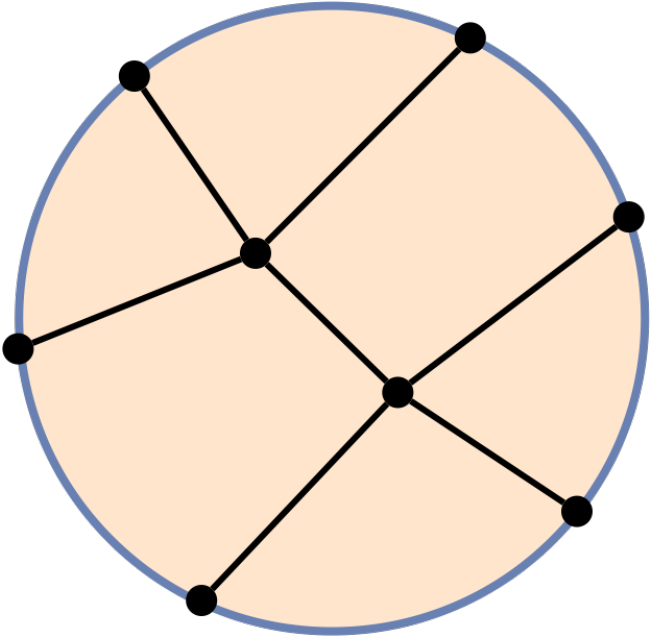}\\
{\bf b.} 
\end{minipage}
\caption{\textbf{a. }The contact diagram. ~~\textbf{b.} The exchange diagram.}
\label{fig:six point static gauge contact and exchange diagrams}
\end{figure}

The contact diagram is shown in Figure~\ref{fig:six point static gauge contact and exchange diagrams}a. The sum of contact diagrams contributing to the correlation function with six identical massless scalars can be written as
\begin{align}
    \langle y(x_1) ... y(x_6) \rangle_{\rm contact} 
= -3 T_s (C_{\Delta=1})^6 Q_{6y},
\label{contact-full}
\end{align}
Here the factor $(C_{\Delta=1})^6$ is from the 
six boundary-bulk propagators, and $Q_{6y}$ denotes a sum of six-point contact integrals 
\begin{align}\label{eq:tyhgertr}
    Q_{6y} = Q_{12,34,56} + \text{14 permutations},
\end{align}
where the first term is given by 
\begin{align}
    & Q_{12,34,56} =\nonumber\\
    =& \int \frac{dz dx}{z^2} 
    [g^{\alpha \beta}\partial_\alpha \Tilde{K}_1 (x_1;x,z) \partial_\beta \Tilde{K}_1 (x_2;x,z)] 
    [g^{\mu \nu}\partial_\mu \Tilde{K}_1 (x_3;x,z) \partial_\nu \Tilde{K}_1 (x_4;x,z)]\nonumber \\
    &\times 
    [g^{\rho \lambda}\partial_\rho \Tilde{K}_1 (x_5;x,z) \partial_\lambda \Tilde{K}_1 (x_6;x,z)],
\label{Qfunc-contact}
\end{align}
and other terms are obtained by permuting the external points. (Note that since the derivatives in the vertices of \eqref{Eq:vertex} combine two external points together, we need to divide the six external points into three pairs, and hence there are 15 combinations, such as $\{12,34,56\}$, $\{12,35,46\}$ and so on). To understand the overall factor in (\ref{contact-full}), note that for six identical scalars there are $6! = 720$ ways to do the Wick contractions. Because there are 15 ways to combine the six points into three pairs, the symmetry factor is  $720/15 = 48$, and from the vertex in \eqref{Eq:vertex} we see that the numerical factor in front of each combination should be $-\dfrac{T_s}{16} \times 48 = -3T_s$.

Using the identity (\ref{Eq:DerivIdentity}), the integral (\ref{Qfunc-contact}) can be reduced to
\begin{align}
Q_{12,34,56} =& I_{111111} - 8 x_{12}^2 x_{34}^2 x_{56}^2 I_{222222} - 2 x_{12}^2 I_{221111} - 2 x_{34}^2 I_{112211} - 2 x_{56}^2 I_{111122}\nonumber\\
    & + 4 x_{12}^2 x_{34}^2 I_{222211} + 4 x_{12}^2 x_{56}^2 I_{221122} + 4 x_{34}^2 x_{56}^2 I_{112222}
\end{align}
Here, the $I$-function is defined as
\begin{align}
    I_{\Delta_1 ... \Delta_n}
    = \int \frac{dz dx}{z^2} \Tilde{K}_{\Delta_1} (x_1;x,z) \ldots \Tilde{K}_{\Delta_n} (x_n;x,z).
\end{align}
A method to analytically evaluate the $I$-functions was given in \cite{Bliard:2022xsm}. For $\Delta_i = 1$ and even $n$, the result is
\begin{align}
	I_{n, \Delta=1} = \frac{\pi}{2(2i)^{n-2}} \sum_{j=1}^n \sum_{k=1, k\neq j}^n \frac{ (x_j - x_k)^{n-4} }{ \prod_{i=1,i\neq j, i \neq k}^n (x_j - x_i)(x_k - x_i) } \ln{(x_j - x_k)^2}.
\end{align}
For $\Delta_i > 1$, one may evaluate these integrals using the pinching method discussed in \cite{Bliard:2022xsm}.  

The exchange diagram is shown in Figure~\ref{fig:six point static gauge contact and exchange diagrams}b. The corresponding contribution to the connected six-point function can be written as 
\begin{align}
    \langle y(x_1) ... y(x_6) \rangle_{\rm exchange}. 
    = T_s (C_{\Delta=1})^6 P_{6y}
\label{exchange-full}
\end{align}
where $P_{6y} = P_{12,34,56}+{\rm 89~permutations}$, and 
\begin{align}
    P_{12,34,56} =& \int \frac{dz_a dx_a}{z_a^2} \int \frac{dz_b dx_b}{z_b^2} 
    [g^{\mu \nu}(a) \partial_\mu ^{(a)}\Tilde{K}_1 (x_1;x_a,z_a) \partial_\nu^{(a)} \Tilde{K}_1 (x_2;x_a,z_a) g^{\rho \lambda}(a) \partial_\lambda^{(a)}\Tilde{K}_1 (x_5;x_a,z_a)] \nonumber \\
    &
    [\partial_\rho^{(a)}\partial_\gamma^{(b)} G(x_a,z_a;x_b,z_b)g^{\alpha \beta}(b) \partial_\alpha^{(b)} \Tilde{K}_1 (x_3;x_b,z_b) \partial_\beta^{(b)} \Tilde{K}_1 (x_4;x_b,z_b) g^{\gamma \delta}(b)   \partial_\delta^{(b)} \Tilde{K}_1 (x_6;x_b,z_b) ]. \label{Eq:p123456}
\end{align}
For exchange diagrams, there are $\binom{4}{1} \binom{4}{1} 6! = 11520$ ways to do the Wick contraction. One $\binom{4}{1}$ factor corresponds to selecting one bulk point from the four-point interaction vertex, and there are two bulk interaction vertices. $6!$ corresponds to arranging the other six bulk points to the six external points. There are still 15 ways to divide the six points in pairs contracted by derivatives. However, from \eqref{Eq:p123456} we see that it matters which group we treat as the third group (e.g. which group we put at the position of 56 in \eqref{Eq:p123456}), and the order of the two points in the third group also matters (e.g. the order of 5 and 6 in \eqref{Eq:p123456}). Thus, we should have $15 \times 3 \times 2 = 90$ independent permutations. For each permutation, there is a symmetry factor $11520/90 = 128$. According to \eqref{Eq:vertex},  
the numerical factor in front of each permutation should be $128 \times \frac{1}{2} \times \left(\frac{1}{8} \right)^2 T_s = 1 \cdot T_s$, as written in (\ref{exchange-full}).  

The integrals \eqref{Eq:p123456} may be simplified using integration by parts and
\begin{align}
    \partial_\mu (\sqrt{g} \partial^\mu K_1) = 0,
\end{align}
as well as the identity (\ref{Eq:DerivIdentity}). 
This leads to a sum of exchange integrals with no derivatives
\begin{align}
\begin{aligned}
    &P_{12,34,56} = \int \frac{dz_a dx_a}{z_a^2} \int \frac{dz_b dx_b}{z_b^2} G(x_a, z_a; x_b, z_b)\times \\
    & [2\Tilde{K}_1 (x_1; x_a, z_a) \Tilde{K}_1 (x_2; x_a, z_a) \Tilde{K}_1 (x_5; x_a, z_a) 
    - 2 x_{15}^2 \Tilde{K}_2 (x_1; x_a, z_a) \Tilde{K}_1 (x_2; x_a, z_a) \Tilde{K}_2 (x_5; x_a, z_a)\\
    &- 2 x_{25}^2 \Tilde{K}_1 (x_1; x_a, z_a) \Tilde{K}_2 (x_2; x_a, z_a) \Tilde{K}_2 (x_5; x_a, z_a)
    - 8 x_{12}^2 \Tilde{K}_2 (x_1; x_a, z_a) \Tilde{K}_2 (x_2; x_a, z_a) \Tilde{K}_1 (x_5; x_a, z_a)\\
    &+8 x_{12}^2 x_{15}^2 \Tilde{K}_3 (x_1; x_a, z_a) \Tilde{K}_2 (x_2; x_a, z_a) \Tilde{K}_2 (x_5; x_a, z_a)
    +8 x_{12}^2 x_{25}^2 \Tilde{K}_2 (x_1; x_a, z_a) \Tilde{K}_3 (x_2; x_a, z_a) \Tilde{K}_2 (x_5; x_a, z_a)]\\
    &\times [2\Tilde{K}_1 (x_3; x_b, z_b) \Tilde{K}_1 (x_4; x_b, z_b) \Tilde{K}_1 (x_6; x_b, z_b) 
    - 2 x_{36}^2 \Tilde{K}_2 (x_3; x_b, z_b) \Tilde{K}_1 (x_4; x_b, z_b) \Tilde{K}_2 (x_6; x_b, z_b)\\
    &- 2 x_{46}^2 \Tilde{K}_1 (x_3; x_b, z_b) \Tilde{K}_2 (x_4; x_b, z_b) \Tilde{K}_2 (x_6; x_b, z_b)
    - 8 x_{34}^2 \Tilde{K}_2 (x_3; x_b, z_b) \Tilde{K}_2 (x_4; x_b, z_b) \Tilde{K}_1 (x_6; x_b, z_b)\\
    &+8 x_{34}^2 x_{36}^2 \Tilde{K}_3 (x_3; x_b, _b) \Tilde{K}_2 (x_4; x_b, z_b) \Tilde{K}_2 (x_6; x_b, z_b)
    +8 x_{34}^2 x_{46}^2 \Tilde{K}_2 (x_3; x_b, z_b) \Tilde{K}_3 (x_4; x_b, z_b) \Tilde{K}_2 (x_6; x_b, z_b)].
\label{exchange-simple}
    \end{aligned}
\end{align}

\begin{figure}[hbt!]
    \centering
    \begin{minipage}{0.49\textwidth}
        \centering
        \includegraphics[width=0.8\textwidth]{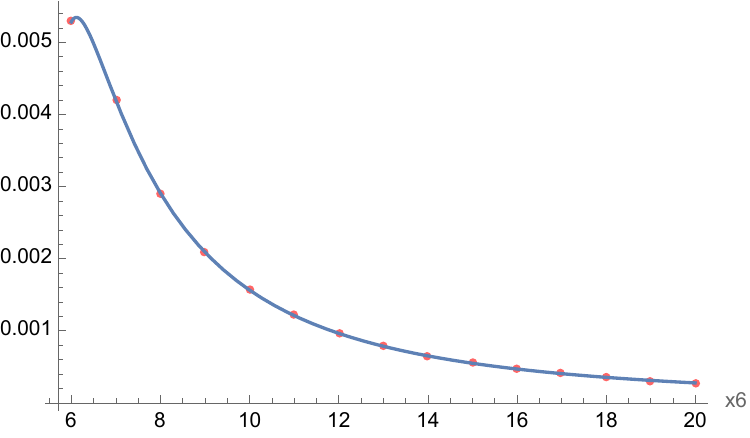}
    \end{minipage}
    \begin{minipage}{0.49\textwidth}
        \centering
        \includegraphics[width=0.8\textwidth]{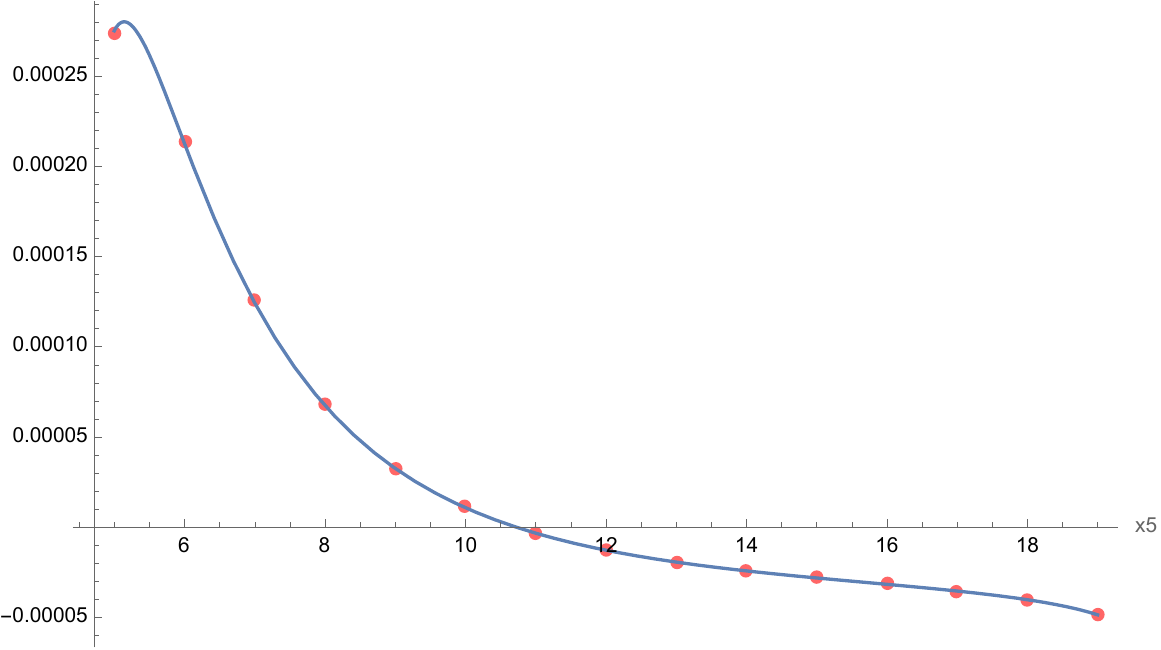}  
    \end{minipage}
    \caption{Numerical comparison between 6-point correlation functions in conformal gauge and static gauge. The solid lines represent the correlation function in conformal gauge, where we have the analytic results, and the dots represent the numerical results computed in static gauge. In the left plot, we vary the external point $x_6$ while keeping $x_1,\ldots, x_5$ fixed to some arbitrary values. In the right plot, we vary $x_5$ while fixing $x_1,\ldots, x_4$ and $x_6$.}
    \label{fig:compare}
\end{figure}

Putting the contact and exchange diagrams together, the connected six-point function in static gauge is
\begin{align}
    \langle y(x_1) ... y(x_6) \rangle _{\rm conn.}
    = T_s (C_{\Delta=1})^6 (P_{6y} - 3 Q_{6y}).
\end{align}
We were not able to analytically evaluate the exchange integrals (\ref{exchange-simple}), but we can perform numerical comparisons with the six-point correlator computed in conformal gauge by fixing the values of the six external points and evaluating the various components of $P_{6y}$ and $Q_{6y}$ numerically. We find very good agreement between the analytic result in conformal gauge and the numerical evaluation in static gauge. 
The results are shown in Figure~\ref{fig:compare} for a sample of different ways of fixing the values of the external points. 

\section{Conclusions}
\label{sec:conclusion}

In this work, we studied the six-point function of identical scalars in the half-BPS Wilson line defect CFT in $\mathcal{N}=4$ SYM at strong coupling. We computed the correlator by studying the dual fundamental string with AdS$_2$ geometry in conformal gauge, in which the interactions between the transverse fluctuations of the string are mediated by a boundary reparametrization mode. The six-point function is sensitive to self-interactions of the reparametrization mode, so the first step in our analysis was to determine the effective action of the reparametrization mode to cubic order. A nice feature of the final result for the six-point function in \eqref{eq:6-pt function result} is that it involves only rational functions and logarithms. We also checked the correctness of the conformal gauge approach by setting up the computation in static gauge. In contrast to conformal gauge, we did not manage to compute the six-point functions analytically in static gauge. Instead we evaluated the relevant six-point contact and four-point exchange Witten diagrams numerically for a representative set of external points for the six operators and found agreement with the analytic formula in conformal gauge. This demonstrates the advantage of using the conformal gauge and the reparametrization path integral in evaluating these correlators. Finally, as a further test of our results and to discuss an interesting physical interpretation of the correlator, we also analytically continued the euclidean six-point function to a Lorentzian out-of-time-order configuration which corresponds to a 3-to-3 scattering process on the worldsheet.

One natural extension of the present work would be to study the six-point function of different (protected) scalars on the Wilson line, for instance focusing on the five $\Delta=1$ scalars transforming in the fundamental of $SO(5)$, which are dual to the fluctuations along $S^5$. The general six-point function is of interest because, unlike the six-point function of identical scalars, it would allow one to test the multi-point superconformal Ward identities conjectured in \cite{Barrat:2021tpn} at the leading order with contributions from fully connected diagrams. Furthermore, one could extract new OPE data from the six-point function that is not readily accessible from the four-point functions at tree level. For example, one could take the coincident limit of three pairs of the six fundamental scalars and extract the OPE coefficient of three copies of the lowest composite scalar transforming in the singlet representation of the $SO(5)$ R-symmetry. Such ``three-particle'' operators are important also for the four-point function at higher loops since they can appear in the OPE expansion. It is important in this procedure to work with the general six-point function in order to disentangle the operators transforming in different representations of $SO(5)$. (Recall, for example, that the operators of lowest scaling dimension appearing in the OPE of two fundamental scalars includes both a singlet and a symmetric traceless composite operator.)\footnote{One can also try to extract some OPE data from the six-point function of identical scalars in \eqref{eq:6-pt function result}, but the data will be averaged over different R-symmetry channels, in addition to involving the usual operator mixing between operators with the same quantum numbers. We were unable to cleanly extract new OPE data from the identical scalar six-point function, even when we supplemented it with input from supersymmetric localization (e.g., the leading rank$-2$ symmetric traceless operator appearing in the OPE of two fundamental scalars is a chiral primary and its OPE data is known exactly \cite{Giombi:2018hsx}).} 

\begin{figure}[t]
\centering
\begin{minipage}{0.33\hsize}
\centering
\includegraphics[clip, height=3cm]{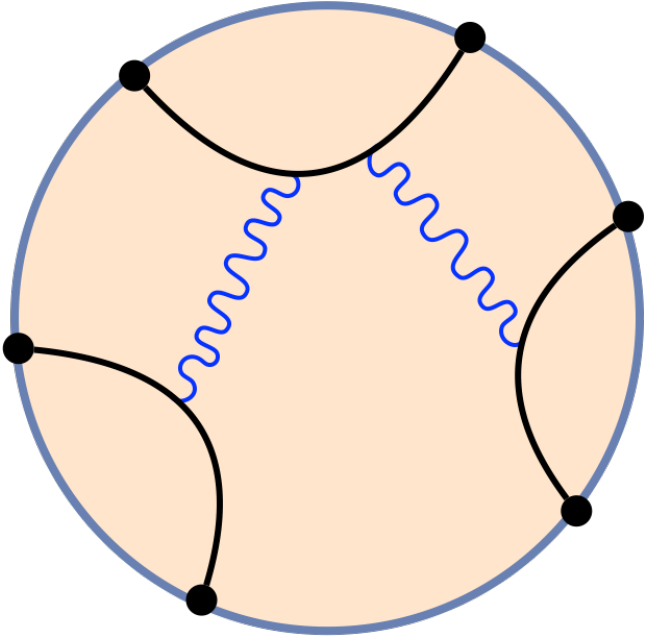}\\
\end{minipage}
\begin{minipage}{0.33\hsize}
\centering
\includegraphics[clip, height=3cm]{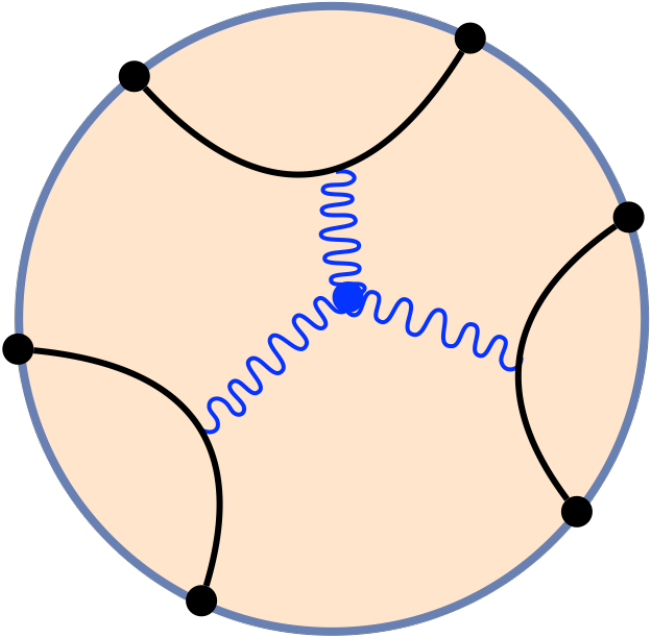}\\
\end{minipage}
\\
\vspace{1cm}
\begin{minipage}{0.3\hsize}
\centering
\includegraphics[clip, height=3cm]{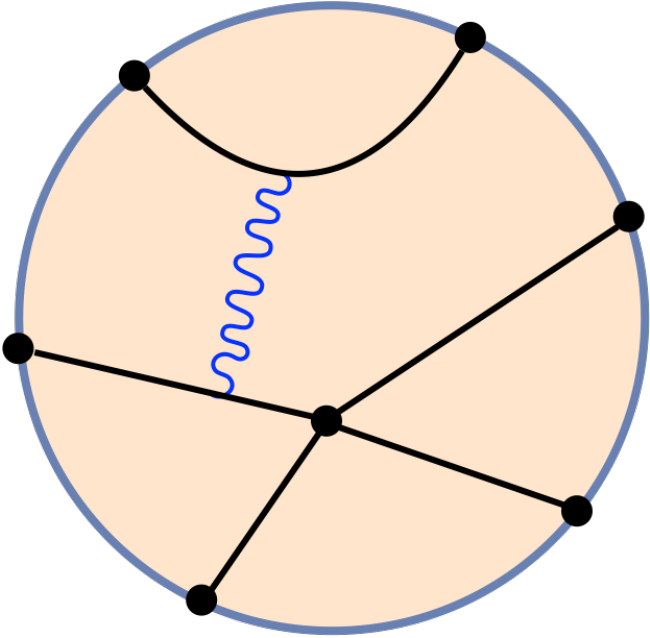}\\
\end{minipage}
\begin{minipage}{0.3\hsize}
\centering
\includegraphics[clip, height=3cm]{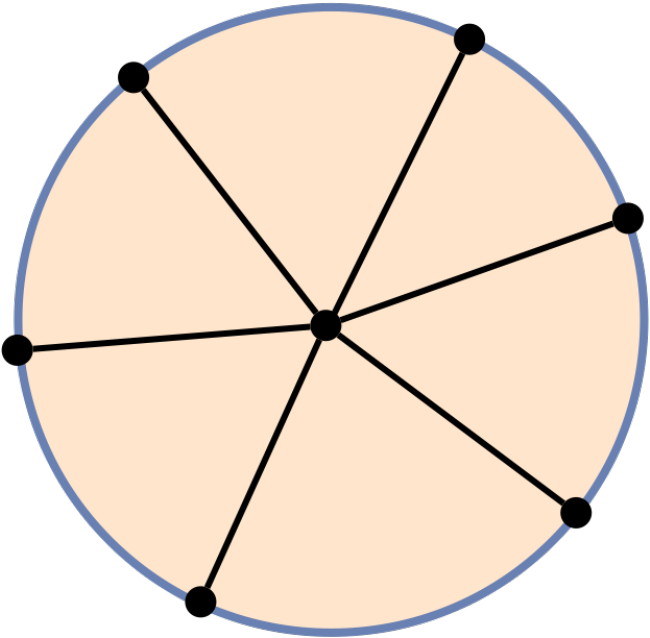}\\
\end{minipage}
\begin{minipage}{0.3\hsize}
\centering
\includegraphics[clip, height=3cm]{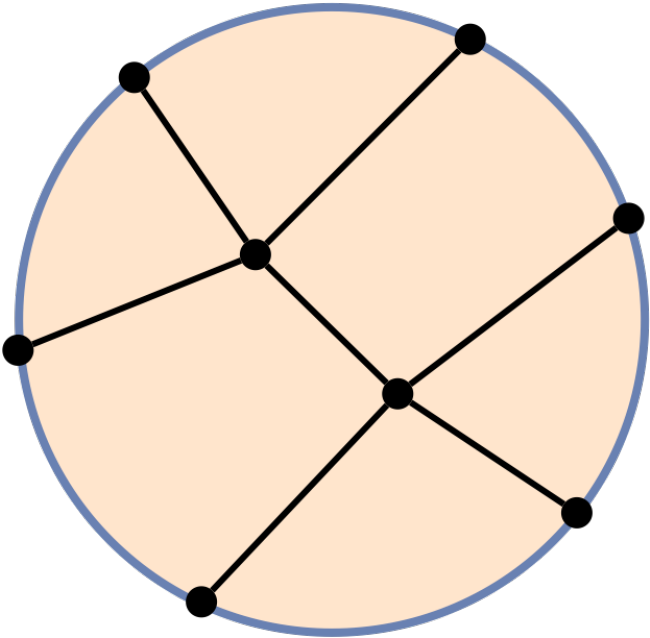}\\
\end{minipage}
\caption{In the conformal gauge approach, there are five different classes of reparametrization diagrams that contribute to the connected correlator of six different scalars. The first two classes of diagrams are qualitatively similar to the diagrams in Figure~\ref{fig:six point reparametrization diagrams} contributing to the six-point function of identical scalars. The remaining three classes of diagrams are new and also involve scalar contact and exchange diagrams.}
\label{fig:six general scalars reparametrization diagrams}
\end{figure}

While we showed in this work that the conformal gauge and the reparametrization path integral simplify drastically the computation of the six-point function of identical scalars
--- essentially because the transverse action is interacting in static gauge and non-interacting in conformal gauge --- 
the generalization to the six-point function involving all of the five $\Delta=1$ scalars is not straightforward. 
To compute the general correlator, we would need to study the AdS$_2$ string in an AdS$_2\times S^5$ subspace, which has five transverse modes that, importantly, are interacting even in conformal gauge.\footnote{For example, in stereographic coordinates $y^m$, $m=1,\ldots,5$ on $S^5$, the tranverse action in conformal gauge is 
\begin{align}
    S_T[y^m]&=\frac{T_s}{2}\int d^2 \sigma \frac{\partial_\alpha y^m \partial^\alpha y^m}{\left(1+\frac{1}{4}y^2\right)^2}.
\end{align}} 
Thus, in addition to the two classes of diagrams we encountered in section~\ref{sec:connected six-point function} involving two reparametrization exchanges and a cubic interaction between reparametrization modes, the computation of the general six point function in conformal gauge will also involve four and six-point contact and exchange diagrams. These are summarized in Figure~\ref{fig:six general scalars reparametrization diagrams}. To make progress in either gauge, it is therefore necessary to find a way of analytically computing the relevant exchange diagrams. (By contrast, for the relevant six-point contact diagrams, one can in principle use the approach described in \cite{Bliard:2022xsm}). The results in this paper can hopefully serve as a stepping stone in that direction.

Another natural direction one can explore further would be to consider six point OTOC configurations besides the one in \eqref{eq:config 1 ds limit}. For example, another double scaling limit that one could take is given by
\begin{align}\label{eq:config 2 ds limit}
    T_{13}\to \infty,\quad &T_{23}\to -\infty &T_s&\to \infty, &\frac{e^{T_{13}}}{T_s},\frac{e^{-T_{23}}}{T_s}:\text{ fixed}.
\end{align}
This can be achieved, for instance, by keeping the Lorentzian time of $W$ in \eqref{eq:6 pt OTOC} fixed $(T_3=0)$ while sending $U$ to late times ($T_1\to \infty$) and $V$ to early times ($T_2\to-\infty$). The OTOC in this configuration can be interpreted as a $W$ two-point function, on a background with two crossing or colliding shockwaves, with one left-moving shockwave produced by $V$ and one right-moving shockwave produced by $U$. This configuration and its shockwave interpretation was studied, in the context of JT gravity and Einstein gravity in AdS$_3$, in \cite{Haehl:2021tft}. However, making this interpretation of the OTOC precise in the context of the string correlators is slightly subtle. This is because a gravitational shockwave is by definition a spacetime geometry with a discontinuity that arises due to the backreaction to a very high energy wave, but the critical string worldsheet does not have gravity in the usual sense and the notion of ``backreaction" is a priori unclear. Instead, as discussed in \cite{Murata:2017rbp} in the case of a string worldsheet with a single shockwave, one can add ``shockwaves'' to the string by creating a sharp kink propagating along the string by suddenly changing the acceleration of the boundary of the string. Thus, ``shockwaves'' on the AdS$_2$ string are related to the ``broken'' or ``segmented'' strings studied in \cite{Callebaut:2015fsa,Vegh:2015ska,Vegh:2015yua}. We hope to revisit these geometries, and their relation to higher point OTOCs, in the near future.

\section*{Acknowledgements}

The work of SG, BO and JS is supported in part by the US NSF under Grant No.~PHY-2209997. SG is grateful to the Mainz Institute for Theoretical Physics (MITP) of the Cluster of Excellence PRISMA* (Project ID 39083149) for its hospitality and its partial support during completion of this work.

\appendix 

\section{Reparametrization mode three-point interaction on the circle}\label{app:rep mode 3-pt function circle}
In this section, we rederive the result for the three-point interaction of the reparametrization modes given in \eqref{eq:<eps(x_1)eps(x_2)eps(x_3)S_{L,3}> 2}, working with the reparametrization action on the circle rather than on the line. Doing the analysis on the circle allows us to handle the $SL(2,\mathbb{R})$ zero modes more explicitly. 

We will focus only on the contribution of the interacting diagrams on the circle, and will not rederive the full connected six-point function. Recall that we write the boundary mode on the circle as $\tilde{\alpha}(\tau)=\tau+\tilde{\epsilon}(\tau)$, where $\tilde{\epsilon}$ is a small perturbation around the saddle point that we can expand in Fourier modes: $\epsilon(\tau)=\sum_{n\in \mathbb{Z}}\tilde{\epsilon}_n e^{-in\tau}$. Here, $n=0,\pm 1$ are zero modes of the action that need to be gauge fixed in the reparametrization path integral.

The two ingredients we need to study the three-point reparametrization interaction diagrams are the reparametrization action to cubic order in $\tilde{\epsilon}$ and the dressed two-point function to linear order. After adding a gauge-fixing term $S_{\rm gf}$, the full action to cubic order is:
\begin{align}\label{eq:S_L + S_gf + O(eps^4)}
    S_L&=S_{\rm gf}+S_{L,2}+S_{L,3}+O(\tilde{\epsilon}^4),
\end{align}
where $S_{L,2}$ and $S_{L,3}$ are given in \eqref{eq:S_L + O(eps^4) fourier circle} and we pick the gauge-fixing term to be:
\begin{align}
    S_{\rm gf}&=T_s\left[\frac{1}{2a}\tilde{\epsilon}_0^2+\frac{1}{b}\tilde{\epsilon}_1\tilde{\epsilon}_{-1}\right].
\end{align}
From the combined quadratic piece $S_{\rm gf}+S_{L,2}$, we read off the Wick contraction of the Fourier modes:
\begin{align}\label{eq:<eps_m eps_n>}
    \braket{\tilde{\epsilon}_m\tilde{\epsilon}_n}_0&=f_m\delta_{m+n,0},&f_m&=\frac{1}{T_s}\left\{\begin{array}{cc} a& m=0\\ b& m=\pm 1\\ \frac{1}{4\pi}\frac{1}{|n|(n^2-1)}& \text{otherwise}\end{array}\right.
\end{align}
A simple gauge choice is to set $a=b=0$, which amounts to excluding the $n=0,\pm 1$ modes from $\tilde{\epsilon}(\tau)$, but we can keep $a$ and $b$ general. The $\epsilon$ position space propagator is then:
\begin{align}
    \braket{\epsilon(\tau_1)\epsilon(\tau_2)}_0&=\sum_{n\in \mathbb{Z}}f_n e^{-in\tau_{12}}=\frac{1}{T_s}\left[a+2b\cos{\tau_{12}}+\frac{1}{4\pi }\sum_{n\neq 0,\pm 1}\frac{e^{-in\tau_{12}}}{|n|(n^2-1)}\right]\label{eq:rtfdrtre}\\&=\frac{1}{T_s}\left[a-\frac{1}{4\pi}+\left(2b+\frac{3}{8\pi}\right)\cos(\tau_{12})+\frac{1}{2\pi}\sin^2\left(\frac{\tau_{12}}{2}\right)\log\left(4\sin^2\left(\frac{\tau_{12}}{2}\right)\right)\right].
\end{align}
Note that (up to a relabelling of $a$ and $b$), this is related to \eqref{eq:<eps(t_1)eps(t_2)> 2} by the usual interchange of euclidean distances on the line and chordal distances on the circle. 

Meanwhile, the dressed-two point function on the circle $B^{\rm circle}(\theta_1,\theta_2)$ is given in \eqref{eq: B on line and circle}, where $\tilde{\beta}$ is the inverse of $\tilde{\alpha}$, which to linear order in $\tilde{\epsilon}$ is $\tilde{\beta}(\theta)=\theta-\tilde{\epsilon}(\theta)+O(\tilde{\epsilon}^2)$. The expansion of the normalized dressed two-point function is then:
\begin{align}
    \mathcal{B}(\theta_1,\theta_2)&\equiv \bigg[2\sin{\frac{\theta_{12}}{2}}\bigg]^2B^{\rm circle}(\theta_1,\theta_2)=1+\mathcal{B}_1(\theta_1,\theta_2)+O(\tilde{\epsilon}^2),
\end{align}
where the linear piece is explicitly
\begin{align}\label{eq:B1 circle}
    \mathcal{B}_1(\theta_1,\theta_2)&=-\dot{\tilde{\epsilon}}_1-\dot{\tilde{\epsilon}}_2+\cot{\frac{\theta_{12}}{2}}\tilde{\epsilon}_{12},
\end{align}
where we use the shorthand $\dot{\tilde{\epsilon}}_n=\dot{\tilde{\epsilon}}(\theta_n)$ and $\tilde{\epsilon}_{ij}=\tilde{\epsilon}(\theta_i)-\tilde{\epsilon}(\theta_j)$. Knowing $\mathcal{B}_1$ is sufficient to study the contribution of the three-point diagrams to the connected six-point function, which in analogy with \eqref{eq:dfy52134rft5} is given by:
\begin{align}\label{eq:gf34543werfgrew}
    -\braket{\mathcal{B}_1(\theta_1,\theta_2)\mathcal{B}_1(\theta_3,\theta_4)\mathcal{B}_1(\theta_5,\theta_6)S_{L,3}}_{0,c},
\end{align}
Here ``c'' stands for ``connected'' and tells us to contract the $\tilde{\epsilon}$'s in the $\mathcal{B}_1$'s with the three $\tilde{\epsilon}$'s in the cubic action. To compute the full connected six-point function, we would also need the quadratic piece $\mathcal{B}_2$.

We will now compute the three-point interacting diagram:
\begin{align}\label{eq:gfrty654esdcg}
    \braket{\tilde{\epsilon}(\theta_1)\tilde{\epsilon}(\theta_2)\tilde{\epsilon}(\theta_3)S_{L,3}}_{0,\text{c}}=\sum_{a,b,c\in \mathbb{Z}}e^{-ia\theta_1-ib\theta_2-ic\theta_3}\braket{\tilde{\epsilon}_a\tilde{\epsilon}_b\tilde{\epsilon}_cS_{L,3}}_{0,\text{c}}
\end{align}
Given the expression for the cubic action in \eqref{eq:S_L + O(eps^4) fourier circle} and using the propagator in \eqref{eq:<eps_m eps_n>} to Wick contract the different Fourier modes, we have
\begin{align}
    \braket{\tilde{\epsilon}_a\tilde{\epsilon}_b\tilde{\epsilon}_cS_{L,3}}_{0,\text{c}}&=-2i\pi T_s\sum_{m,n\in \mathbb{Z}}\text{sgn}(p)p^2(p^2-1)\big[\braket{\tilde{\epsilon}_a\tilde{\epsilon}_m}_0 \braket{\tilde{\epsilon}_b\tilde{\epsilon}_n}_0\braket{\tilde{\epsilon}_c\tilde{\epsilon}_p}_0+a\leftrightarrow c+b\leftrightarrow c\big]\\&=-2i \pi T_s\sum_{m,n\in \mathbb{Z}}\text{sgn}(p)p^2(p^2-1)\big[f_m f_n f_p \delta_{m+a,0}\delta_{n+b,0}\delta_{p+c,0}+a\leftrightarrow c+b\leftrightarrow c\big]\nonumber
\end{align}
(Recall that $p=-m-n$). Substituting this into \eqref{eq:gfrty654esdcg} and using the three Kronecker deltas to fix $a$, $b$ and $c$, we find:
\begin{align}
    \braket{\tilde{\epsilon}(\theta_1)\tilde{\epsilon}(\theta_2)\tilde{\epsilon}(\theta_3)S_{L,3}}_{0,\text{c}}&=-2i \pi T_s \sum_{m,n\in \mathbb{Z}}\text{sgn}(p)p^2(p^2-1)f_m f_n f_p \left[e^{im\theta_{13}+in\theta_{23}}+\theta_1\leftrightarrow \theta_2+\theta_2\leftrightarrow \theta_3\right]\nonumber\\&=-\frac{i}{2}\sum_{m,n\in \mathbb{Z}}f_m f_n p(1-\delta_{p,1}-\delta_{p,-1}) \left[e^{im\theta_{13}+in\theta_{23}}+\theta_1\leftrightarrow \theta_3+\theta_2\leftrightarrow \theta_3\right].\label{eq:tg45rtrew343}
\end{align}
To get to the second line, we used $f_p\text{sgn}(p)p^2=\frac{p}{4\pi T_s}(1-\delta_{p,1}-\delta_{p,-1})$, which follows from \eqref{eq:<eps_m eps_n>}. Let's focus on one term inside the square bracket in \eqref{eq:tg45rtrew343} at a time. There are three different contributions to the sum over $m$ and $n$, corresponding to the ``1,'' the ``$-\delta_{p,1}$'' and the ``$-\delta_{p,-1}$'' in $1-\delta_{p,1}-\delta_{p,-1}$. The first is the simplest:
\begin{align}\label{eq:tfdsert6543wsdft}
    \frac{i}{2}\sum_{m,n\in \mathbb{Z}}f_m f_n (m+n) e^{-im\theta_{13}-in\theta_{23}}&=-\frac{1}{2}\partial_3\biggr[\sum_m f_m e^{im\theta_{13}}\sum_n f_n e^{in\theta_{23}}\biggr],\nonumber\\&=-\frac{1}{2}\partial_3[\braket{\tilde{\epsilon}(\theta_1)\tilde{\epsilon}(\theta_3)}_0\braket{\tilde{\epsilon}(\theta_2)\tilde{\epsilon}(\theta_3)}_0].
\end{align}
To get to the second line we used \eqref{eq:rtfdrtre}. In direct analogy with \eqref{eq:<eps(x_1)eps(x_2)eps(x_3)S_{L,3}> 2}, we see that the double sum breaks into a product of single sums. This implies that the three-point interaction diagram effectively ``factorizes'' in terms of products of the $\epsilon$ propagator, as long as the contributions from the ``$-\delta_{p,1}$'' and the ``$-\delta_{p,-1}$'' to \eqref{eq:tg45rtrew343} can be neglected. We now show that is indeed the case.

We need to compute:
\begin{align}
    \sum_{m,n\in \mathbb{Z}}f_m f_n p \delta_{p,1} e^{im\theta_{13}+in\theta_{23}}=e^{-i\theta_{23}}\sum_{m\in \mathbb{Z}}f_m f_{-m-1}e^{im\theta_{12}}\label{eq:sum1}\\
    \sum_{m,n\in \mathbb{Z}}f_m f_n p \delta_{p,-1} e^{im\theta_{13}+in\theta_{23}}=-e^{i\theta_{23}}\sum_{m\in \mathbb{Z}}f_m f_{-m+1}e^{im\theta_{12}}\label{eq:sum2}
\end{align}
Given that $f_m=f_{-m}$, \eqref{eq:sum2} is minus the complex conjugate of \eqref{eq:sum1}. The last piece that we need to evaluate is then:
\begin{align}
    k(\theta)\equiv \sum_m f_m f_{-m-1}e^{im\theta}=\frac{ab}{T_s^2}\left(1+e^{-i\theta}\right)&+\frac{b}{24\pi T_s^2}(e^{i\theta}+e^{-2i\theta})\label{eq:fdwe43w24}\\&+\frac{1}{16\pi^2 T_s^2}\sum_{m\neq -2,-1,0,1}\frac{e^{im\theta}}{|m||m+1|m(m^2-1)(m+2)}.\nonumber
\end{align}
The remaining sum in \eqref{eq:fdwe43w24} can be evaluated and simplifies to:
\begin{align}\label{eq:MhfBB5Y8ec}
    \sum_{m\neq -2,-1,0,1}\frac{e^{im\theta}}{|m||m+1|m(m^2-1)(m+2)}=\frac{e^{-\frac{i\theta}{2}}}{36}\biggr(&3(33-4\pi^2+12\pi |\theta|-6\theta^2)\cos{\frac{\theta}{2}}+20\cos{\frac{3\theta}{2}}\nonumber\\&+6(|\theta|-\pi)(9\sin{\frac{|\theta|}{2}}+\sin{\frac{3|\theta|}{2}})\biggr).
\end{align}
Combining \eqref{eq:tfdsert6543wsdft}-\eqref{eq:fdwe43w24}, eq.~\eqref{eq:tg45rtrew343} becomes
\begin{align}\label{eq:yTTKleY8SG}
    \braket{\tilde{\epsilon}(\theta_1)\tilde{\epsilon}(\theta_2)\tilde{\epsilon}(\theta_3)S_{L,3}}_{0,\text{c}}&=-\frac{1}{2}\partial_{\theta_1}\left[\braket{\tilde{\epsilon}(\theta_1)\tilde{\epsilon}(\theta_2)}\braket{\tilde{\epsilon}(\theta_2)\tilde{\epsilon}(\theta_3)}\right]+1\leftrightarrow 2+1\leftrightarrow 3 +R (\theta_1,\theta_2,\theta_3)
\end{align}
where 
\begin{align}
    R(\theta_1,\theta_2,\theta_3)&=\frac{i}{2}\left(e^{-i\theta_{23}}k(\theta_{12})-e^{i\theta_{23}}k(\theta_{12})^*\right)+1\leftrightarrow 2+1\leftrightarrow 3.
\end{align}
Eq.~\eqref{eq:yTTKleY8SG} almost matches eq.~\eqref{eq:<eps(x_1)eps(x_2)eps(x_3)S_{L,3}> 2}, which we derived working on the line and using analytic regularization to handle the zero modes, except that \eqref{eq:yTTKleY8SG} has the extra term $R(\theta_1,\theta_2,\theta_3)$. This term is not zero, but we claim that it does not contribute to \eqref{eq:gf34543werfgrew}, and therefore can be neglected in the computation of the connected six-point function on the circle. Concretely, the claim is that if we expand each $\mathcal{B}_1$ in \eqref{eq:gf34543werfgrew} into four $\tilde{\epsilon}$ terms according to \eqref{eq:B1 circle}, apply \eqref{eq:yTTKleY8SG} to the $64$ different terms, and then sum up all the terms, the contribution from $R$ is zero. This appears tedious to check analytically, but we confirmed it numerically within machine precision for a random sample of values for $\theta_1,\ldots,\theta_6$. 

The rest of the computation of the connected six-point function on the circle proceeds in direct analogy with the computation on the line presented in section~\ref{sec:connected six-point function}. The final result--- after mapping chordal distances to euclidean distances--- matches \eqref{eq:6-pt function result}.

\section{A conformal bilocal reparametrization action}\label{app:conf bilocal rep action}
In this appendix, we give a closed form expression for a bilocal reparametrization action that has the same symmetries as the AdS$_2$ string reparametrization action defined in \eqref{eq:longitudinal action extremization}. Curiously, the action also precisely matches the string reparametrization action at quadratic and cubic order in the saddle point expansion (but apparently not at higher orders; see footnote~\ref{fn:bilocal action}). 

Consider the following bilocal reparametrization action:
\begin{align}\label{eq:bilocal action}
    S_{\rm bilocal}[\alpha]&=-C\int_{-\infty}^\infty dt_1 dt_2 \left[\left[\frac{\dot{\alpha}(t_1) \dot{\alpha}(t_2)}{(\alpha(t_1)-\alpha(t_2))^2}\right]^h \frac{1}{|t_{12}|^{2-2h+\eta}}-\frac{1}{|t_{12}|^{2+\eta}}\right].
\end{align}
Here, $\alpha(t)$ is a reparametrization of the line, $h$ and $\eta$ are real parameters, and $C$ is a coupling constant with mass dimension $[C]=\eta$. This action was studied for the case $2-2h+\eta=0$ in \cite{Milekhin:2021cou,Milekhin:2021sqd}, and discussed for the case $\eta=0$ towards the end of appendix G of \cite{Maldacena:2016hyu}. We have added a $-1/|t_{12}|^{2+\eta}$ counterterm to make the action finite in the coincident limit $t_2\to t_1$ as long as $\eta<1$;\footnote{The finiteness of \eqref{eq:bilocal action} follows from the fact that in the coincident limit 
\begin{align}
    \frac{\dot{\alpha}(t_1)\dot{\alpha}(t_2)}{(\alpha(t_1)-\alpha(t_2))^2}&=\frac{1}{t_{12}^2}+\frac{1}{6}\{\alpha,t_1\}+O(t_1-t_2),&t_2\to t_1,
\end{align}
where $\{\alpha,t\}=-\frac{3}{2}\frac{\ddot{\alpha}^2}{\dot{\alpha}^2}+\frac{\dddot{\alpha}}{\dot{\alpha}}$ is the Schwarzian derivative.} the counterterm is zero in analytic regularization.

We are interested in \eqref{eq:bilocal action} when $\eta\to 0$. In this special case, the action has two $SL(2,\mathbb{R})$ symmetries, just like the string reparametrization action in \eqref{eq:longitudinal action extremization}, and it is therefore ``conformal.'' To see this, let $f(t)=\frac{at+b}{ct+d}$ denote a general $SL(2,\mathbb{R})$ transformation and let $\alpha_r(t)=\alpha(f(t))$ and $\alpha_\ell(t)=f(\alpha(t))$. Given that
\begin{align}
    \frac{\dot{\alpha}_\ell(t_1)\dot{\alpha}_\ell(t_2)}{(\alpha_\ell(t_1)-\alpha_\ell(t_2))^2}&=\frac{\dot{\alpha}(t_1)\dot{\alpha}(t_2)}{(\alpha(t_1)\alpha(t_2))^2},\label{eq:7Wlog1hLrA}\\
    \frac{\dot{\alpha}_r(t_1)\dot{\alpha}_r(t_2)}{(\alpha_r(t_1)-\alpha_r(t_2))^2}&=\dot{f}(t_1)\dot{f}(t_2)\frac{\dot{\alpha}(f(t_1))\dot{\alpha}(f(t_2))}{(\alpha(f(t_1))-\alpha(f(t_2)))^2},\\\frac{\dot{f}(t_1)\dot{f}(t_2)}{(f(t_1)-f(t_2))^2}&=\frac{1}{t_{12}^2},\label{eq:lqOvQ3ckCM}
\end{align}
one can readily check that $S_{\rm bilocal}$ is invariant under $\alpha\to \alpha_\ell$ for any $\eta$ and invariant under $\alpha\to \alpha_r$ when $\eta\to 0$.

Now consider the expansion of the action in \eqref{eq:bilocal action} about the saddle point by letting $\alpha(t)=t+\epsilon(t)$ and expanding in $\epsilon$. Using \eqref{eq:analytic reg. integral line 1} and integration by parts to simplify the result, we find 
\begin{align}\label{eq:bilocal O(eps^4)}
    S_{\rm bilocal}&=2C h(1-h)\bigg[\int dt_1 dt_2 \frac{\epsilon(t_1)\epsilon(t_2)}{|t_{12}|^{4+\eta}}-2\int dt_1 dt_2 \frac{\epsilon(t_1)^2\epsilon(t_2)}{|t_{12}|^{4+\eta}t_{12}}\nonumber\\&\quad +\frac{1}{2}\int dt_1 dt_2 \biggr((1+2h)(3-2h)\frac{\frac{4}{3}\epsilon(t_1)^3\epsilon(t_2)-\epsilon(t_1)^2\epsilon(t_2)^2}{|t_{12}|^{6+\eta}}\nonumber\\&\quad+h(1-h)\frac{-2\epsilon(t_1)\dot{\epsilon}(t_1)^2\epsilon(t_2)+\dot{\epsilon}(t_1)^2\epsilon(t_2)^2-\frac{1}{4}t_{12}^2\dot{\epsilon}(t_1)^2\dot{\epsilon}(t_2)^2}{|t_{12}|^{4+\eta}}\biggr)+O(\epsilon^5)\biggr].
\end{align}
To put this in the Fourier representation, we write $\epsilon(t)=\int \frac{d\omega}{2\pi}e^{-i \omega t}\epsilon(\omega)$ and evaluate the $t_1$ and $t_2$ integrals using \eqref{eq:analytic reg. integral line 2}-\eqref{eq:analytic reg. integral line 3}. This leads to:
\begin{align}
    S_{\rm bilocal}&=\frac{Ch(1-h)}{6}\biggr[\int d\omega\epsilon(\omega)\epsilon(-\omega)|\omega|^3-\frac{i}{4\pi}\int d\omega_1 d\omega_2 \epsilon(\omega_1)\epsilon(\omega_2)\epsilon(-\omega_1-\omega_2)|\omega_2|^4 \text{sgn}(\omega_2)\nonumber\\&-\pi\left(\int \prod_{i=1}^4 \frac{d\omega_i}{2\pi}\epsilon(\omega_i)\right) 2\pi \delta\big(\sum_{i=1}^4 \omega_i\big)\biggr[\frac{(1+2h)(3-2h)}{20}\big(\frac{4|\omega_4|^5}{3}-|\omega_3+\omega_4|^5\big)\nonumber\\&+ h(1-h)\bigg(-2|\omega_4|^3\omega_1\omega_2+|\omega_3+\omega_4|^3\omega_1\omega_2-\frac{3}{2}|\omega_3+\omega_4|\omega_1\omega_2\omega_3\omega_4\bigg)\biggr]+O(\epsilon^5)\biggr]
\end{align}
As promised, the action in \eqref{eq:bilocal O(eps^4)} precisely matches the string reparametrization action to cubic order given in \eqref{eq:quadratic rep action} and \eqref{eq:reparametrization action to cubic order}, as long as we set $C=\frac{3T_s}{\pi h(1-h)}$. 

The analog of \eqref{eq:bilocal action} on the circle (with the counterterm set to zero for simplicity) is:
\begin{align}\label{eq:bilocal action circle}
    S_{\rm bilocal}[\alpha]&=-C\int_{-\pi}^\pi d\tau_1 d\tau_2 \bigg[\frac{\dot{\alpha}(\tau_1) \dot{\alpha}(\tau_2)}{\big[2\sin{\frac{\alpha(\tau_1)-\alpha(\tau_2)}{2}}\big]^2}\bigg]^h \frac{1}{\left|2\sin{\frac{\tau_{12}}{2}}\right|^{2-2h+\eta}}.
\end{align}
If we set $\alpha(\tau)=\tau+\tilde{\epsilon}(\tau)$, expand in $\tilde{\epsilon}$, and use \eqref{eq:analytic reg. integral circle 1} and integration by parts to simplify terms, we again find that the bilocal action on the circle matches the string reparametrization action given in \eqref{eq:SL circle cubic order} to cubic order.

\paragraph{Conformal $k$-local conformal reparametrization.} Note that \eqref{eq:bilocal action} is constructed by multiplying a reparametrization-dressed conformal two-point function with an undressed two-point function. It is easy to generalize \eqref{eq:bilocal action} and write down a $k$-local action invariant under two $SL(2,\mathbb{R})$ symmetries starting from a pair of conformal $k$-point functions. In particular, let
\begin{align}
    G(t_1,\ldots,t_k)=\braket{O_{h_1}(t_1)\ldots O_{h_k}(t_k)},
\end{align}
be any correlator in a 1d CFT of $k$ operators $O_{h_i}$ with conformal dimensions $h_i$, and let
\begin{align}
    \bar{G}(t_1,\ldots,t_k)=\braket{\bar{O}_{1-h_1}(t_1)\ldots \bar{O}_{1-h_k}(t_k)},
\end{align}
be any correlator in a 1d CFT of $k$ operators $\bar{O}_{1-h_i}$ with conformal dimensions $1-h_i$. Then we define the reparametrization action
\begin{align}\label{eq:S k-local}
    S_{k-\text{local}}[\alpha(t)]&=C\int \prod_{i=1}^k dt_i \prod_{i=1}^k \dot{\alpha}(t_i)^{h_i}G(\alpha(t_1),\ldots,\alpha(t_n))\bar{G}(t_1,\ldots,t_n).
\end{align}
This action is also invariant under both $\alpha(t)\to f(\alpha(t))$ and $\alpha(t)\to \alpha(f(t))$, where $f$ is a general $SL(2,\mathbb{R})$ transformation. This follows from \eqref{eq:7Wlog1hLrA}-\eqref{eq:lqOvQ3ckCM} together with the conformal Ward identity, which says that
\begin{align}
    G(t_1,\ldots,t_k)&=\prod_{i=1}^k \dot{f}(t_i)^{h_i}G(f(t_1),\ldots,f(t_k)),
\end{align}
and with an analogous statement for $\bar{G}$.

Within this construction, \eqref{eq:S k-local} is unique to the extent that the $k$-pt function in 1d CFT is unique. For instance, the bilocal action is unique up to normalization $C$ and choice of conformal dimension $h$, and the $3$-local action is unique up to normalization and choice of three conformal dimensions $h_1$, $h_2$, $h_3$.

\section{Useful integrals for analytic regularization}\label{app:useful integrals}
In analytic regularization, the following identities hold on the line:
\begin{align}
    \int_{-\infty}^\infty dt \frac{1}{|t|^\Delta}&=0,\label{eq:analytic reg. integral line 1}
    \\\int_{-\infty}^\infty dt \frac{e^{i\omega t}}{|t|^\Delta}&=2\sin\left(\frac{\pi}{2}\Delta\right)\Gamma(1-\Delta)|\omega|^{\Delta-1},\label{eq:analytic reg. integral line 2}\\
    \int_{-\infty}^\infty dt \frac{e^{i\omega t}}{|t|^\Delta t}&=-2i\sin\left(\frac{\pi}{2}\Delta\right)\Gamma(-\Delta)\text{sgn}(\omega)|\omega|^{\Delta}\label{eq:analytic reg. integral line 3}.
\end{align}
(The third identity is essentially the derivative of the second identity with respect to $\omega$). Likewise, the following identities hold on the circle:
\begin{align}
    \int_{-\pi}^\pi d\tau \frac{1}{|2\sin{\frac{\tau}{2}}|^\Delta}&=2^{1-\Delta}\sqrt{\pi}\frac{\Gamma\left(\frac{1-\Delta}{2}\right)}{\Gamma\left(1-\frac{\Delta}{2}\right)},\label{eq:analytic reg. integral circle 1}\\
    \int_{-\pi}^\pi d\tau \frac{e^{in\tau}}{|2\sin{\frac{\tau}{2}}|^\Delta}&=\frac{2\pi (-1)^n \Gamma(1-\Delta)}{\Gamma(1-n-\frac{\Delta}{2})\Gamma(1+n-\frac{\Delta}{2})},&\forall n\in \mathbb{Z}.\label{eq:analytic reg. integral circle 2}
\end{align}

\bibliographystyle{ssg}
\bibliography{mybib}

\end{document}